\documentclass[epj-spec]{svjour}

\usepackage{graphics,epsfig,psfrag,mathbbol,verbatim}
\usepackage{my_mcite}

\def\tr{{\rm tr}}

\newcommand{\ve}[1]{\ensuremath{\mbox{\boldmath{$#1$}}}}

\newcommand\fra[2]{\frac{#1}{#2}\,}

\begin{document}

\title{Topological objects in QCD}

\author{
Falk Bruckmann\thanks{\email{falk.bruckmann@physik.uni-regensburg.de}} }

\institute{Universit\"at Regensburg, Institut f\"ur Theoretische Physik, D-93040 Regensburg}

\abstract{
Topological excitations are prominent candidates for 
explaining nonperturbative effects in QCD like confinement. 
In these lectures, I cover both formal
treatments and applications of topological objects. 
The typical phenomena like BPS bounds, topology, the
semiclassical approximation 
and chiral fermions are introduced by virtue of kinks. 
Then I proceed in higher dimensions with magnetic monopoles and instantons 
and special emphasis on calorons. 
Analytical aspects are discussed and an overview over models based on these objects 
as well as lattice results is given. 
}

\maketitle

\section{Appetiser}
\label{sect_appetizer}

\enlargethispage{\baselineskip}

\begin{figure}[h]
\includegraphics[width=0.3\linewidth]{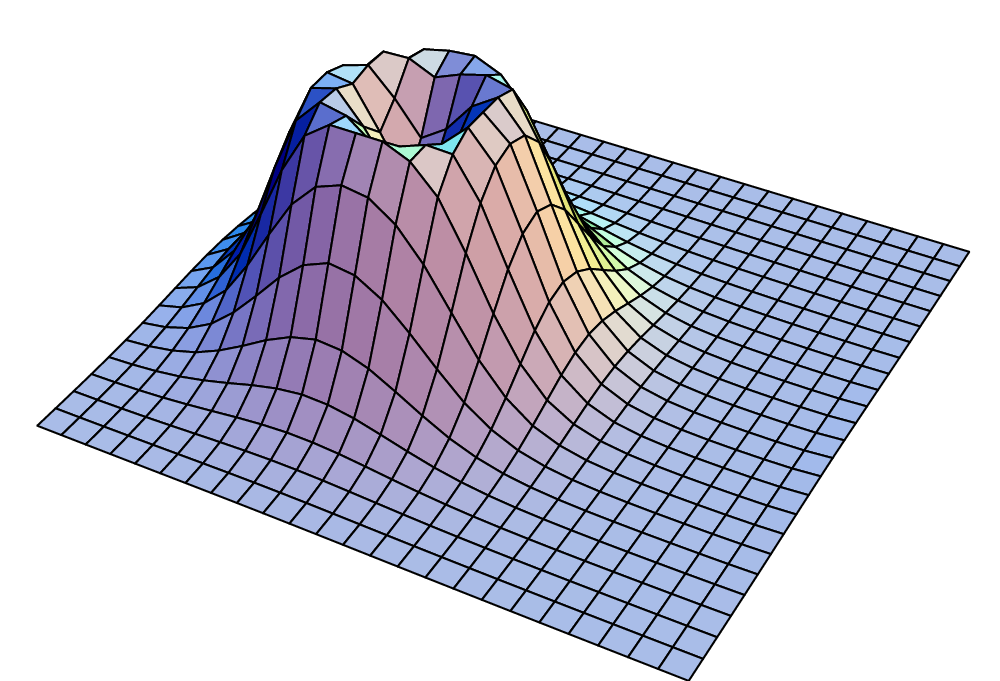}
\includegraphics[width=0.3\linewidth]{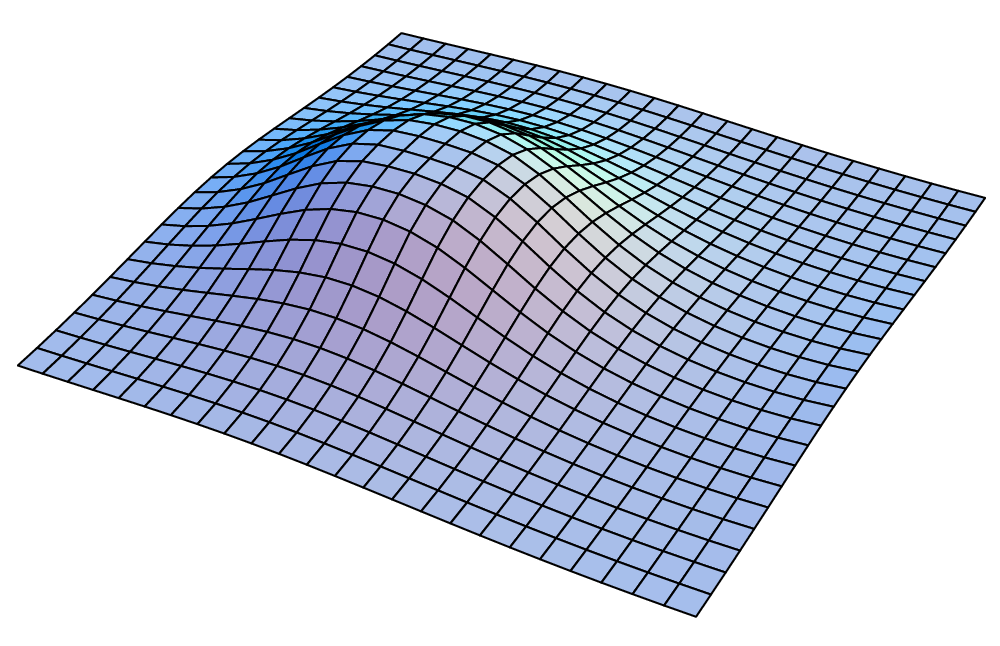}
\includegraphics[width=0.3\linewidth]{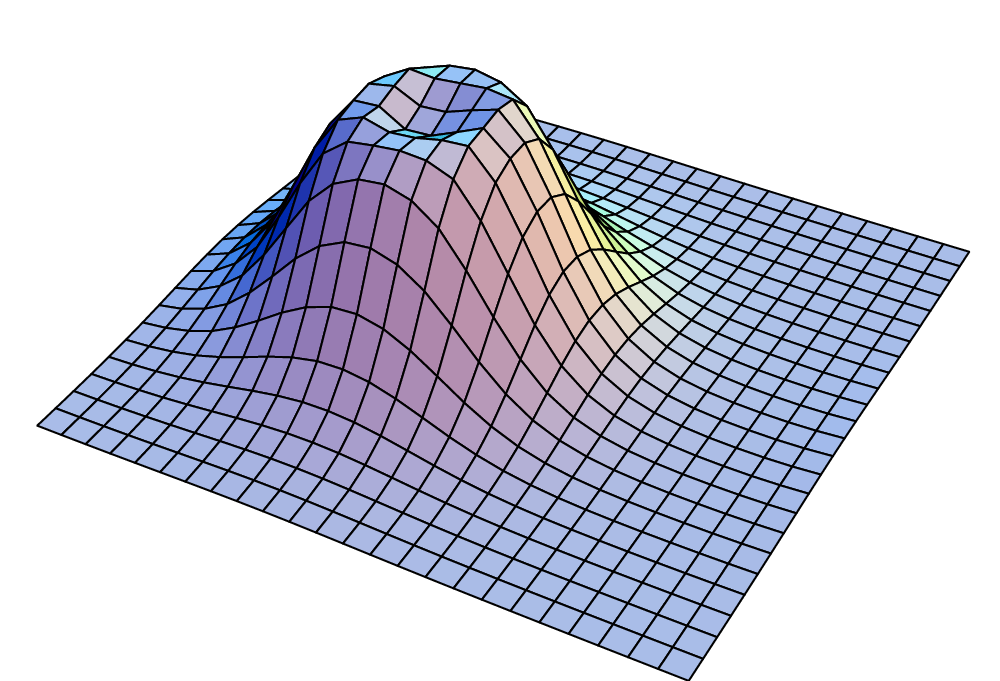}
\caption{Action density, Polyakov loop and zero mode profile 
of an $SU(2)$ configuration on an asymmetric lattice after long over-improved cooling
in some lattice plane, from \protect\cite{bruckmann:04b},
cf. Sect.~\protect\ref{sect_cooling} and Fig.~\protect\ref{fig_cal_higher}.}
\label{fig_appetizer}
\end{figure}

When a lattice gauge configuration is subject to cooling, 
the result can be as depicted in Fig.~\ref{fig_appetizer}. 
Without going into details here, 
this figure reveals typical aspects of 
a {\em soliton stabilised by topology}. 
In this lecture I will introduce definitions and properties of these
beautiful objects and discuss their relevance for particle physics. 
Before I come to gauge 
objects like monopoles and instantons,
I will demonstrate the main features by virtue of an
example in a scalar theory. 

\section{The kink}
\label{sect_kink}

\begin{figure}[t]
\begin{center}
\psfrag{mv}{$-v$}
\psfrag{v}{$v$}
\psfrag{Vp}{$V(\phi)$}
\psfrag{p}{$\phi$}
\includegraphics[width=0.45\linewidth]{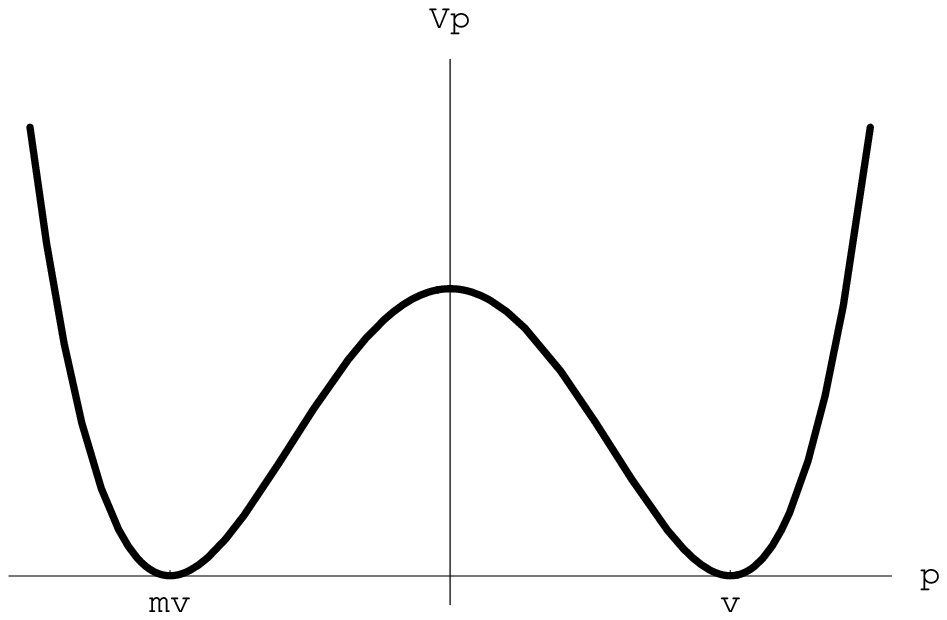}
\psfrag{mv}{$\!-v$}
\psfrag{v}{$v$}
\psfrag{Vp}{$\bar{V}(x)$}
\psfrag{p}{$x$}
\includegraphics[width=0.45\linewidth]{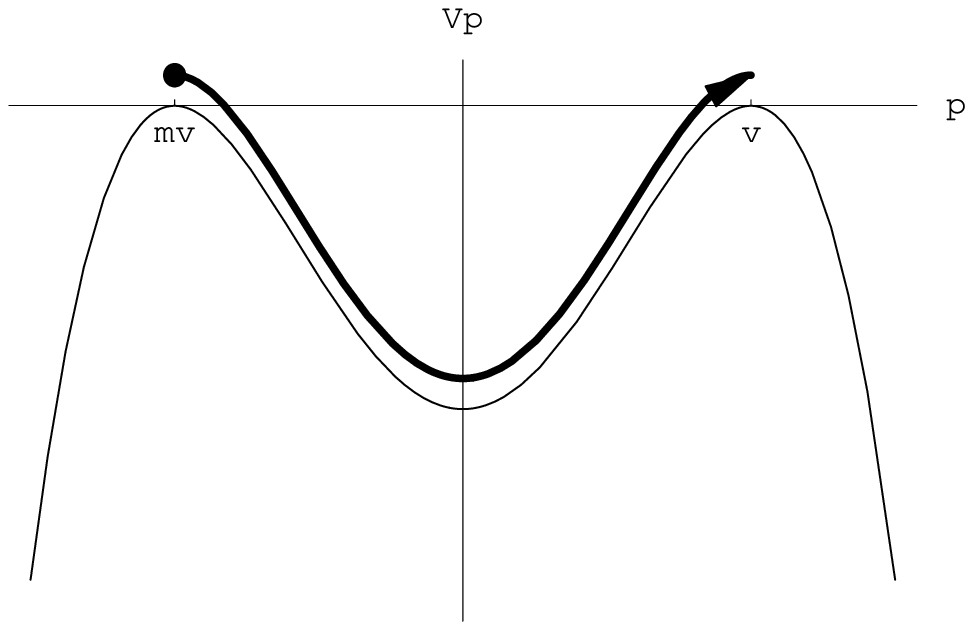}
\caption{Left: the mexican hat potential. 
Right: In the particle picture the potential is inverted.
The soliton `rolling' from the hill at $-v$ in the infinite past
to the hill at $v$ in the infinite future is depicted.}
\label{fig_mexican_hat_01}
\end{center}
\end{figure}

As the first model let us take one of the simplest quantum field theoretical systems, a real scalar field $\phi$ 
in a $1\!+\!1$ dimensional Minkowski space with metric $\eta_{\mu\nu}={\rm diag}(+1,-1)$.
The Lagrangian,
\begin{equation}
\mathcal{L}=\fra{1}{2}\partial_\mu\phi\,\partial^\mu\!\phi-V(\phi)\,,
\end{equation}
shall contain a potential $V(\phi)$ that has several minima of same height, set to $V=0$. 
For definiteness I will choose the famous mexican hat potential\,,
\begin{equation}
V(\phi)=\fra{\lambda}{4!}(\phi^2-v^2)^2\,,
\end{equation}   
see Fig.~\ref{fig_mexican_hat_01} left.
An alternative would be the so-called `sine-Gordon' model $V(\phi)\simeq(1-\cos\phi)$ 
which is periodic and thus has infinitely many minima.

The mexican hat potential contains the well-known $\phi^4$ interaction,
whereas its mass term comes with a negative sign
rendering $\phi=0$ unstable.
Obviously, this potential also has two stable vacua at $\phi=\pm v$ 
($v$ stands for vacuum expectation value, see below) with second derivatives
\begin{equation}
V''(\phi=\pm v)=\fra{\lambda}{3}v^2\equiv m^2\,.
\end{equation} 
$m$ is the mass of perturbative excitations.

In addition, tunnelling occurs as a typical nonperturbative effect.
Related to this is {\em the existence of a static solution of the Euclidean equations of motion with finite action 
connecting the vacua}, which I will derive now.  

First of all, the transition to Euclidean space is performed by going to imaginary time
\begin{equation}
\mathcal{L}_E(x_1,x_2)\equiv -\mathcal{L}(x_0=ix_2,x_1)=
\fra{1}{2}(\partial_{x_2}\phi)^2+
\fra{1}{2}(\partial_{x}\phi)^2+V(\phi)\,,\qquad 
x\equiv x_1\,.
\label{eqn_kink_L_E}
\end{equation}
This Lagrangian is very similar to a Hamiltonian, its first term will vanish for static solutions.
Steadily moving solutions can be obtained easily by Lorentz boosting the static one.

The positivity of the Lagrangian implies that configurations with finite action must have the potential term vanish asymptotically, 
i.e. the field has to go to one of the vacua:
\begin{equation}
\phi(x=\pm\infty)\in\{\pm v\}\,.
\end{equation}  

\subsection{Particle mechanics analogy}

The Lagrangian (\ref{eqn_kink_L_E}) 
without the first term reminds 
of the one familiar in 
particle mechanics,
\begin{equation}
\bar{L}(x(t))=\fra{1}{2}\dot{x}^2-\bar{V}(x)\,,\qquad
\bar{V}=-V\,,
\end{equation} 
upon substituting $\phi(x)\to x(t)$.
The bar $\bar{..}$  denotes quantities in the particle picture
(and I have set the mass in the kinetic term to unity). 
Notice that for the analogy to work, 
{\em the particle moves in the inverted potential}. 
In particular, the region between the vacua becomes a classically allowed one.  
The boundary conditions translate into
$
x(t=\pm\infty)\in\{\pm v\}
$.

It is clear that this system has two trivial solutions, 
where the particle stays at the hill $-v$ or $v$ forever,
plus {\em nontrivial solutions `rolling' from one hill to the other},
see Fig.~\ref{fig_mexican_hat_01} right.
We can use energy conservation to write 
\begin{equation}
\fra{1}{2}\dot{x}^2+\bar{V}(x)=\bar{E}=0\,,\qquad
\dot{x}=\pm\sqrt{2(-\bar{V})}\,,
\end{equation}
which enables us to give the solution explicitly.

\subsection{The explicit solution and its energy}

\begin{figure}[t]
\begin{center}
\psfrag{mvv}{$-v$}
\psfrag{v}{$v$}
\psfrag{p}{$\phi$}
\psfrag{x}{$x$}
\psfrag{y}{$y$}
\includegraphics[width=0.45\linewidth]{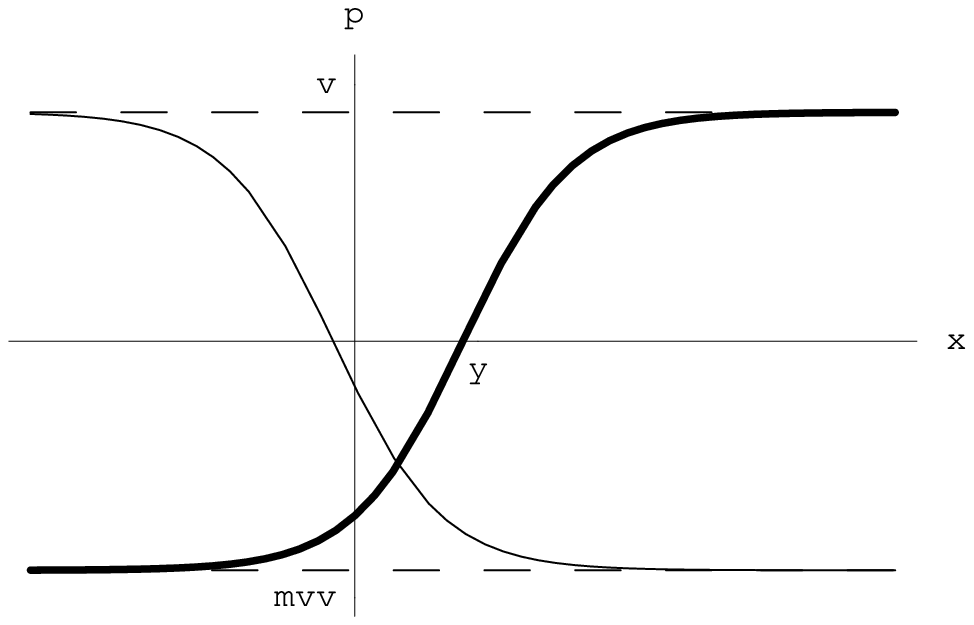}
\psfrag{x}{$x$}
\psfrag{y}{$y$}
\psfrag{E}{\hspace{-1cm}{\rm energy density}}
\includegraphics[width=0.45\linewidth]{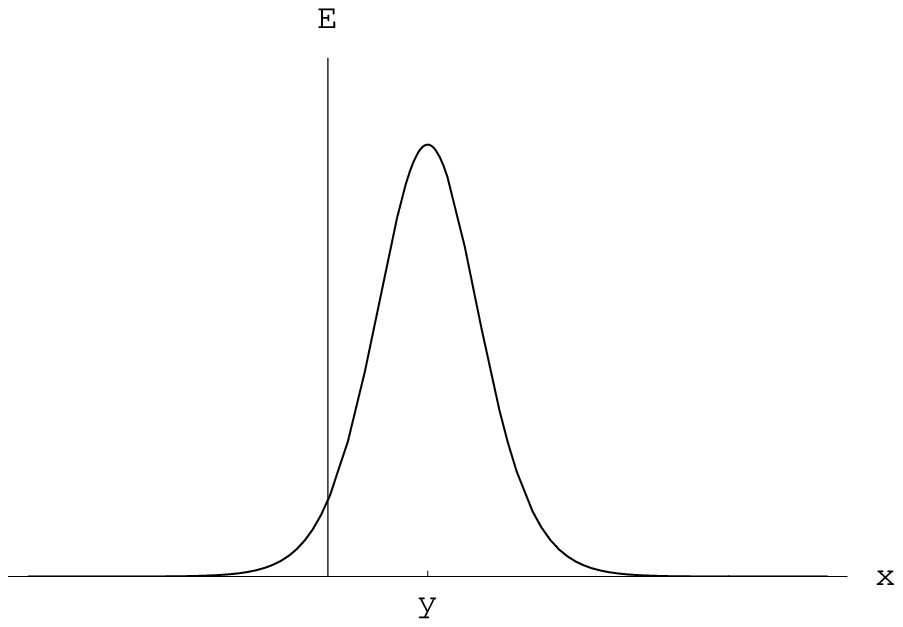}
\caption{The kink solution (left, fat curve, together with an antikink) 
and its energy density (right).}
\label{fig_kink}
\end{center}
\end{figure}

Going back to our original theory we can deduce that there exists a static solution 
that obeys
\begin{equation}
\partial_x\phi=\pm\sqrt{2V}=\pm \sqrt{\fra{\lambda}{12}} (v^2-\phi^2)\qquad
({\rm for}\: |\phi|\leq v)\,,
\label{eqn_kink_deqn}
\end{equation}
which is a {\em nonlinear} 
(but in contrast to the equations of motion only) first order differential equation.
It can actually be solved analytically to
\begin{equation}
\phi=\pm v \tanh(\fra{m}{2}(x-y))
\label{eq_kink_soln}
\end{equation}
and is plotted in Fig.~\ref{fig_kink} left.
The solution with plus sign evolves (in space) from $-v$ to $v$ 
and is named {\em kink} or {\em soliton}. 
Since it does not spread with time, 
it is also called {\em solitary wave}\footnote{
See \cite{rajaraman:82} for a very good account of physical criteria for the nomenclature.}.
Correspondingly, the solution with minus sign is the antikink or antisoliton.

Both approach the limiting values $\pm v$ exponentially 
with decay constant proportional to $1/m$.  
In other words, the transition takes place in a space region of finite size, 
this is why the kink can also be seen as a domain wall between different vacua.

Of particular interest is the parameter $y$. 
It comes about by solving the differential equation (\ref{eqn_kink_deqn})
and is due to the translational invariance of the system.
All solutions with different shift $y$ have the same action and energy.
The energy (or action) density (\ref{eqn_kink_L_E}) is maximal at $x=y$, 
both because the field $\phi$ is in the `false vacuum' $\phi=0$ 
and because it changes most there.
Since the energy density decays with the same constant $1/m$ as the field itself,
the kink is {\em localised} in space.

The calculation of the total energy is very instructive. Starting from
\begin{equation}
E=\int_{-\infty}^{\infty}\!\!\!\!\!dx\,[\fra{1}{2}(\partial_x\phi)^2+V]
\end{equation} 
and using the `virial theorem' Eq.~(\ref{eqn_kink_deqn}), one can write the integrand as $2V$
respectively $\partial_x\phi\,\sqrt{2V}$,
I stick to the kink for the moment. Hence
\begin{equation}
E=\int_{x=-\infty}^{x=\infty}\!\!\!\!\!dx\,\fra{\partial\phi}{\partial x}\sqrt{2V(\phi(x))}
=\int_{\phi=-v}^{\phi=v}\!\!\!\!\!d\phi\,\sqrt{2V(\phi)}\,.
\end{equation}
At this point one can recognise the WKB formula, which relates the transition amplitude 
to the integral over the square root of the potential between the turning points, 
thus the transition amplitude is given as $\exp(-E_{\rm kink})$. 
With the antiderivative $W$,
\begin{equation}
\fra{\partial W}{\partial\phi}=\sqrt{2V(\phi)}\,,\qquad
W=\fra{\lambda}{12}(v^2-\fra{\phi^2}{3})\phi\qquad
({\rm for}\: |\phi|\leq v)\,,
\end{equation} 
and identifying the energy of the static kink as its mass, we can finally write
\begin{equation}
m_{\rm kink}=W|_{x=-\infty}^{x=\infty}=2\,\fra{m^3}{\lambda}\,.
\end{equation}
Thus these `dual particles' are very massive in the perturbative limit.

\subsection{Bogomolnyi bound and topology}

In view of its applications in more complicated theories I will now redo this computation, 
this time with the Euclidean action:
\begin{equation}
S_E=\int \!dx_2 dx\, \{\fra{1}{2}(\partial_{x_2}\phi)^2+
\fra{1}{2}(\partial_{x}\phi)^2+V\}
=\int \!dx_2 dx\, \{\fra{1}{2}(\partial_{x_2}\phi)^2+
\fra{1}{2}[\partial_{x}\phi\mp \sqrt{2V}]^2
\pm \partial_{x}\phi\sqrt{2V}\}\,.
\label{eqn_kink_BPS}
\end{equation} 
This way of expressing the integrand (also known as BPS trick) 
leads to the Bogomolnyi bound,
because the action is bounded from below by
\begin{equation}
S_E\geq |\int \!dx_2 dx\, \partial_{x}\phi\sqrt{2V}|\,,
\end{equation} 
where the equality holds iff both squares in (\ref{eqn_kink_BPS}) vanish, that is just for the kink or antikink.
With the help of the function $W$ and denoting the Euclidean time interval by $T$ we obtain
\begin{equation}
S_E\geq T \,|\!\int_{x=-\infty}^{x=\infty}\!\!\!\!\! dx\, \partial_{x}W(\phi(x))|\,,
\end{equation} 
which clearly is a boundary term independent of the shape of $\phi(x)$ in the bulk.

Therefore, the action is bounded by a {\em topological quantum number} $q$, namely
\begin{equation}
S\geq T\,\fra{2m^3}{\lambda}|q|\,,\qquad
q=\fra{1}{2v}[\phi(x=+\infty)-\phi(x=-\infty)]
=\left\{
\begin{array}{cl}
0 & {\rm trivial\: vacua}\\
1 & {\rm kink}\\
-1 & {\rm antikink}
\end{array}
\right.\,.
\label{eqn_kink_topological}
\end{equation} 
This statement applies to every configuration with finite action, not just classical solutions.

As an integer $q$ cannot be deformed continuously 
(because this would require 
$\phi(x=\pm\infty)\neq \pm v,\: 
V(x=\pm\infty)\neq 0$ and $S\to \infty$).
Hence, {\em the space of finite action solutions splits into sectors 
labelled by the topological quantum number $q$
and separated by infinite barrieres, 
and the action is bounded by a constant times $|q|$,
 where the equality holds for classical solutions}.
We will encounter such a situation again and again for topological objects, also in higher dimensions. 
For the sine-Gordon model, for instance, 
$q$ can just take any integer value.


There exist a topological current for this charge, it is simply
\begin{equation}
J^\mu=\fra{1}{2v}\epsilon^{\mu\nu}\partial_\nu\phi\,,\qquad
\partial_\mu J^\mu=0\,.
\end{equation} 
As a typical phenomenon this current is conserved without using the equations of motion, 
thus it is not a Noether current. The charge then emerges in the usual way,
\begin{equation}
\int_\infty^{-\infty}\!\!\!\!\! dx\, J^0=\fra{1}{2v}\int_\infty^{-\infty}\!\!\!\!\! dx\,\partial_x\phi=q\,.
\end{equation} 

Let us have a look at $\phi$ as a mapping, 
provided finite action. 
When restricted to the boundary of space it maps into the vacuum manifold, 
both being sets of two points here,
\begin{equation}
\phi|_{x=\pm\infty}:\partial R\simeq Z_2 \longrightarrow 
\{\pm v\}\simeq Z_2\,.
\end{equation} 
The topological quantum number $q$ characterizes the mapping $\phi|_{x=\pm\infty}$ 
in that it measures whether the image is fully covered\footnote{
The trivial sector $q=0$ splits further into two disconnected components around the distinct vacua.}
and `in which direction'.  

\begin{figure}[t]
\begin{center}
\psfrag{mvv}{$-v$}
\psfrag{v}{$v$}
\psfrag{p}{$\phi$}
\psfrag{x}{$x$}
\psfrag{y1}{$y_1$}
\psfrag{y2}{$y_2$}
\psfrag{y3}{$y_3$}
\psfrag{y4}{$y_4$}
\includegraphics[width=0.45\linewidth]{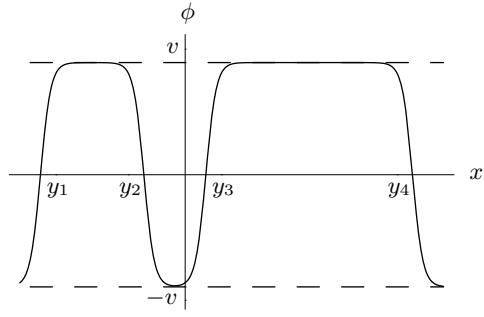}
\caption{A multi-soliton consisting of two kinks and two antikinks.}
\label{fig_kink_multi}
\end{center}
\end{figure}

Chains of kinks and antikinks are called multi-solitons,
an example is shown in Fig.\ \ref{fig_kink_multi}.
These are approximate solutions when diluted, 
i.e.\ when the difference of their locations $y_A$ is much bigger than the width $1/m$.

\subsection{Application: semiclassical calculation of the path integral}
\label{sect_kink_semi}

Let me go back to the particle picture for a moment and consider the Euclidean time evolution
(for reasons of better convergence and its analogy to statistical mechanics)
\begin{equation}
G_\pm (t) \equiv \langle x=\pm v | e^{-Ht/\hbar} | y=v \rangle
=N\int_{x(0)=\pm v}^{x(t)=v}\!\!\!\!\!D x\, e^{-S[x]/\hbar}\,.
\end{equation} 
The path integral\footnote{Its problems like the infinite normalisation $N$ cannot be discussed here.} 
weights all pathes $x(t)$ starting at $\pm v$ and ending at $v$ by the exponent of their action. 
In the semiclassical approximation one expands the exponent around (approximate) solutions
that obey these boundary conditions, too, 
and performs a Gaussian integration:
\begin{equation}
G_\pm (t) =  \#\int \!\! Dx\ \sum_A \exp\left(-S[x^A]/\hbar +0-\fra{1}{2}
\int\! dt\, \fra{\delta^2 L}{\delta x^2}(x^A)[x(t)-x^A(t)]^2/\hbar\right)\,.
\end{equation} 
In the eigensystem of the second variation of $L$, one can decompose the path integral 
(with Jacobian unity since this is a unitary transformation $L_2 \to l_2$) 
and perform all integrations, 
\begin{equation}
x(t)-x^A(t)=\sum_nc_nx_n(t)\,,\quad
\int\!Dx = \int\prod_n dc_n\cdot 1\,,\quad
\int\!dc_n\,e^{-\fra{1}{2}\lambda_n c_n^2}\sim\fra{1}{\sqrt{\lambda_n}}.
\end{equation} 
Hence to this order we arrive at a formula involving the determinant of the fluctuation operator,
\begin{equation}
G_\pm (t) =  \# \sum_A e^{-S[x^A]/\hbar}\fra{1}{\sqrt{\det[-\partial_t^2+V''(x^A(t))]]}}\,.
\end{equation}

However, $S$ is independent of the parameters $y$ of the classical solution 
(for the kink just the location parameter $y$) 
and this will lead to zero modes of the form $\partial_y x^A(t)$ 
of the fluctuation operator $-\partial_t^2+V''(x^A(t))$, 
as can easily be checked by using the equatons of motion.
These flat directions have to be split off from the determinant 
and can be treated by an integration over the collective coordinates
with a Jacobian $J$,
\begin{equation}
x(t)-x^A(t)=c_0 x_0(t)+\mbox{rest}\,,\qquad
\int\!\! D x =\int \!\! dc_0\cdot \mbox{rest}
=\int\!\! d y \,J\cdot \mbox{rest}\,.
\end{equation}
So the final expression for the propagator in the semiclassical approximation is
\begin{equation}
G_\pm (t) =  \# \sum_A \int\!\! d y_A\, J_A\, e^{-S[x^A]/\hbar}
\frac{1}{\sqrt{\det'[-\partial_t^2+V''(x^A(t))]}}\,,
\end{equation} 
where $\det'$ denotes the determinant without zero eigenvalues.

Let us apply this technique to a dilute gas of multi-solitons. 
The index $A$ runs over all even or odd integers $n$ of (alternating) solitons and antisolitons, 
depending on the boundary condition.
The parameters $y_A$ are the locations $\{y_1,\ldots,y_n\}$.
The integral over them (nothing else depends on the $y$'s) gives a factor $t^n/n!$.
The Jacobian can be shown to be $J=\sqrt{S_{\rm kink}}$ raised to the $n$th power.

In the first approximation without soliton interactions the total action is just $n$ times the individual kink action.
The hardest part of semiclassics is usually the (regularisation of the) fluctuation determinant, 
which we parametrise by $K^n\exp(-mt/2)$, 
the latter term being the determinant of the trival vacuum (harmonic oscillator).

Collecting all terms we have
\begin{eqnarray}
G_\pm (t) &=&  \# \sum_{n
\begin{array}{c}
\mbox{\footnotesize{even}}\\
\mbox{\footnotesize{odd}}
\end{array}}
\fra{t^n}{n!}J^ne^{-nS_{\rm kink}/\hbar}K^ne^{-mt/2}\nonumber\\
&=& \#\, e^{-mt/2} [\exp(JKe^{-S_{\rm kink}/\hbar}t)\pm
\exp(-JKe^{-S_{\rm kink}/\hbar}t)]\,.
\end{eqnarray}
Comparing this to the propagator decomposed into energy eigenstates,
\begin{equation}
G_\pm (t) = \psi_0^\ast(\pm v)\psi_0(v)e^{-E_0t/\hbar}
+\psi_1^\ast(\pm v)\psi_1(v)e^{-E_1t/\hbar}+\mbox{rest}\,,
\end{equation}
we conclude that the ground and first excited state are as expected spatially even and odd, respectively, 
and that their  energy is
\begin{equation}
E_{0,1}=\frac{\hbar m}{2}\mp \hbar JK e^{-S_{\rm kink}/\hbar}\,.
\end{equation}
The first term on the r.h.s.\ is the perturbative contribution of ground states of harmonic oscillators at each vacuum.
In the semiclassical approximation the multisolitons give the first nonperturbative contribution
to the energy level splitting caused by tunnelling, which because of $S_{\rm kink}\sim1/\lambda$ 
is not seen in ordinary perturbation theory.

\subsection{Fermions in the kink background}

One can add fermions to the scalar theory (back in $1+1$ dimensions) by a Yukawa coupling\footnote{
Under certain circumstances such a system becomes supersymmetric.},
\begin{equation}
\mathcal{L}=\mathcal{L}_{\rm bosonic}-\bar{\psi}(i\gamma^\mu\partial_\mu+g\phi)\psi\,,
\end{equation}
where the (Euclidean) $\gamma$'s can be chosen to be Pauli matrices.
The kink background $\phi=\phi_{\rm kink}$ will act like a space-dependent mass.

The Dirac-Hamiltonian reads
\begin{equation}
H=\gamma^2(\gamma^1\partial_x+g\phi)
\end{equation}
and anticommutes with one of the $\gamma$-matrices, $\{H,\gamma^1\}=0$, 
which therefore relates eigenstates with opposite energy
(almost like chiral symmetry $\gamma_5$). On $E=0$ one can diagonalise $\gamma^1$ simultaneously with $H$. 
Denoting the $\gamma^1$-eigenstates with eigenvalue $\pm 1$ by $s_\pm$, the zero modes
found by Jackiw and Rebbi \cite{jackiw:76d} 
have the form
\begin{equation}
\psi_{E=0}(x)=\chi_\pm(x) s_\pm\,,\qquad
\chi_\pm(x)=\exp\left(\mp \int_0^x\!\!\!\! dx'\phi(x')\right)\,.
\end{equation}
Asymptotically they go like $\exp(\pm\phi(x=\pm\infty)x)$ 
and therefore decay at both ends, 
iff the asymptotic values of $\phi$ are of different sign.

We conclude that {\em a normalisable zero mode exists and is exponentially localised}
(with vev $v$ and around $x=y$) for the kink with $q=1$, for which the $\gamma^1$-eigenvalue is $-1$,
and for the antikink $q\!=\!-1$, for which $\gamma^1$ gives $+1$. 
The trivial sector $q\!=\!0$ has no normalisable zero mode. 

Following our argument such zero modes exist for all configurations 
not just for classical solutions, 
which is the content of the index theorem by Bott and Seeley \cite{bott:78}.
These zero modes have applications as domain wall fermions.

\section{Derrick's Theorem}

A simple argument about the existence of solitonic solutions in higher dimensions 
has been given long ago by Derrick \cite{derrick:64}. 
Consider an action functional of a scalar field in $d$ dimensions,
\begin{equation}
S=\int\!\! d^dx\,[\fra{1}{2}(\partial_\mu\phi)^2+V(\phi)]
\equiv I_{\rm kin}+I_{\rm pot}\,,
\end{equation} 
and look for solutions $\delta S=0$ which are stable, i.e.\ $\delta^2 S\geq 0$.

A specific variation of such a solution is the rescaling $\phi_\lambda(x)=\phi(\lambda x)$.
It can easily be seen (by variable redefinition) 
that the functional depends on $\lambda$ as 
\begin{equation}
S(\lambda)=\lambda^{2-d}I_{\rm kin}(\lambda=1)+\lambda^{-d}I_{\rm pot}(\lambda=1)
\end{equation}
and that the two requirements above lead to $(2-d)\geq 0$.

Hence such scalar systems do not admit interesting solutions for dimensions higher than two 
(for the kink $d=1$). 
Therefore we will from now on concern ourselves with {\em gauge theories}.

\section{Magnetic monopoles}


Next we consider a gauge-Higgs system in $3+1$ dimensions, named after Georgi and Glashow 
(close to the electroweak theory, the notion of which will be used):
\begin{equation}
\mathcal{L}=-\fra{1}{2}\tr\, F^{\mu\nu}F_{\mu\nu}
+\tr\, D_\mu\phi\, D^\mu\phi-\fra{\lambda}{8}(2\,\tr\, \phi^2-v^2)^2\,.
\end{equation}
The field strength of the $SU(2)$ gauge field $A_\mu$ 
and the covariant derivative of the scalar field $\phi$ are
\begin{equation}
F_{\mu\nu}\equiv\partial_\mu A_\nu - \partial_\nu A_\mu-i g [A_\mu,A_\nu]\,,\qquad
D_\mu\phi\equiv\partial_\mu\phi-ig[A_\mu,\phi]\,,
\end{equation}
which display the self-interaction of the gauge bosons 
and the fact that $\phi$ is in the adjoint representation.

All the (hermitean) matrices can be written by virtue of the Pauli-matrices $\sigma^a$,
the generators of the algebra $su(2)$:
\begin{equation}
\phi=\phi^a\,\fra{\sigma^a}{2},\qquad
A_\mu=A_\mu^a\,\fra{\sigma^a}{2},\qquad
a=1,2,3\qquad
\mbox{(sum convention)}\,.
\end{equation}

The vacua 
are clearly at $\phi^a\phi^a=v^2$, 
which is a whole two-sphere $S^2_{\rm colour}$ in three-dimensional colour space.
A particular realisation is $\phi^a=v n^a$ 
with a normalised colour vector $n^a$, $n^an^a=1$.

{\em Symmetry breaking} occurs 
because the Lagrangian has an $SU(2)/Z_2\simeq SO(3)$ symmetry of rotations of $\phi$,
whereas the vacuum only has an $SO(2)$ symmetry of those colour rotations that leave $n$ invariant.
In the perturbative expansion around a vacuum, the {\em Higgs effect} 
is the generation of mass for the gauge field along the colour two-sphere $S^2_{\rm colour}$
and for the scalar fields  perpendicular to that sphere,
while the gauge field of the unbroken $U(1)$ symmetry remains massless,
\begin{equation}
m_{\rm W-boson}=vg\,,\qquad
m_{\rm Higgs}=v\sqrt{\lambda}\,,\qquad
m_{\rm photon}=0\,.
\end{equation}

As to solitonic solutions, finite action needs asymptotically $\phi^a\phi^a\to v^2$. 
We will again look for static solutions and gauge $A_0=0$,
 such that the field strength is given by the (coloured) magnetic field $\vec{B}$ alone.

\subsection{BPS trick and explicit solution}

Like for the kink, two squares can be separated off in the calculation of the energy
\begin{equation}
E=\int \!d^3\!x\,\{\tr\,(\vec{D}\phi)^2+\tr\,\vec{B}^2+V(\phi)\}
=\int \!d^3\!x\,\{\tr\,[\vec{D}\phi\mp\vec{B}]^2+V(\phi)\pm 2\,\tr\,\vec{B}\vec{D}\phi\}\,.
\label{eqn_mon_bps}
\end{equation} 
The energy is bounded by the last term, which can actually be rewritten as a surface term,
\begin{equation}
E\geq 2\, |\!\int \!d^3\!x\,\tr\,\vec{B}\vec{D}\phi|
=2\, |\!\int \!d^3\!x\,\ve{\partial}\,\tr\,(\vec{B}\phi)|
=v\,|\!\int_{S^2_\infty} \!\!\!\!d^2\!\ve{\sigma} (\vec{B}^a n^a)|\,.
\end{equation}
This is nothing but the magnetic flux projected onto the Abelian direction given by $n$
and gives rise to 
the magnetic charge
\begin{equation}
E\geq 4\pi v \,|q_{\rm mag}|\,.
\label{eqn_mon_bound}
\end{equation}


\begin{figure}[t]
\begin{center}
\includegraphics[width=0.3\linewidth]{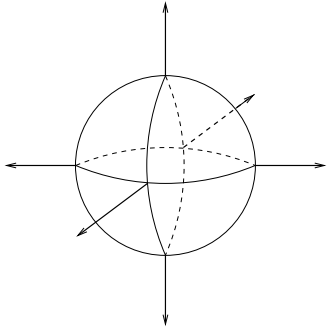}
\caption{The hedgehog as a prototype of a mapping $S^2\to S^2$
with winding number 1.}
\label{fig_hedgehog}
\end{center}
\end{figure}

In order to obtain the magnetic monopole solution by 't Hooft and Polyakov
\cite{thooft:74,*polyakov:74}
one makes a radial ansatz
\begin{equation}
A_i^a=\epsilon_{iaj}\fra{x_j}{|\vec{x}|}A(|\vec{x}|)\,,\qquad
\phi^a=\fra{x_a}{|\vec{x}|}\phi(|\vec{x}|)\,.
\end{equation}
Note the typical mixing of space and colour space (indices) for such symmetric ans\"atze. 
Analytic solutions are available in the limit of vanishing potential \cite{bogomolnyi:76,*prasad:75}, 
solving $\vec{D}\phi=\pm\vec{B}$, see Eq.~(\ref{eqn_mon_bps}).

Asymptotically, the function $\phi(|\vec{x}|)$ approaches 
the vacuum expectation value $v$, 
while $A(|\vec{x}|)$ will go like $1/g|\vec{x}|$. 
Consequently, the projected magnetic field 
\begin{equation}
|\vec{x}|\to\infty\,:\quad\vec{B}^a n^a\to\fra{\vec{x}}{g|\vec{x}|^3}
\end{equation}
behaves like a Coulomb field with $q_{\rm mag}=1/g$.

The colour structure of the scalar field is such
that it points in colour space in the same direction 
as the unit vector in coordinate space, 
the famous {\em hedgehog shape} shown in Fig.~\ref{fig_hedgehog}.

\subsection{Topology}

\begin{figure}[t]
\begin{center}
\includegraphics[scale=0.8]{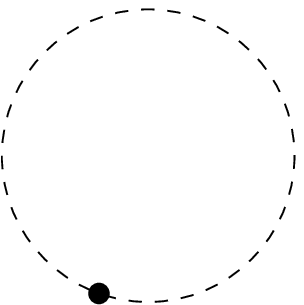}
\includegraphics[scale=0.8]{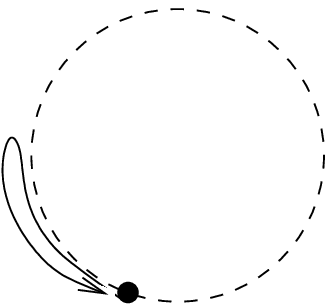}
\includegraphics[scale=0.8]{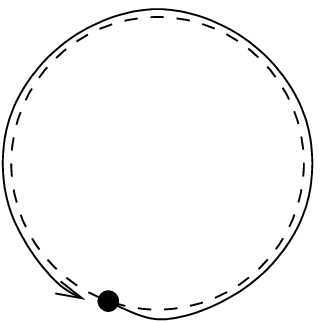}
\includegraphics[scale=0.8]{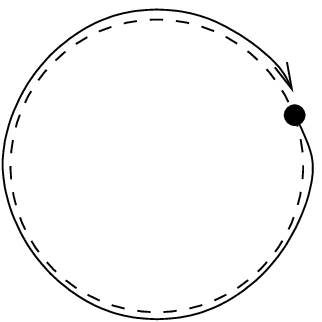}
\includegraphics[scale=0.8]{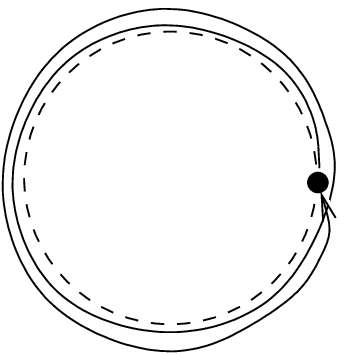}
\caption{The one-minute topologist. 
From left to right the panels show mappings $S^1\to S^1$: 
the first two with vanishing winding number
(the constant mapping and a deformation thereof), 
then mappings with winding number $1$, $-1$ 
and finally $2$.
}
\label{fig_windings}
\end{center}
\end{figure}

The topology of this object reveals itself in the asymptotic scalar field $\phi^a$, 
which can be divided by $v$ to obtain the normalised field $n^a$:
\begin{equation}
n^a(\vec{x}):\partial R^3\simeq S^2_\infty \longrightarrow
S^2_{\rm colour}=SO(3)/SO(2)\,.
\end{equation}
It is a mapping from the boundary of space onto the vacuum manifold
(being a coset space). Such mappings are characterised by a winding number 
or degree deg$(n)$ in the {\em second homotopy group} $\pi_2(S^2)=Z$. 
Without many details, this quantity can be best understood by visualising the corresponding mappings $S^1\to S^1$,
which are governed by the first homotopy group $\pi_1(S^1)=Z$.
As is clear from Fig.\ \ref{fig_windings}, this integer is not changed by small deformations 
and counts, how many times the image sphere is covered by the preimage sphere and in which direction
(similar to $q$ in the kink case).
In the same way the hedgehog covers the image sphere $S^2_{\rm colour}$ just once 
and thus has deg$(n_{\rm hedgehog})=1$.

In fact one can prove (with the help of the so-called 't Hooft field strength tensor) that 
$q_{\rm mag}={\rm deg}(n)/g$
which shows that the magnetic charge is a topological quantum number.
There also exists a topological current $J^\mu$ such that $\int d^3x\,J^0={\rm deg}(n)$.

Moreover, a winding $n$ at spatial infinity cannot be extended smoothly into the bulk, 
rather one encounters a zero in the scalar field $\phi$, 
where $n$ is not defined.
The vector $\vec{y}$ in $\phi^a(\vec{y})=0$,
where the symmetry is restored locally, 
is called the location of the monopole 
and is a free parameter of the solution 
(for the radial ansatz above it coincides with the origin).

Many of these features I have exemplified in the simpler scalar system with kinks. 
Likewise fermionic zero modes exist in the monopole background \cite{jackiw:76d},
they will be discussed after instantons have been introduced.

\subsection{Physical consequences}

An attractive property of the magnetic monopole is that it quantises electric charges, namely
\begin{equation}
q_{\rm mag}\cdot g={\rm deg}(n)\in Z
\end{equation}
(in the unbroken Abelian theory $g$ plays the role of the electric charge).

However, magnetic monopoles have not been observed experimentally nor 
are they contained in the Standard Model, 
since the electroweak symmetry breaking
\begin{equation}
SU(2)_{\rm isospin}\times U(1)_{\rm hypercharge}\longrightarrow U(1)_{\rm electromagnetism}
\end{equation}
is nontrivial, as parametrised by the Weinberg angle.
On the other hand, monopoles are generic in Grand Unified Theories 
(and it is a constraint on cosmological models to sufficiently dilute them).

The mass of the magnetic monopoles follows from the bound, Eq.~(\ref{eqn_mon_bound}), 
\begin{equation}
m_{\rm mon}=v\cdot \fra{4\pi}{g}=\fra{4\pi}{g^2} m_W\,.
\end{equation}
It is again proportional to the vev, but monopoles are much heavier than the W-bosons
(in the weak coupling regime). 

\subsection{Can one `abelianise' the monopole?}
\label{sect_mon_abel}

In the so-called unitary gauge (used to extract the field content after symmetry breaking), 
$\phi$ is rotated onto a fixed colour direction, say $\sigma^3$.
For the magnetic monopole this procedure fails, not only at the monopole location, 
where $\phi$ vanishes and an Abelian direction is not well-defined,
but also around it, 
since a trivial field $n^a=(0,0,1)$ would have vanishing winding number in contrast to the original hedgehog.
Still the configuration is magnetically charged, because the projection $\vec{B}^an^a$ is gauge invariant.

This puzzle is resolved by the fact that the gauge transformed $\vec{A}(\vec{x})$ 
is the gauge field of the Dirac monopole 
with location $\vec{y}$ and a Dirac string emanating from $\vec{y}$ to spatial infinity, 
to provide the magnetic influx\footnote{
Alternatively, one can evade singularities by using the language of fibre bundles.}.
Plus there are exponentially decaying (`massive') parts of the gauge field,
fine-tuned as to avoid
the singularities in the full theory. 
The latter make it clear that a superposition of these solitonic objects is difficult.

\section{Instantons}

In the remainder of these lectures I will mostly consider {\em four-dimensional Euclidean Yang-Mills theory},
the purely gluonic part of QCD. For simplicity, I will restrict myself to gauge group $SU(2)$.
The Lagrangian, 
\begin{equation}
\mathcal{L}=\fra{1}{2}\tr\, F_{\mu\nu}F_{\mu\nu}=\fra{1}{2}((\vec{E}^a)^2+(\vec{B}^a)^2)\,,
\end{equation}
contains the nonlinear dynamics of gluons,
which 
results in a special running of the coupling 
(including asymptotic freedom)
and the existence of topological objects.

For the latter we start by considering the asymptotic behaviour 
in the four-dimensional radius $r$, where finite action requires
\begin{equation}
r=\sqrt{x_\mu^2}\to\infty:\quad
F_{\mu\nu}\to 0\,,\quad
A_\mu\to\mbox{pure gauge.}
\end{equation} 
Under these circumstances configurations can be lifted to the compact four-sphere
$\dot{R}^4\simeq S^4$ \cite{uhlenbeck:78}.

\subsection{BPS trick and topology}

As should be familiar by now, 
the aim is to split the action  into a sum of squares 
and a surface term. One starts by writing
\begin{equation}
S=\int d^4x\,\fra{1}{2}\tr F_{\mu\nu}^2
=\int d^4x\,\fra{1}{4}(\tr F_{\mu\nu}^2+\tr \tilde{F}_{\mu\nu}^2)\,,\qquad
\tilde{F}_{\mu\nu}=\fra{1}{2}\epsilon_{\mu\nu\rho\sigma}F_{\rho\sigma}\,.
\end{equation}
The quantity $\tilde{F}$ is the dual field strength, 
it has electric and magnetic field interchanged. 
In the next step one basically rewrites 
$\vec{E}^2+\vec{B}^2$ as $(\vec{E}\mp \vec{B})^2\pm 2 \vec{E}\vec{B}$,
\begin{equation}
S=\int d^4x\,\{\fra{1}{2}\tr (F_{\mu\nu}\mp \tilde{F}_{\mu\nu})^2
\pm \fra{1}{2}\tr F_{\mu\nu}\tilde{F}_{\mu\nu}\}
\label{eqn_inst_bps}\,,
\end{equation}
to derive a lower bound, 
 \begin{equation}
S\geq|\int d^4x\,\fra{1}{2}\tr F_{\mu\nu}\tilde{F}_{\mu\nu}|
\equiv \fra{8\pi^2}{g^2}|Q|
\label{eqn_def_Q}\,,
\end{equation}
with $Q$ the {\em instanton number or topological charge}\footnote{ 
or Pontryagin index or second Chern class}, 
that can be computed via a surface integral
\begin{equation}
Q=\int \!\!d^4 x\,\partial_\mu K_\mu(A)\,,\qquad
K_\mu(A)=\fra{g^2}{16\pi^2}\epsilon_{\mu\nu\rho\sigma}
(A_\nu^a\partial_\rho A_\sigma^a
+\fra{g}{3}\epsilon_{abc}
A_\nu^a A_\rho^b A_\sigma^c)\,.
\label{eq_inst_K}
\end{equation}
At the boundary of space-time $A_\mu$ is a pure gauge with a gauge transformation\footnote{
Gauge transformations with winding number are sometimes called `large gauge transformations'.} $\Omega$
and the topological charge $Q$ equals its winding number,
\begin{equation}
Q=\int_{S^3_\infty} \!\!d^3\sigma K_\perp(A_\mu=\fra{i}{g}\Omega^\dagger\partial_\mu\Omega)
={\rm deg}(\Omega)\,,
\end{equation}
which is now governed by the third homotopy group
\begin{equation}
\Omega:\,\partial R^4\simeq S^3_\infty\longrightarrow SU(2)\simeq S^3\,,\qquad
{\rm deg}(\Omega)\in\pi_3(S^3)=Z\,.
\end{equation}
For gauge groups $SU(N)$ the same considerations hold, 
since although those are higher dimensional manifolds, 
the `number of three-dimensional holes' in them is the same, $\pi_3(SU(N))=Z$.

\subsection{Selfdual solutions}

\begin{figure}[t]
\begin{center}
\psfrag{r}{size $\rho$}
\psfrag{x}{$x_\mu$}
\psfrag{y}{$y_\mu$}
\psfrag{rd}{$r^{-8}$ decay}
\psfrag{q}{$\!\!\!\!\tr F_{\mu\nu}^2$}
\includegraphics[width=0.5\linewidth]{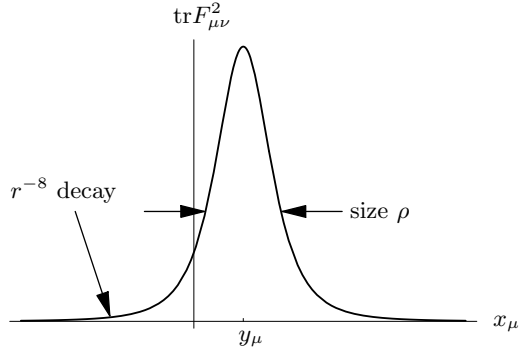}
\caption{The action density profile of an instanton.}
\label{fig_inst_profile}
\end{center}
\end{figure}

As familiar from the lower-dimensional examples, 
the BPS trick, Eq.~(\ref{eqn_inst_bps}), reduces the differential equation on $A_\mu(x)$ 
from second order in the equations of motion $D_\mu\tilde{F}_{\mu\nu}=0$ 
to first order in the (anti)selfduality equation
\begin{equation}
F_{\mu\nu}=\pm \tilde{F}_{\mu\nu}\rightleftharpoons \vec{E}=\pm\vec{B}\,.
\end{equation}
Conversely, the equations of motion are fulfilled for (anti)selfdual fields
by virtue of the Bianchi identity $D_\mu F_{\mu\nu}=0$.

The explicit solution of unit topological charge was found by 
Belavin, Polyakov, Schwartz and Tyupkin \cite{belavin:75}. 
Make a radial ansatz for $\Omega$, 
\begin{equation}
\Omega(x)=\fra{x_0}{r}\Eins_2+i\,\fra{x_a}{r}\sigma^a\,,
\end{equation}
which as the identical mapping has winding number 1. 
Then the asymptotic gauge field is
\begin{equation}
A_\mu^a\to \fra{2}{g}\eta_{\mu\nu}^a\,\fra{x_\nu}{r^2}\,,\qquad 
\eta_{\mu\nu}^a\in\{-1,0,1\}\ldots \mbox{the 't Hooft tensor,}
\end{equation}
which quite simply extends into the bulk,
\begin{equation}
A_\mu^a = \fra{2}{g}\eta_{\mu\nu}^a\,\fra{x_\nu}{r^2+\rho^2}\,.
\end{equation}
The antiinstanton with $Q=-1$ is obtained upon changing some signs,
$\eta_{\mu\nu}^a \to \bar{\eta}_{\mu\nu}^a$.

The action density, 
which for instantons equals the topological charge density,
\begin{equation} 
\tr F_{\mu\nu}^2 = \frac{1}{g^2}\frac{\rho^4}{(r^2+\rho^2)^4}\,,
\end{equation}
is concentrated in space and time, 
which lead to the name instanton. 
It decays algebraically as shown in Fig.\ \ref{fig_inst_profile}
and integrates to the unit $S=8\pi^2/g^2$.

The instanton profile is parametrised by a size $\rho$. 
Other parameters of the most general charge 1 solution,
\begin{equation}
A_\mu = U^\dagger \left(\fra{2}{g}\,\eta_{\mu\nu}^a\,\fra{(x-y)_\nu}{(x-y)^2+\rho^2}
\frac{\sigma^a}{2}\right)U\,,
\label{eq_inst_soln}
\end{equation}
are the four-dimensional location $y_\mu$ and the colour orientation\footnote{
which for charge 1 can be compensated by a gauge transformation} $U$.

For instantons of higher charge a subclass is known explicitly \cite{thooft:76d,*corrigan:77,*wilczek:77},
\begin{equation}
A_\mu = \eta_{\mu\nu}^a \partial_\nu\log\left(
1+\sum_{p=1}^Q \frac{\rho^{(p)}}{(x_\mu-y_\mu^{(p)})^2}\right)\,,
\label{eqn_inst_cftw}
\end{equation}
which contains $Q$ lumps of topological charge
with $Q$ locations $y_\mu^{(p)}$ and sizes $\rho^{(p)}$, respectively.
However, these lumps all have got the same colour orientations, 
hence this ansatz yields $5Q$ out of $8Q-3$ moduli.

\subsection{Massless fermions coupled to instantons}
\label{sect_inst_zero}

The Lagrangian coupling fermions 
(in the fundamental representation) 
to the gauge field, 
\begin{equation}
\mathcal{L}_\psi=\bar{\psi}(i\gamma^\mu D_\mu+im)\psi\,,
\end{equation}
has a chiral symmetry in the case of vanishing mass, $\{i\gamma^\mu D_\mu,\gamma_5\}=0$. 
Consequently -- and like for the kink --
eigenvalues of the Dirac operators come in pairs $\pm\lambda$,
while on the zero modes $\gamma_5$  can be diagonalised distinguishing modes of definite chirality.

To be concrete, I chose the (Euclidean) Weyl representation
\begin{equation}
\gamma^\mu=\left(
\begin{array}{cc}0&\sigma_\mu\\~\bar{\sigma}_\mu&0\end{array}
\right)=\left(
\begin{array}{cc}
0&(i\Eins_2,\ve{\sigma})\\~
(-i\Eins_2,\ve{\sigma})&0
\end{array}
\right)\,,\quad
\gamma_5=\gamma^0\gamma^1\gamma^2\gamma^3=\left(
\begin{array}{cc}
-\Eins_2&0\\0&\Eins_2
\end{array}
\right)\,,
\end{equation}
where left-handed and right-handed modes 
correspond to upper and lower components, respectively (by convention).

It turns out that {\em in an instanton background there is 1 left-handed
zero mode, but no right handed one}.
That is the equation $\bar{\sigma}_\mu D_\mu \psi_L=0$ has a solution
\begin{equation}
\psi_L\sim\frac{\rho}{((x_\mu-y_\mu)^2+\rho^2)^{3/2}}
\end{equation}
centered at the instanton and spherically symmetric like the latter,
while $\sigma_\mu D_\mu \psi_R=0$ allows no normalisable solution
(basically because $-D_\mu^2$ is positive and $\bar{\eta}_{\mu\nu}^a\eta_{\mu\nu}^b=0$).

Analogously, the antiinstanton has 1 right-handed zero mode. 
This is in agreement with the Atiyah-Singer index theorem \cite{atiyah:71},
which equates the index, the difference of numbers of left-handed versus
right-handed modes, and the topological charge,
\begin{equation}
\mbox{index}\equiv n_L-n_R=Q\,,
\end{equation}
for any configuration. 
For the instanton solution this equation is fulfilled with the minimal 
number of zero modes, $1-0=1$.

\subsection{Tunnelling picture, spectral flow and the axial anomaly}

Instantons can also be seen as tunnelling events.
In the Weyl or temporal gauge $A_0=0$ the Yang-Mills (Minkowskian) Lagrangian density
und Hamiltonian read
\begin{equation}
\mathcal{L}=\fra{1}{2}((\partial_0\vec{A}^{\!a})^2-\vec{B}^{a2})\,,\qquad
H=\int\!\!d^3x\,\fra{1}{2}(\ve{\Pi}^{a2}+\vec{B}^{a2})\,,\quad
\ve{\Pi}=\partial_0\vec{A}=\vec{E}\,.
\end{equation}
Clearly, vacua $E=0$ emerge for pure gauges 
$\vec{A}=\fra{i}{g}\Omega^\dagger(\vec{x})\ve{\partial}\Omega(\vec{x})$.
These are characterised by the {\em Chern-Simons number},
\begin{equation}
CS(\vec{A})=\int\!\!d^3x\,K_0(\vec{A})\,,
\end{equation}
see Eq.~(\ref{eq_inst_K}),
which reduces to the degree of $\Omega(\vec{x})$ 
as a mapping from the compactified three-space $\dot{R}^3\simeq S^3$ into the gauge group $SU(2)$.

The immediate conclusion is that vacua with different Chern-Simons number 
cannot be deformed into each other within vacua. 
The configuration space of gauge theories, however, is connected.
That means that the energy must be positive inbetween.

\begin{figure}[h]
\psfrag{E}{$E$}
\psfrag{1}{\vspace{1cm}$CS=1$}
\psfrag{2}{$2$}
\psfrag{3}{$3$}
\psfrag{A}{
$\vec{A}$-space\newline
}
\includegraphics[width=0.45\linewidth]{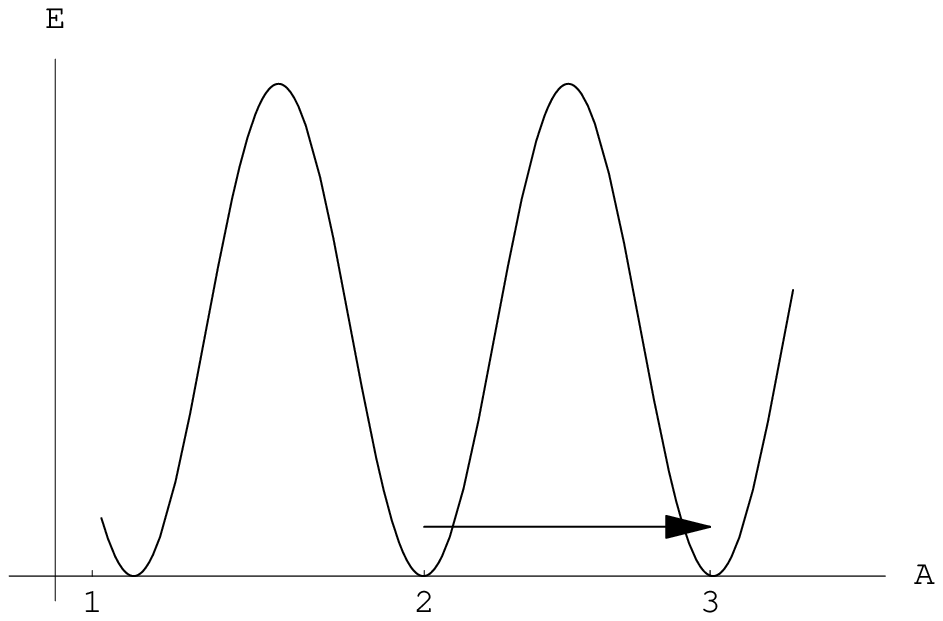}\hfill
\psfrag{E}{$\!\!\!\!E(x_0)$}
\psfrag{x0}{$x_0$}
\includegraphics[width=0.45\linewidth]{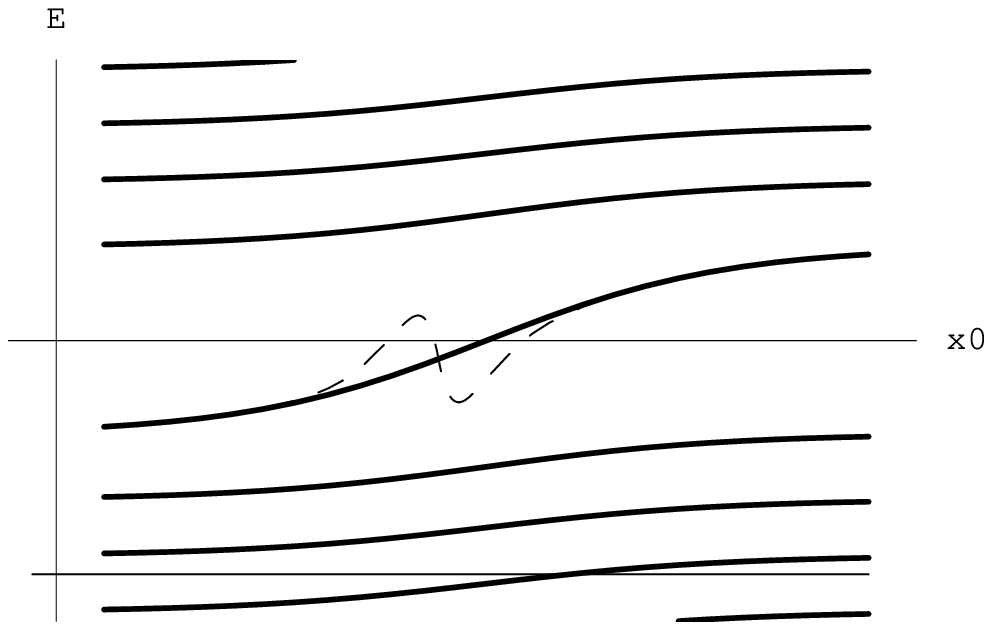}
\caption{Left: the energy as a function of the gauge configurations
with the tunnelling caused by an instanton. 
Right: the spectral flow of the fermionic energy as a function of time in an instanton background (bold curves), 
a deformation thereof (dashed curve)
and a cut-off.}
\label{fig_inst_tunnelling}
\end{figure}

Hence, in the space of $\vec{A}$-fields there are infinitely many vacua of same energy,
see Fig.~\ref{fig_inst_tunnelling} left,  
and the instanton is a tunnelling process between consecutive ones. 
It has nonvanishing field strength in the bulk, 
but approaches vacua $\vec{A}_{\pm\infty}$ in the infinite past and infinite future,
the difference of their Chern-Simons number being just the topological charge
\begin{equation}
Q=\int\!\!d^4x\, \partial_\mu K_\mu(A)=CS(\vec{A}_{\infty})-CS(\vec{A}_{-\infty})+0
\end{equation}
as follows from Eq.~(\ref{eq_inst_K}) 
and the fact that the cylindrical surface $|\vec{x}|\to\infty$ does not contribute 
due to $A_0=0$.

This aspect of instantons has an important physical counterpart for fermions.
The ($x_0$-dependent) Dirac-Hamiltonian,
\begin{equation}
H_{D}=-\gamma_0\ve{\gamma}\,\vec{D}(\vec{A})\,,
\end{equation}
has the same spectrum in the background of the vacua $\vec{A}_{\infty}$ and $\vec{A}_{-\infty}$.
The instanton transition inbetween, however, is connected with a rearrangement of all eigenvalues
including crossings of $E=0$, see Fig.~\ref{fig_inst_tunnelling} right.
The number of these crossings is called the {\em spectral flow} 
and can be shown to be equal to  the index of $i\gamma^\mu D_\mu$ and hence $Q$ \cite{atiyah:80}.
To illustrate this fact, one can involve the adiabatic approximation, 
then the normalisability of the four-dimensional zero mode 
requires $E_{\pm\infty}$ to be of opposite sign.
The crossings $E=0$ contribute to the spectral flow with the sign of the slope, 
such that this quantity is invariant under deformations 
that might cause oscillations around $E=0$, as shown in Fig.~\ref{fig_inst_tunnelling} right. 

The spectral flow is related to the {\em axial anomaly}. 
The axial current $j_\mu^5=\bar{\psi}\gamma_\mu\gamma^5\psi$, which is classically conserved,
$\partial_\mu j_\mu^5=0$, needs to be renormalised in the quantum theory,
resulting in 
\begin{equation}
\partial_\mu j_\mu^5=2\fra{1}{16\pi^2}\tr\, F_{\mu\nu}\tilde{F}_{\mu\nu}\,.
\label{eqn_inst_anomaly}
\end{equation}
When a cut-off in the Dirac sea is applied to the fermion spectrum of Fig.~\ref{fig_inst_tunnelling},
modes will reappear after tunnelling.
This amounts to $\Delta Q^5=Q_{\infty}-Q_{-\infty}=2Q$, 
because in an instanton-like background $\Delta Q_L=1\,,\Delta Q_R=-1$, 
i.e.\ one fermion flips its chirality.
This is nothing but the integrated version of the 0-th component of the anomaly.

\subsection{(Some) Analytical aspects of instantons}

In the following I will present some very interesting relations among instantons
on torus-like manifolds (close to the lattice) 
which will help to find exact solutions and their properties.

\subsubsection{The Nahm transform}

The Nahm transform is a mapping between instantons on four-tori \cite{nahm:80,*braam:89}:\\

\begin{tabular}{ccc}
$SU(N)$, charge $Q$, on $T^4$ &
$\longleftrightarrow$&
$SU(Q)$, charge $N$, on $\tilde{T}^4$
\end{tabular}\\

\noindent It interchanges the topological charge and the rank of the gauge group 
and also inverts the extension of the torus, 
since $x_\mu\sim x_\mu+L_\mu$ on $T^4$, 
whereas $z_\mu\sim z_\mu+1/L_\mu$ on $\tilde{T}^4$.
Whenever necessary $Q$ will be assumed positive, 
for antiselfdual gauge fields completely analogous relations hold.

The Nahm transform squares to the identity and is a hyperK\"ahler isometry
(meaning it keeps the metric on the moduli spaces\footnote{
The Nahm transformation also has an interpretation as a duality in string theory.}).
What is so useful for our purposes is 
that it is actually a constructive procedure, via the fermionic zero modes.

The new gauge field generated by the Nahm transform,
\begin{equation}
\hat{A}_\mu^{pq}(z) \equiv \int \!\! d^4x\,\psi^{(p)}_z(x)^\dagger
i\partial_{z_\mu}\psi^{(q)}_z(x)\,,
\label{eqn_nahm_gaugefield}
\end{equation}
is a bilinear in the chiral zero modes $\psi(x)$ 
of the original field $A_\mu(x)$,
\begin{equation}
\sigma_\mu(\partial_\mu\Eins_N - i A_\mu + 2\pi i z_\mu\Eins_N)
 \psi^{(p)}_z(x)=0\,,\qquad
p=1,\ldots,Q\,,
\label{eqn_nahm_pre}
\end{equation}
which exist due to the index theorem 
(without zero modes of wrong chirality).

The old coordinate $x$ is integrated out in Eq.\ (\ref{eqn_nahm_gaugefield}),
likewise the colour and spin indices are saturated.
The not so straightforward part of the expression for the new gauge field
is the introduction of the new coordinate $z$.
As can be seen in Eq.~(\ref{eqn_nahm_pre}), 
$z_\mu$ is added to the original field $A_\mu(x)$ with an identity in colour space.
This transfers the $SU(N)$ gauge field into a $U(N)$ one with a constant trace part,
which does not change the field strength 
nor the topological charge nor the number of zero modes.
The term $2\pi i z_\mu \Eins_N$ can be gauged away 
by a gauge transformation $\exp(2\pi i x_\mu z_\mu \Eins_N)$, 
which however is periodic only if $L_\mu z_\mu = 1$ (no sum).

This explains the extensions of the dual torus $\tilde{T}\ni z_\mu$. 
Moreover, the new gauge field $\hat{A}_\mu(z)$
\begin{itemize}

\item[-] is invariant under gauge transformations of the original gauge field,

\item[-] transforms like a gauge field under a ($z$-dependent) base change 
of the original zero modes,

\item[-] is a hermitean $Q \times Q$ matrix, actually it can be restricted to be $su(Q)$-valued,

\item[-] is (anti)selfdual, iff the original gauge field $A_\mu(x)$ is (anti)selfdual,

\item[-] has topological charge $N$.

\end{itemize}
The last two points prove that what has been generated is a charge $N$ instanton 
and this completes the introduction of the Nahm transform.

To solve for the instantons on the dual side can be simpler than the original problem.
For topological charge 1 the dual field is $U(1)$ which amounts to a 
{\em linear problem}.
As a byproduct there are no charge 1 instantons on the four-torus, 
since there no $U(1)$ instantons on the dual torus 
(unless twisted boundary conditions are applied;
$Q=1$ configurations do exist on $T^4$).

I will also consider the Nahm transform on 
related manifolds $T^{4-n}\times R^n,
\:n\in\{1,2,3,4\}$, where some of the directions have been decompactified. 
The dual manifolds are $\tilde{T}^{4-n}$, because the infinite line is dual to a point.
Then the (anti)selfduality equations on the dual side have {\em less derivatives},
which is another simplifaction of the problem.
On $\tilde{T}^{4-n}$ one still deals with (anti)selfdual $SU(Q)$ gauge fields,
but the topological charge $N$ is replaced by $N$ singularities.

\subsubsection{The ADHM formalism}
 
The formalism by Atiyah, Drinfeld, Hitchin and Manin \cite{atiyah:78,*christ:78}
gives a recipe to 
obtain in principle all $SU(N)$ instantons on $R^4$. 
It can be best understood as an inverse Nahm transform for the extreme case $n=4$.
The dual space shrinks to a point, 
in other words the dual problem is {\em purely algebraic} 
(but non-linear for higher charge).

The ADHM data, 
\begin{equation}
\Delta_x=\left(
\begin{array}{c}
\lambda\\
B-x
\end{array}
\right)\,,
\end{equation}
consist of a vector $\lambda$ parametrising the aforementioned singularities 
and a matrix $B$ containing the dual gauge `field'.
Both, $\lambda$ and $B$ have quaternionic entries (for $SU(2)$) and their dimensions scale with the charge.

The formalism requires that $\Delta_x^\dagger\Delta_x$ is real and invertible,
which expresses the (anti)selfduality on the dual side.
The remaining steps are very close to those in the inverse Nahm transform.
One solves for the chiral zero `modes',
\begin{equation}
\Delta_x^\dagger v_x=0\,,
\end{equation}
which are actually vectors not depending on any $z$, but on $x$ parametrically.
Finally the original gauge field looks very similar to that in Eq.~(\ref{eqn_nahm_gaugefield}),
\begin{equation}
A_\mu(x)=i v_x^\dagger \partial_\mu v_x=0\,.
\end{equation}
The ADHM formalism can rather simply be solved to obtain the class given by Eq.~(\ref{eqn_inst_cftw}): 
$\lambda$ is real and contains the sizes $\rho^{(p)}$ 
and $B$ is diagonal with the locations $y_\mu^{(p)}$ as entries.

For the intermediate cases $T^3\times R\: (n=1)$ and $T^2\times R^2\: (n=2)$
particular instanton results have been obtained \cite{jardim:99,*ford:02a,*vanbaal:96}. 
The case $n=3$ will be discussed at length now, because it has the physical
interpretation of finite temperature.

\section{Calorons}
\label{sect_cal}

Calorons are {\em instantons at finite temperature}, that is over the manifold $S^1\times R^3$.
As usual the compact direction has circumference $\beta=1/k_B T$, which like the coupling $g$ will be set to 1 below (mostly).

\subsection{Infinite instanton chains}

\begin{figure}[t]
\begin{center}
\psfrag{b}{$\beta$}
\psfrag{r}{$\rho$}
\includegraphics[scale=0.8]{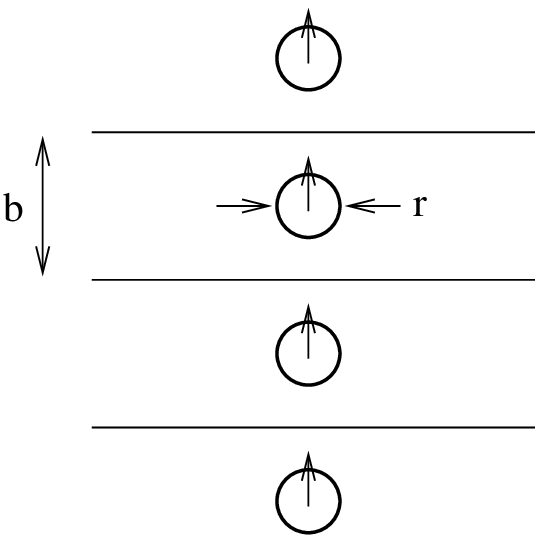}
\hspace{3cm}
\includegraphics[scale=0.8]{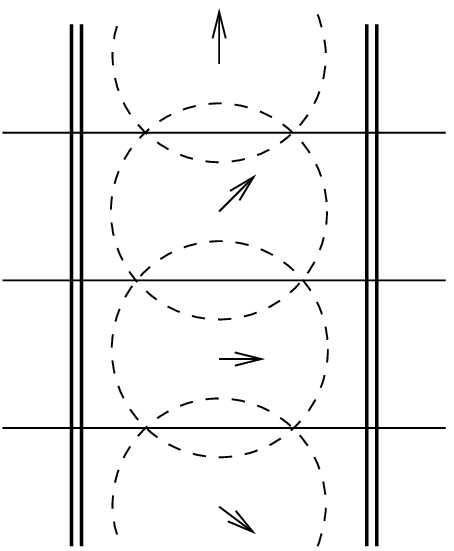}
\caption{Calorons as infinite chains of instantons. 
Left the HS caloron with same colour orientations of the copies and in the instanton limit $\rho\ll\beta$.
Right a caloron of nontrivial holonomy indicated by the rotating colour orientations of the copies.
This caloron is large and thanks to overlap effects 
there is no action density at the location of the instanton (dashed circles),
but at two monopole worldlines (double lines indicating a finite profile).}
\label{fig_cal_chains}
\end{center}
\end{figure}

Before I come to the Nahm transform in this setting,
let me collect some physical intuition approaching the problem from instantons over $R^4$.
The compactification of the time-like direction amounts to infinitely many copies along $x_0$,
i.e.\ charge infinity instantons.

In the simplest case all these instantons have the same colour orientations. 
The ansatz (\ref{eqn_inst_cftw}) can be pushed to the extreme of charge infinity
and yields the Harrington-Shepard caloron \cite{harrington:78}, 
\begin{equation}
A_\mu=\eta_{\mu\nu}^a \partial_\nu\log\left(
1+\frac{\pi\rho^2}{|\vec{x}-\vec{y}|}
\frac{\sinh 2\pi |\vec{x}-\vec{y}|}{\cosh 2\pi|\vec{x}-\vec{y}|-\cos 2\pi (x_0-y_0)}\right)\,.
\end{equation}
In the course of this (partial) dimensional reduction, two scales compete: 
the instanton size $\rho$ and the time-like extension $\beta$. 
If the latter is large, the copies do not feel much of their neighbours 
and the caloron has properties of an instanton.
In the other case of large size $\rho$ strong overlap effects occur. 
The configuration becomes static and it has been noticed first by Rossi \cite{rossi:79}
that {\em the HS caloron turns into the magnetic monopole}.
This is actually not too surprising, as the dimensionally reduced 
($\partial_0=0$) Yang-Mills theory is the Georgi-Glashow model, 
where the scalar field $\phi$ is the Yang-Mills field $A_0$
such that $\vec{D}\phi=\vec{E}$, plus of course $V(\phi)=0$, the PS limit.

The caloron becomes much more intricate in the case 
of different colour orientations of the copies.
This requires to solve the full ADHM formalism at infinite charge,
which has been accomplished by Kraan and van Baal \cite{kraan:98a} 
and Lee and Lu \cite{lee:98b} (for higher charge see \cite{bruckmann:02b,bruckmann:04a}).
It led to {\em calorons of nontrivial holonomy}, 
the term will become clear in a minute.
The gauge field constructed this way is periodic only up to a gauge transformation
(this dictates that the relative colour rotations between neighbours 
is the same along the entire chain, see Fig.~\ref{fig_cal_chains} right).
It can be made periodic by an $x_0$-dependent gauge transformation, 
which in turn induces a nonvanising $A_0$ at spatial infinity. 
This means that the Polyakov loop,
\begin{equation}
\mathcal{P}(\vec{x})\equiv \mathcal{P}\exp\left(i\int_0^\beta\!\! d x_0\,A_0\right)\,,\qquad
\mathcal{P}\ldots \mbox{path (here time) ordering,}
\end{equation}
taken to spatial infinity\footnote{
For magnetically neutral configurations the asymptotic Polyakov loop 
can be gauged to an angle-independent value.}, 
called holonomy,
\begin{equation}
\mathcal{P}_\infty\equiv\lim_{|\vec{x}|\to\infty}\mathcal{P}(\vec{x})
\neq \pm \Eins_2\,,
\end{equation}
is nontrivial. This holonomy plays the role of the Higgs field fixing a colour direction,
albeit in the gauge group $\log\mathcal{P}_\infty\sim\phi$ 
(see also the relation between $A_0$ and $\phi$ above).
It could be thought of as a background or environment for the caloron
and as we will see gives rise to a radical deviation from the instanton picture.

\subsection{Nahm picture and substructure}

The dual gauge field $\hat{A}_\mu(z)$ in the Nahm transform depends on one 
compact coordinate $z\in\tilde{S}^1$ with circumference $1/\beta$ 
(still the index $\mu$ runs over 0 to 3). 
As already mentioned above, this simplifies the dual problem, 
namely to an {\em ordinary} differential equation.
In order to obtain a caloron of charge $Q$, 
the dual gauge field has to be a $Q\times Q$ matrix
and in order to have gauge group $SU(N)$,
$N$ singularities will appear on the dual side. 
The selfduality equation reads
\begin{equation}
\hat{E}_i(z)-\hat{B}_i(z)
=\partial_z \hat{A}_i(z)-i[\hat{A}_0(z),\hat{A}_i(z)]
-i\epsilon_{ijk}[\hat{A}_j(z),\hat{A}_k(z)]=\sum_{a=1}^N(\ldots)_a\delta(z-\mu_a)\,.
\end{equation}
The eigenvalues of the holonomy $\mathcal{P}_\infty=\mbox{diag}(e^{2\pi i \mu_a})$ 
appear on the r.h.s. giving the location of the singularities
and dividing the dual circle $\tilde{S}^1$ into $N$ intervals 
(or less if some $\mu$'s are equal). 
The structure of the singularities was not written out here, 
details can be found in e.g.\ \cite{bruckmann:03c}.

\begin{figure}[t]
\begin{center}
\psfrag{-1/2}{$-1/2$}
\psfrag{1/2}{$1/2$}
\psfrag{A(z)}{$\hat{\vec{A}}(z)$}
\psfrag{z}{$\:z$}
\psfrag{-o}{$-\omega$}
\psfrag{o}{$\omega$}
\psfrag{y1}{$\vec{y}^{(1)}$}
\psfrag{y2}{$\vec{y}^{(2)}$}
\includegraphics[width=0.7\linewidth]{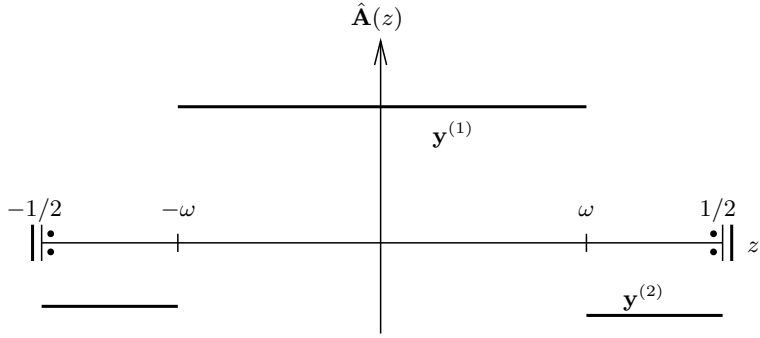}
\caption{Nahm picture for the caloron: piecewise constant dual gauge fields give monopole locations.}
\label{fig_cal_nahm}
\end{center}
\end{figure}

For topological charge 1 (i.e.\ `$1\times 1$-matrices'), 
there are no commutator terms.
Consequently, $\hat{A}_\mu(z)$ is piecewise constant,
the dual zero modes $\hat{\psi}_x(z)$ are piecewise exponential, 
and the caloron gauge field $A_\mu(x)$ can be written down in closed form.

Let us stick to gauge group $SU(2)$ and parametrise the holonomy as 
$\mathcal{P}_\infty=\exp(2\pi i \omega\sigma^3/2)$ (i.e.\ $-\mu_1=\mu_2=\omega\in[0,1/2]$).
Then the dual gauge field\footnote{
$\hat{A}_0$ can be gauged away up to the dual holonomy and contains the time locations of the caloron, 
as it goes together with $x_0$.} $\hat{\vec{A}}(z)$ is constant, 
say $\vec{y}^{(1)}$, between $-\omega$ and $\omega$ 
and also between $\omega$ and $1-\omega$, say there equal to $\vec{y}^{(2)}$,
see Fig.\ \ref{fig_cal_nahm}.
From Nahm's original transform it is known that these are data of one BPS monopole each. 

Hence the dual data indicate a substructure, namely that
the charge $1$ $SU(2)$ caloron has two magnetic monopoles. 
More generally, {\em the charge $1$ $SU(N)$ caloron `dissociates' into $N$ constituent monopoles}.
Hence the caloron realises the old idea of `instanton quarks' \cite{belavin:79} of fractional charge.
The lengthes of the intervals on the dual circle give the monopole masses as
$8\pi^2\nu_i/\beta$, $\nu_i\equiv\mu_{i+1}-\mu_i$, 
which upon integration over $x_0$ make up the unit action of an instanton. 
If some of the holonomy eigenvalues coincide, the corresponding constituents become massless and infinitely spread.

\begin{figure}[t]
\includegraphics[width=0.32\linewidth]{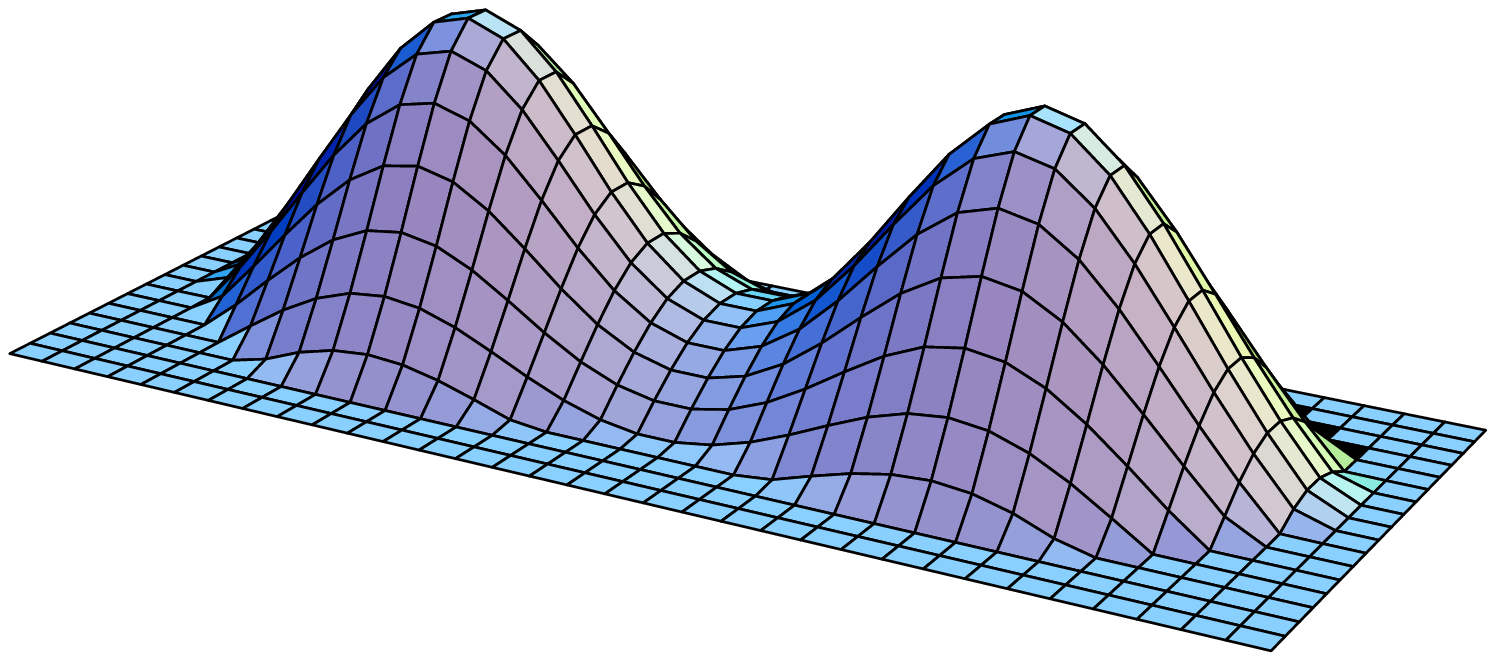}
\includegraphics[width=0.32\linewidth]{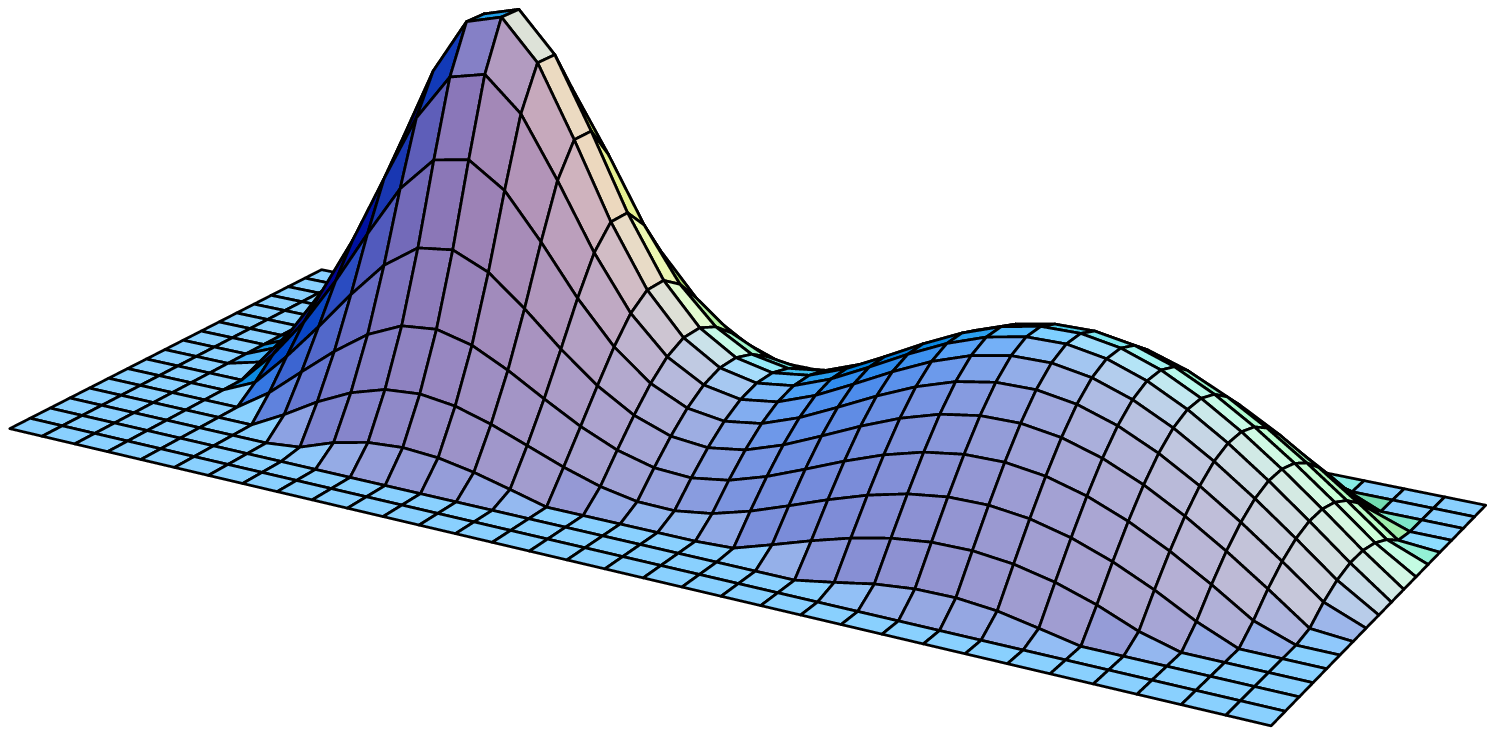}
\includegraphics[width=0.32\linewidth]{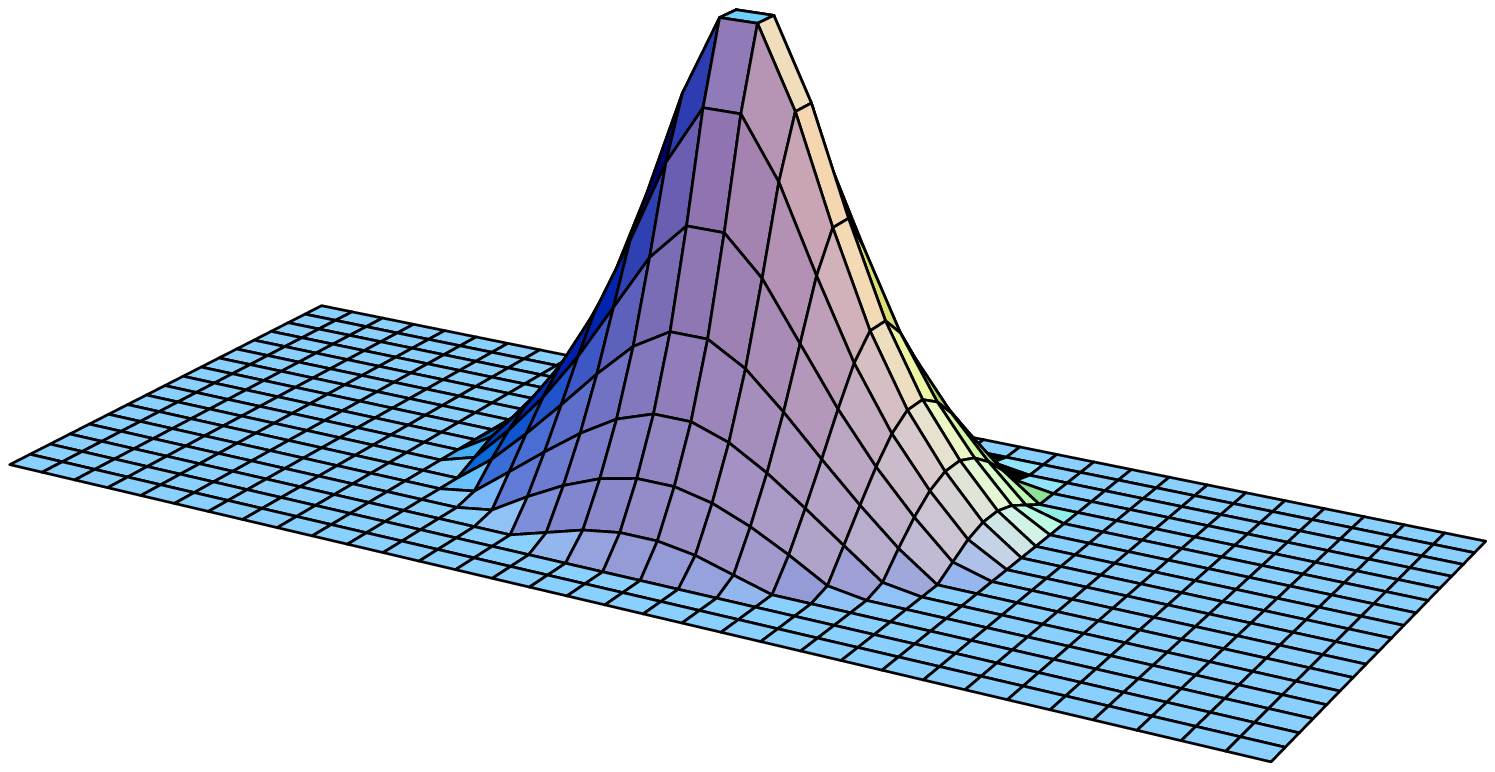}
\includegraphics[width=0.32\linewidth]{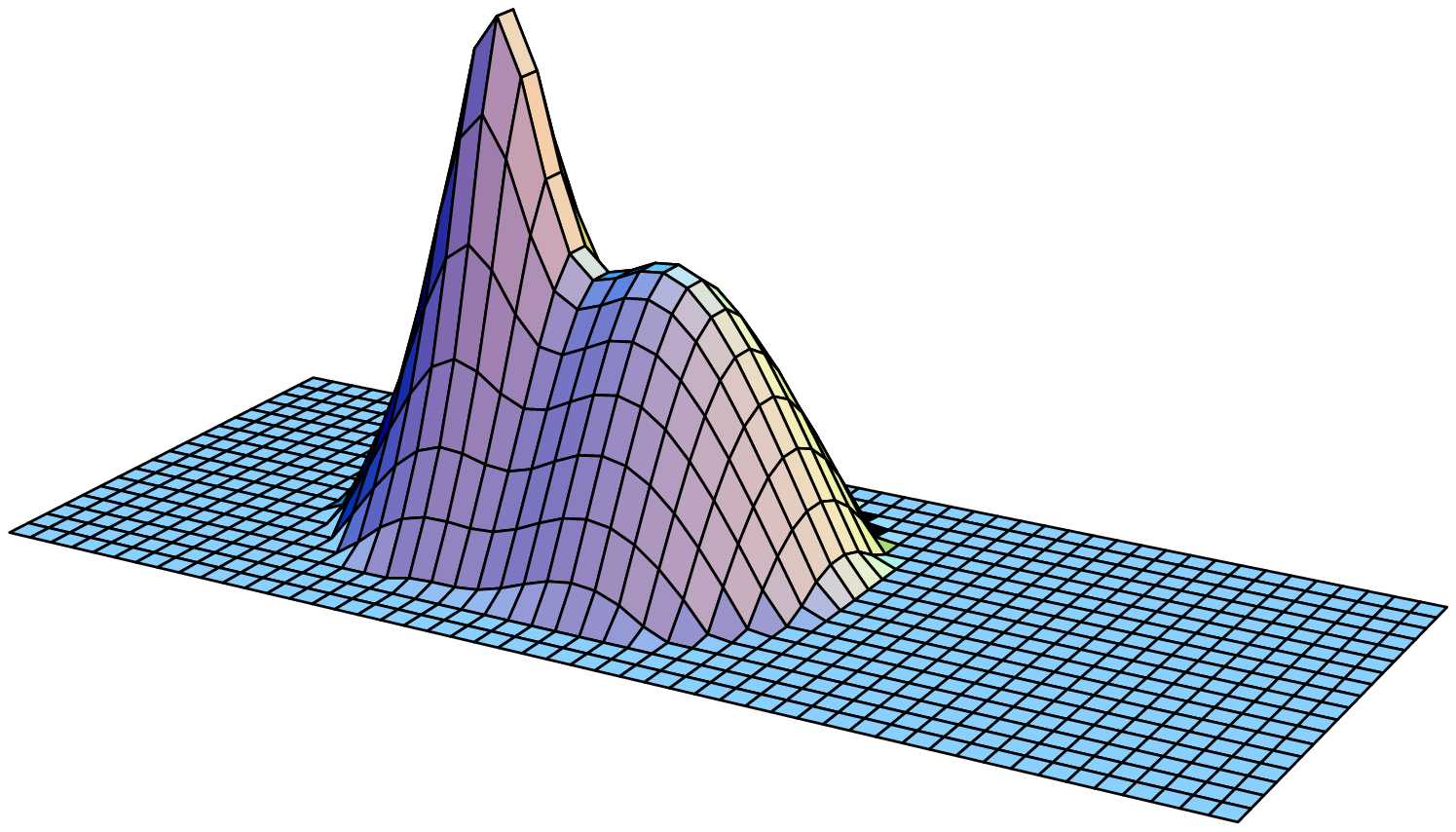}
\includegraphics[width=0.32\linewidth]{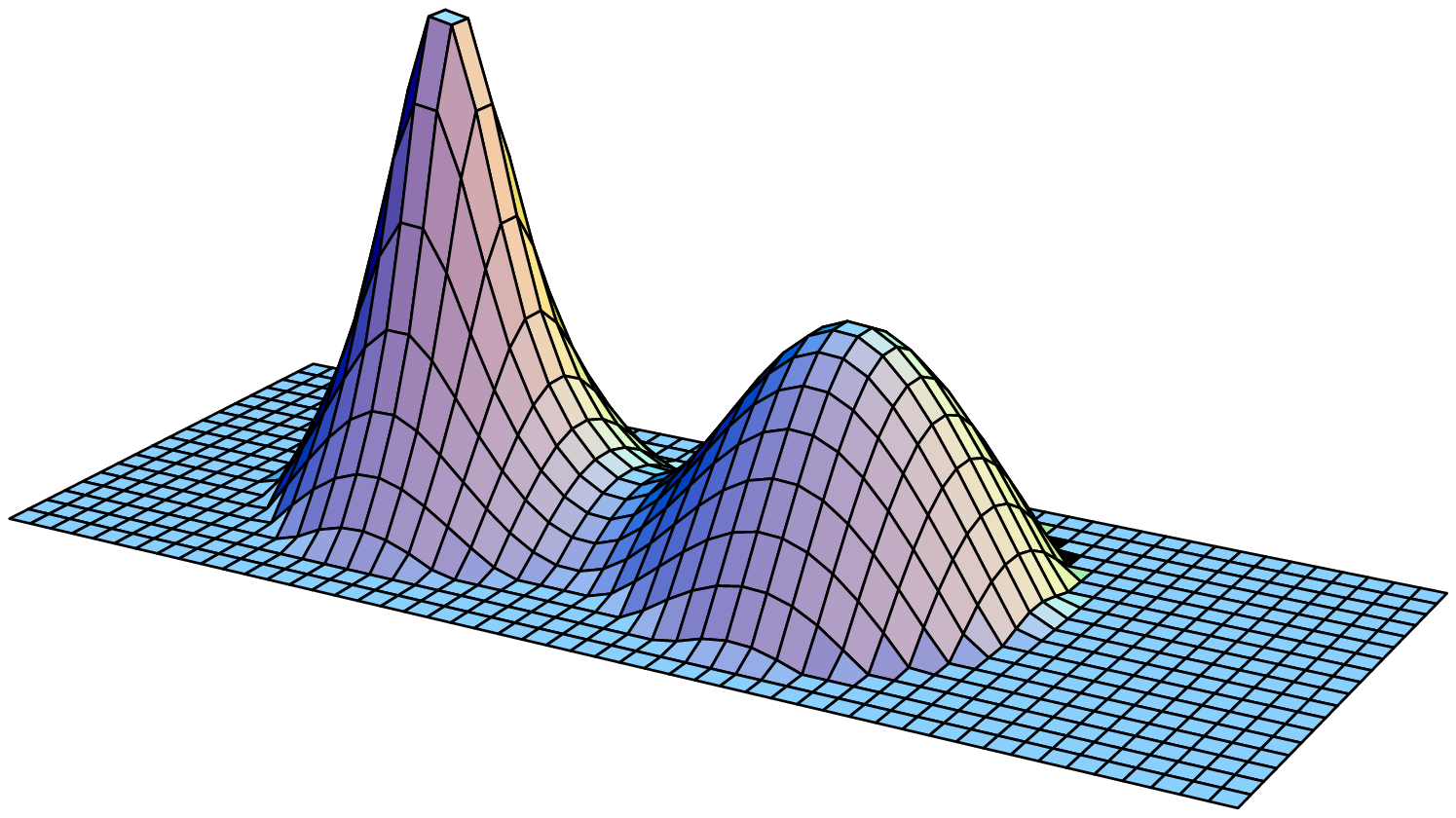}
\includegraphics[width=0.32\linewidth]{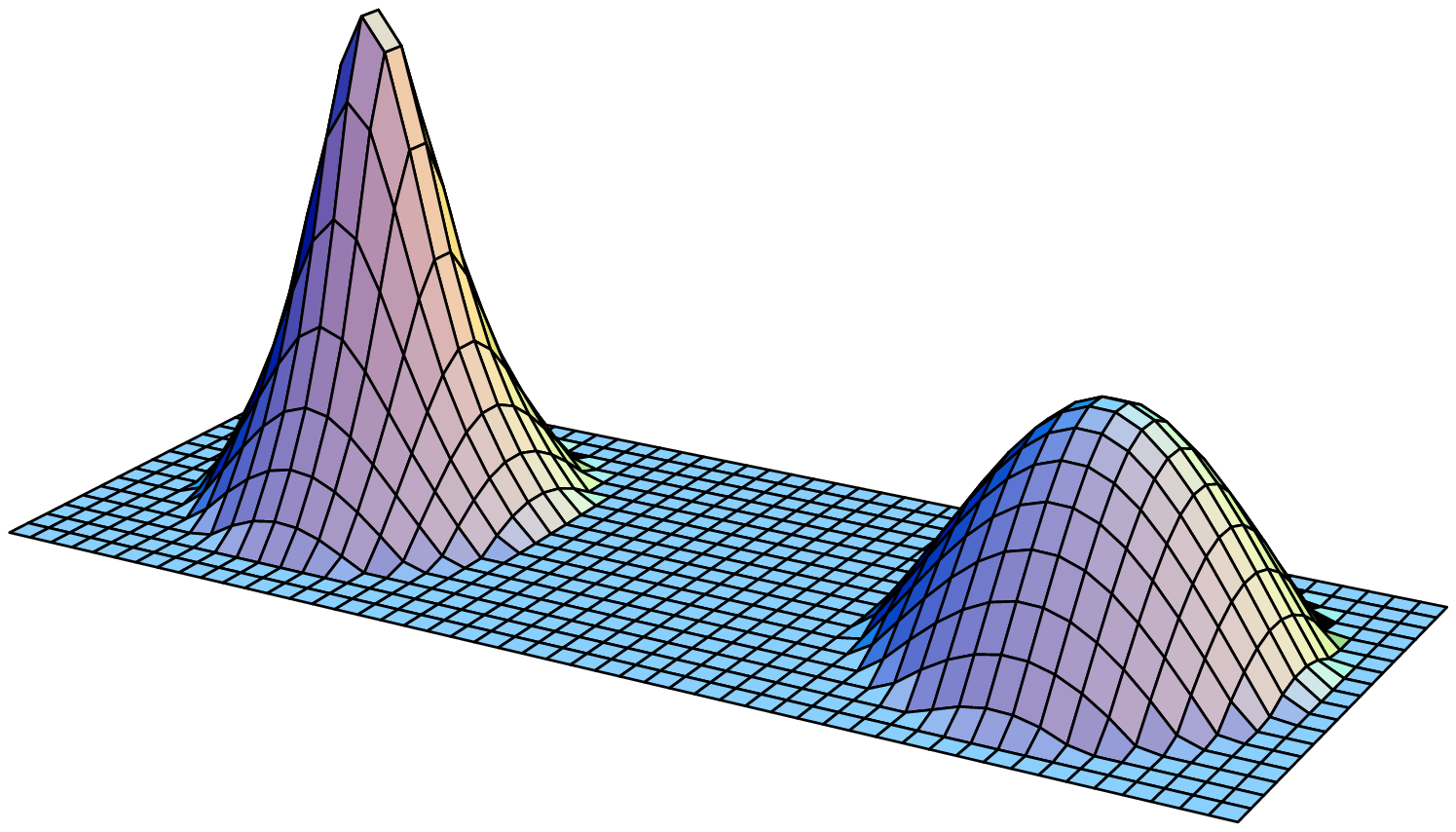}
\caption{Plots of the action density of various calorons as a function of two spatial directions, 
from \protect\cite{kraan:98a}. 
The upper panels show the influence of the holonomy parameter $\omega$ 
(from left to right $1/4$, $1/8$ and $0$)
on the monopole masses. The lower panels show the influence of the size parameter $\rho$ 
(from left to right $1.6$, $1.2$ and $0.8$ in units of $\beta$, $\omega=1/8$) on the separation of the monopoles.}
\label{fig_cal_moduli}
\end{figure}

\subsection{Gluonic features}

Let us discuss the properties of $SU(2)$ caloron solutions in more detail. 
The constituent masses are given by $2\omega$ and $2\bar{\omega}\equiv 1-2\omega$,
see Fig.\ \ref{fig_cal_moduli} top.
The extreme cases $\omega=0,\,1/2$ give trivial holonomy $\mathcal{P}_\infty =\pm \Eins_2$ 
and thus no symmetry breaking (the Higgs field $\log \mathcal{P}_\infty$ vanishes), 
that is the HS caloron with only one magnetic monopole.

For the symmetric case $\omega=\bar{\omega}=1/4$ both monopoles have equal mass 
and are identical from the point of view of action density. 
The holonomy is on the equator of $SU(2)$, $\mathcal{P}_\infty = i\sigma_3$ and traceless,
which makes this case attractive for the confined phase,
see Sect.~\ref{sect_cal_new}.

The action density becomes static\footnote{
The gauge field itself cannot be static due to Taubes' winding \cite{kraan:98a}.}
in the regime of well-separated monopoles $|\vec{y}^{(1)}-\vec{y}^{(2)}|\gg\beta$.
This distance takes over the meaning of the caloron's size.
The monopoles themselves are of fixed size proportional to $\beta$.
When close together the constituents develop a time-dependence 
and merge to an instanton-like lump of size $\rho=\sqrt{|\vec{y}^{(1)}-\vec{y}^{(2)}|\beta/\pi}\ll\beta$,
see Fig.\ \ref{fig_cal_moduli}.

Furthermore, the fields far away from the cores become Abelian along $\mathcal{P}_\infty$. 
In this limit, where exponentially decaying parts of the fields are neglected, 
what is left are dipole fields with sources at the monopole locations. 
This means in particular, that both monopoles have got opposite magnetic and
opposite `electric'\footnote{
The electric charge may also be called `scalar'. 
In any case it is in Euclidean space and quantised due to selfduality. 
That should be kept in mind when one prefers to call the constituents `dyons'.}
charge and the force between them compensates.

Another interesting observable is the Polyakov loop in the bulk. 
It actually passes through $\Eins_2$ and $-\Eins_2$ near the monopoles. 
This strong signature of the substructure is present even when the action density has only one lump. 
The stability of this feature points to a topological origin. 
Indeed, the Polyakov loop has a winding number equal to the topological charge 
and thus must visit both poles to fully cover the gauge group \cite{ford:98,*reinhardt:97b,*jahn:98}.

\subsection{Higher charge calorons and moduli counting}

Calorons of higher charge can help to study the overlap of monopole constituents 
and their superposition problem,
since these calorons consist of $Q$ monopoles of each kind.

\begin{figure}[h]
\includegraphics[width=0.32\linewidth]{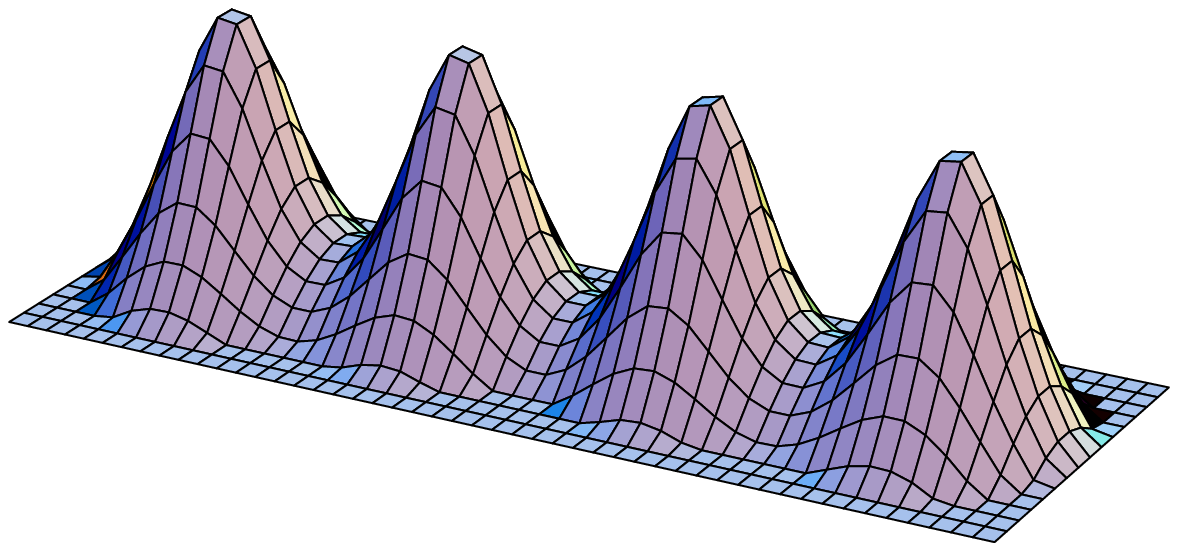}
\includegraphics[width=0.32\linewidth]{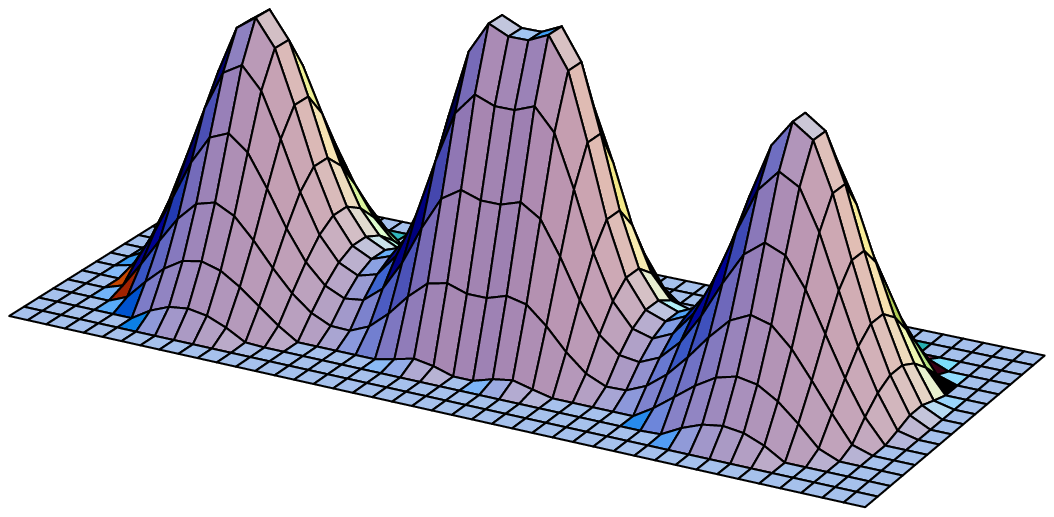}
\includegraphics[width=0.32\linewidth]{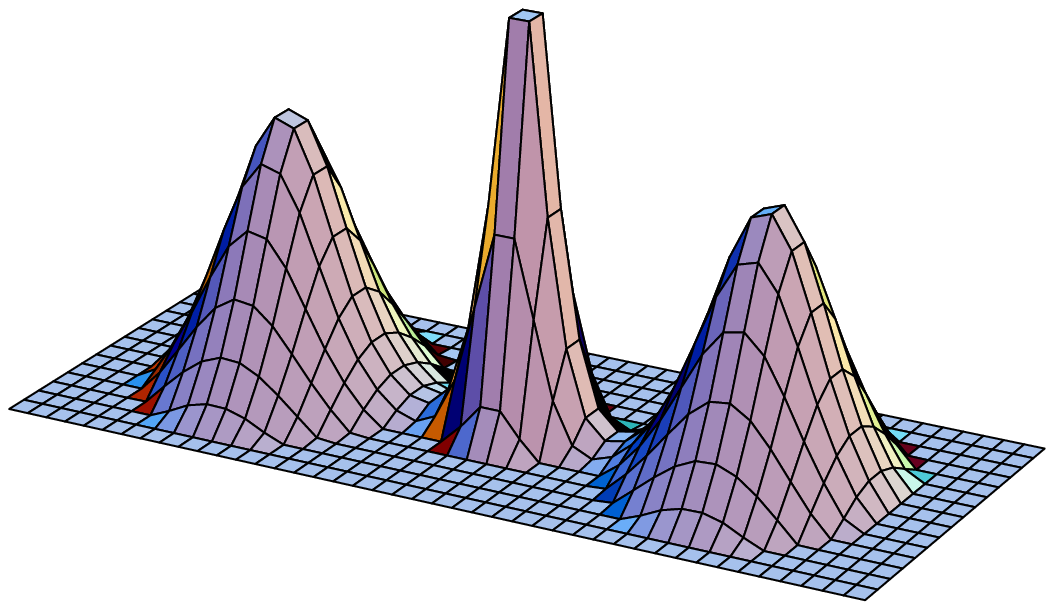}
\includegraphics[width=0.17\linewidth,angle=270]{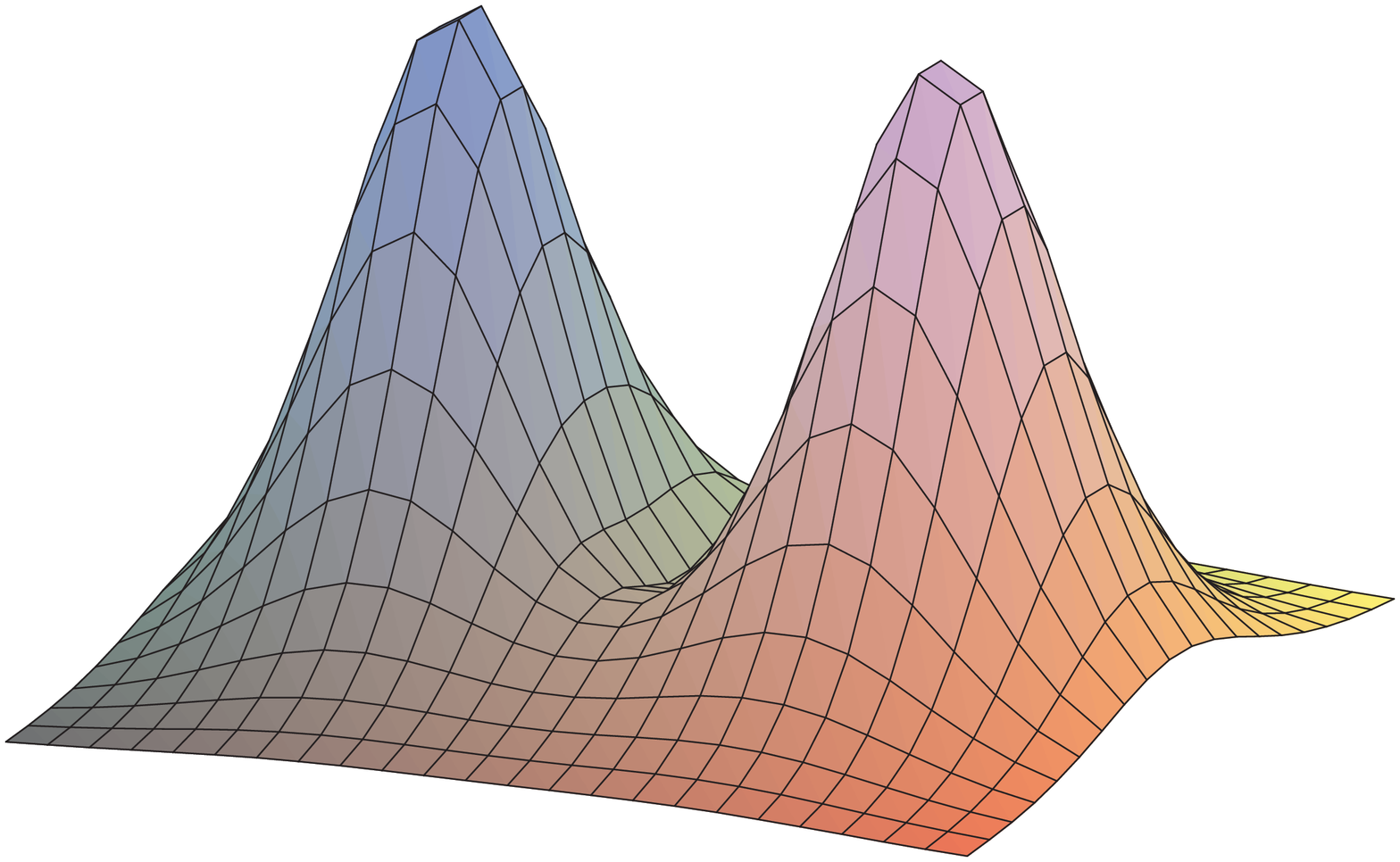}\hfill
\includegraphics[width=0.17\linewidth,angle=270]{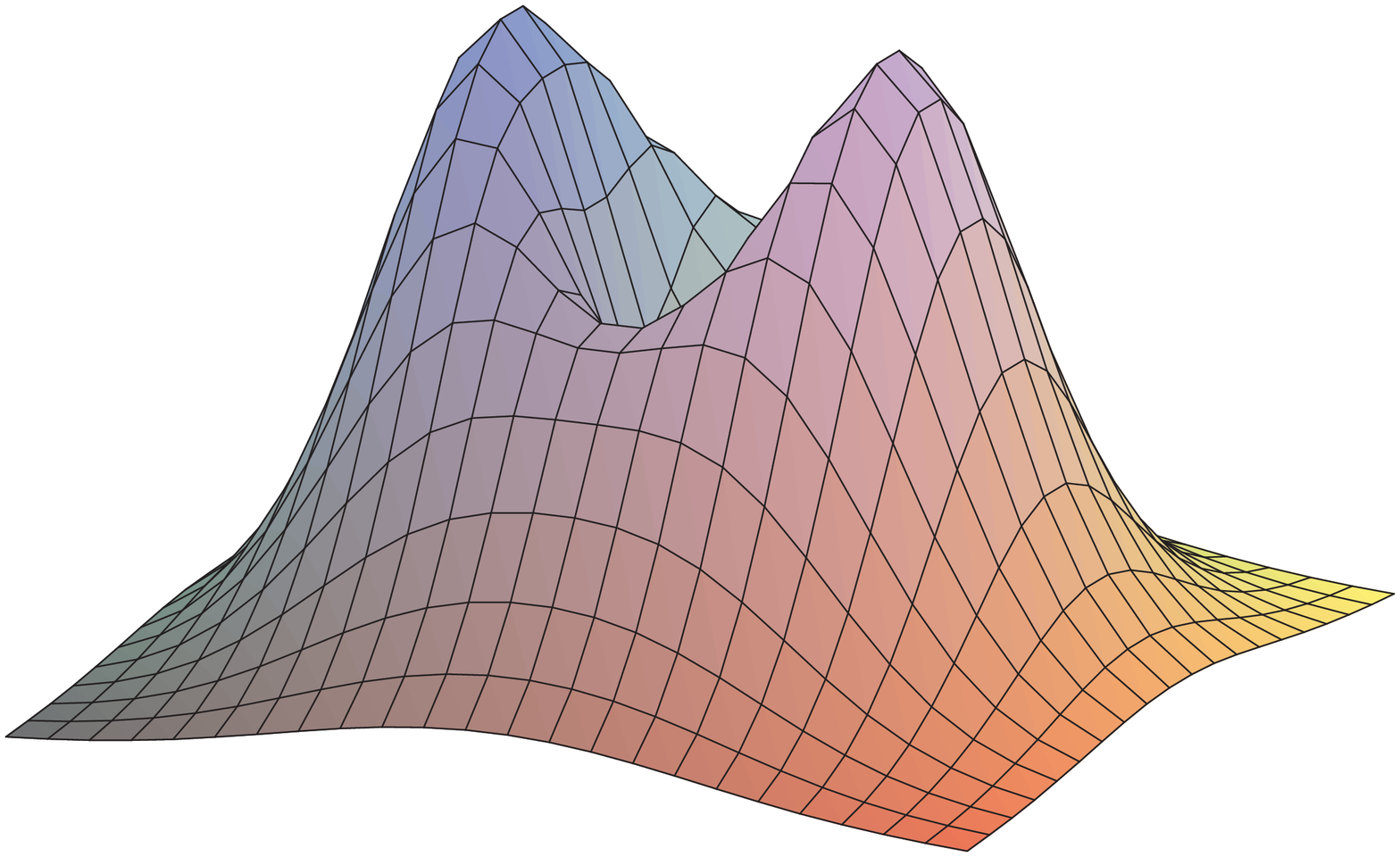}\hfill
\includegraphics[width=0.17\linewidth,angle=270]{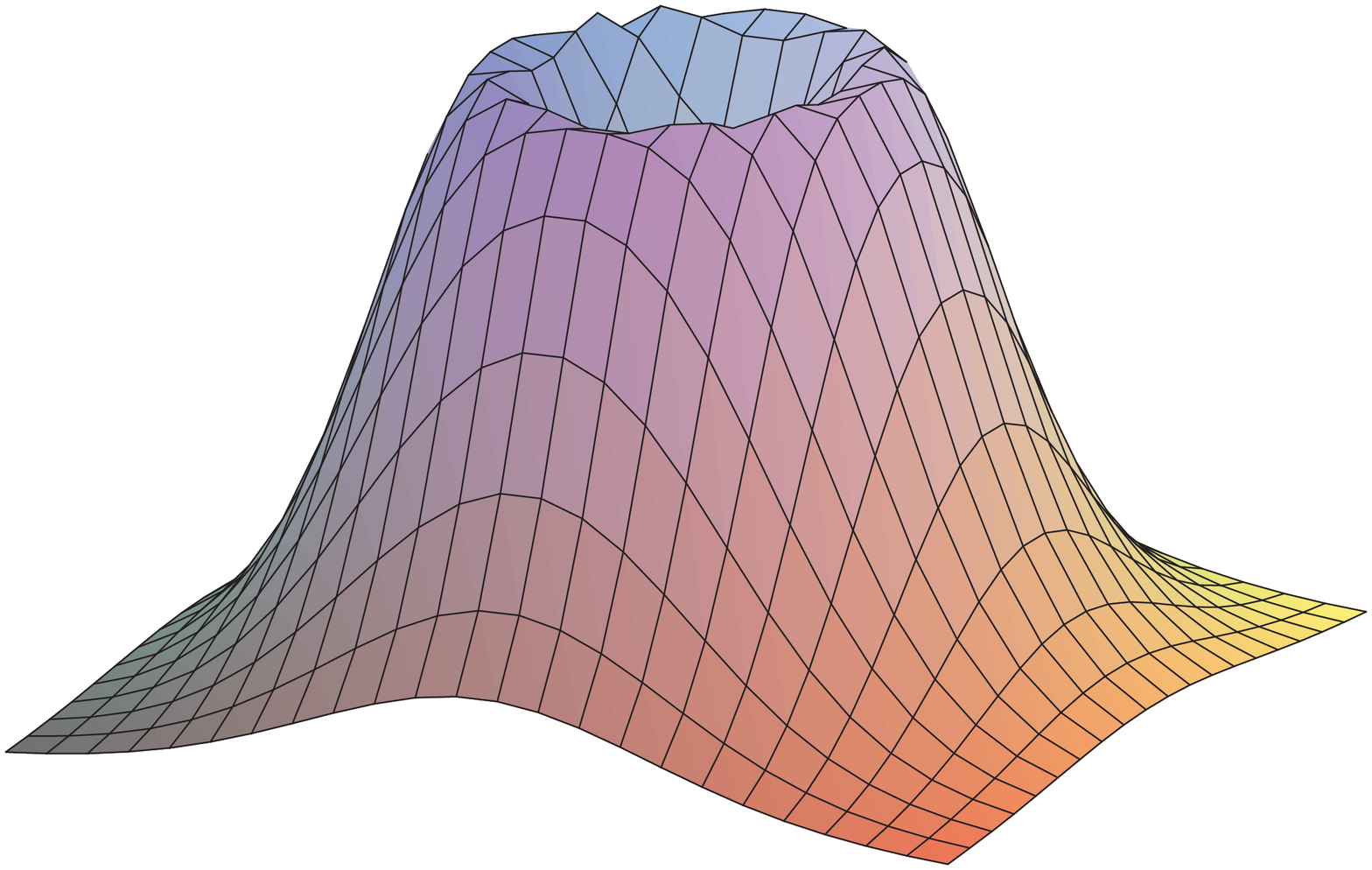}
\caption{Two subclasses of charge 2 calorons 
from \protect\cite{bruckmann:02b} (top) and \protect\cite{bruckmann:04a} (bottom)
described in the text.}
\label{fig_cal_higher}
\end{figure}

The dual gauge fields are now $Q\times Q$ matrices and in general do not commute.
We have found a subclass of caloron solutions of any charge, where the monopoles alternate on a line 
(this restriction came from the trick of avoiding commutators 
by setting two vector components of $\hat{\vec{A}}$ to zero).
In Fig.\ \ref{fig_cal_higher} top I show examples from this class of charge 2.
The locations of the two constituents in the middle are at hand 
and can be varied to form an instanton lump again.

We were also able to explicitly solve for other subsets of charge 2 solutions, where like charges can overlap.
In Fig.\ \ref{fig_cal_higher}  
we zoom into those constituents and find a vulcano structure.
This highly nontrivial shape is caused by a charge $2$-monopole \cite{forgacs:81}.
The extraction of monopole locations from the dual gauge field matrices is quite involved here.


The $8Q$ ($-3$ for global gauge rotations) moduli of charge $Q$ instantons are usually obtained 
from $4Q$ four-dimensional locations, $Q$ sizes and $3Q$ colour orientations.
For dissociated calorons the counting is different: $3Q+3Q$ moduli come from the three-dimensional locations of 
monopoles and antimonopoles, and there are $Q+Q$ moduli of time locations or $U(1)$ phases (again minus global phases).
Infact, the metric on the caloron moduli space is flat 
in these locations in the large separation limit \cite{kraan:98a}.

The holonomy parameter $\omega$ does not count as moduli, rather as a superselection parameter.
The reason is that although it gives rise to a flat direction in the action, 
the corresponding zero mode 
of the fluctuation operator is not normalisable, 
because $A_0$ has to change asymptotically.

\subsection{Fermionic zero modes in the caloron background}

The fermionic zero mode in the caloron background is confronted with a dilemma:
The index theorem for this setting \cite{nye:00} calls for just one zero mode. 
On the other hand, there are 2 (for $SU(N)$ even $N$) monopoles to localise to!

\begin{figure}[t]
\includegraphics[width=0.32\linewidth]{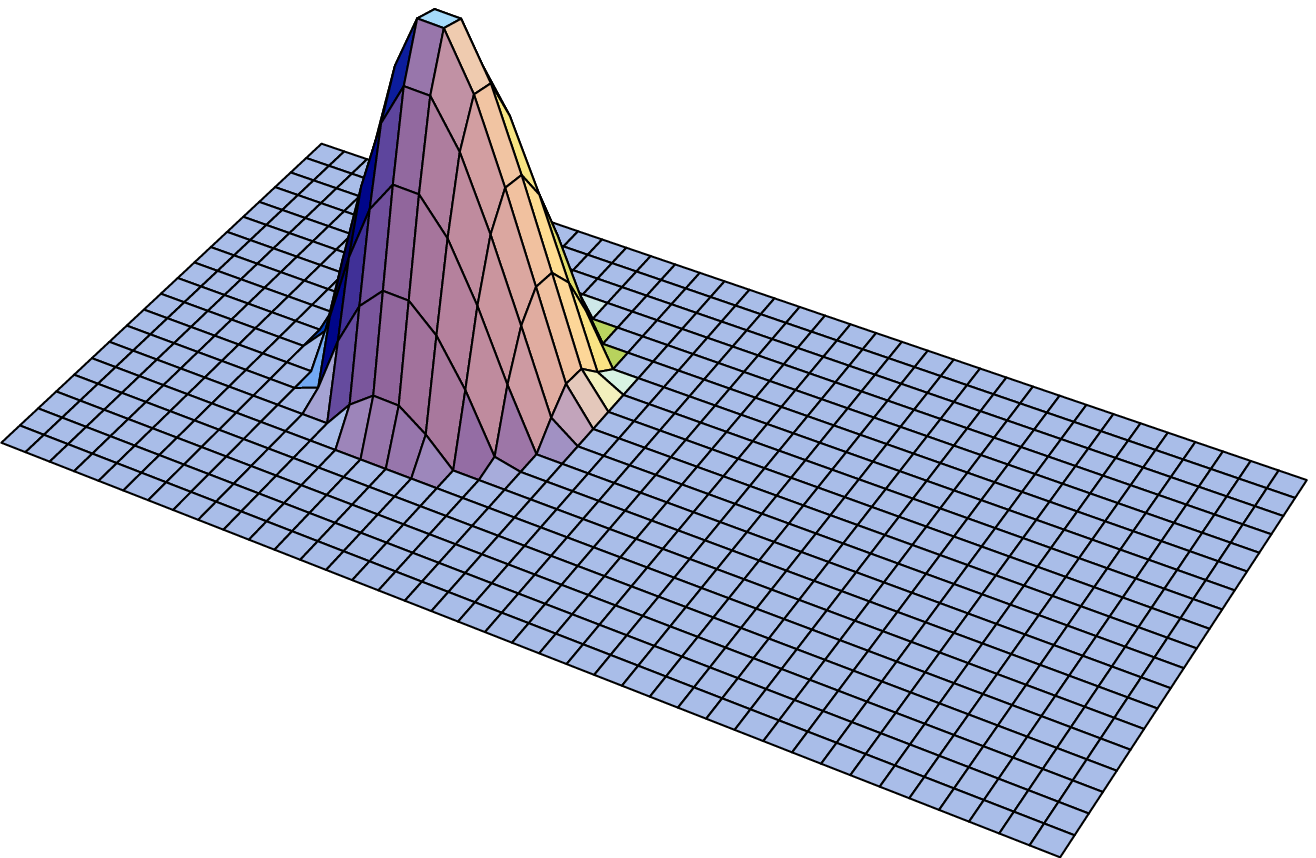}
\includegraphics[width=0.32\linewidth]{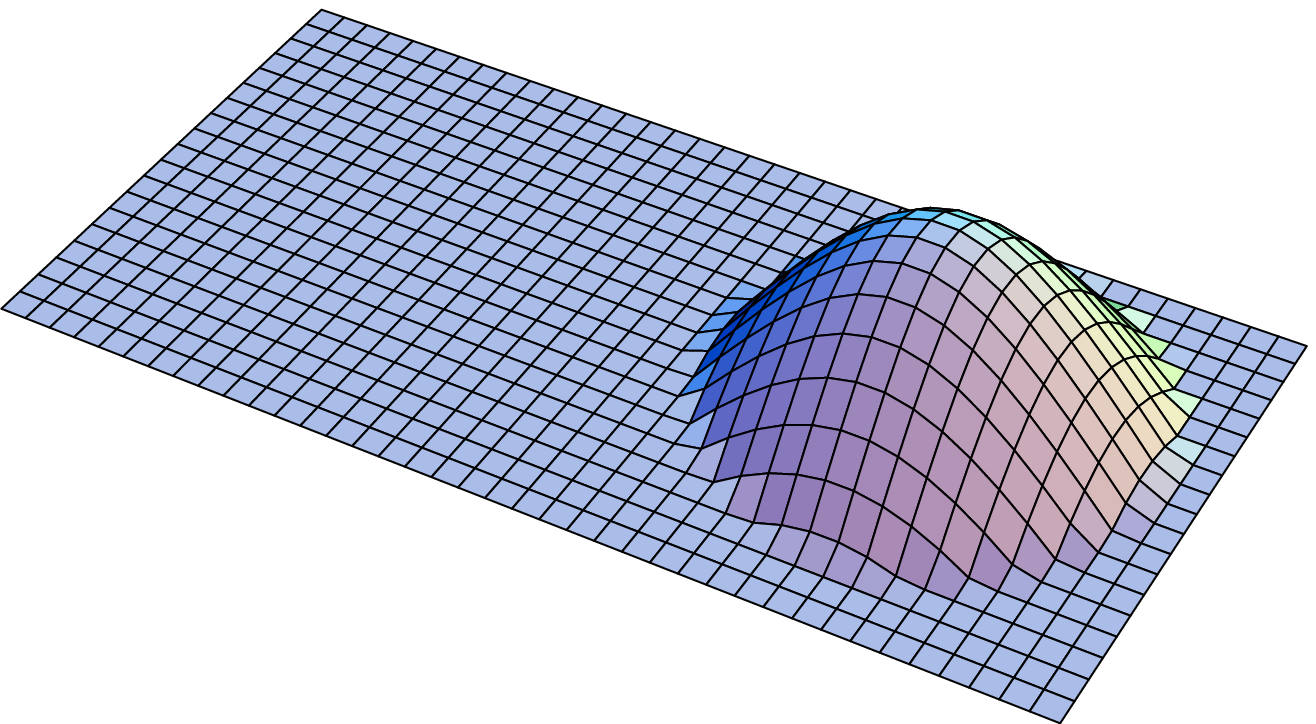}\hfill
\includegraphics[width=0.32\linewidth]{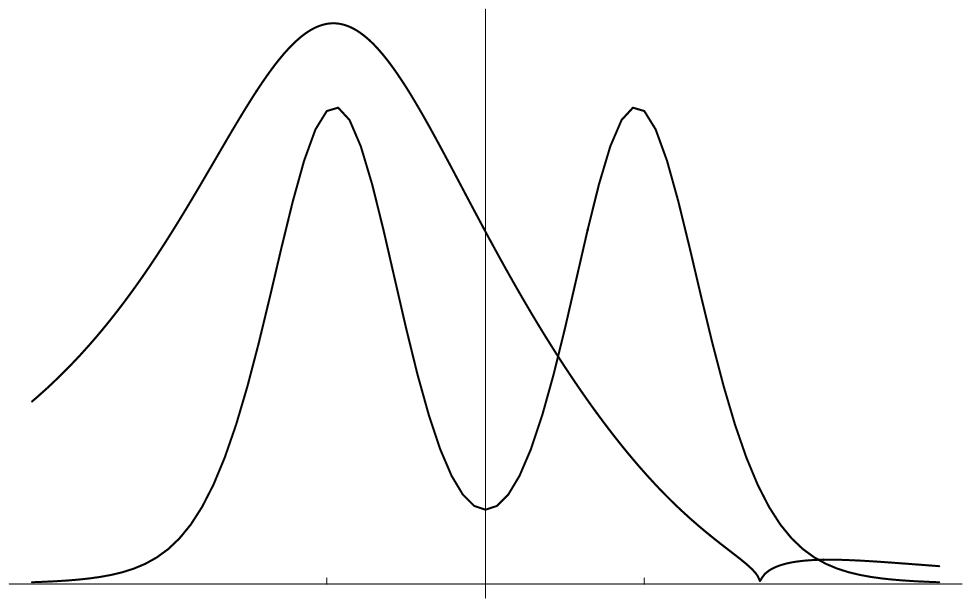}
\caption{Zero modes of calorons: left and middle the profile $|\psi_0(x)|^2$ (logarithmically) 
of the antiperiodic and periodic mode 
for the caloron of Fig.~\protect\ref{fig_cal_moduli}, middle bottom panel, from \protect\cite{garciaperez:99c}.
The right plot shows the periodic zero mode including its zero 
and the action density of an $\omega=1/4$ caloron, from \protect\cite{bruckmann:05a}.}
\label{fig_cal_zero}
\end{figure}

The resolution of the puzzle is that 
{\em the zero mode hops with the boundary conditions in the compact direction}
\cite{garciaperez:99c}.
Let the zero mode be periodic up to a complex phase 
\begin{equation}
\Psi_z(x+\beta e_0)=e^{2\pi i z}\,\Psi_z(x)\,,\qquad
|\Psi_z(x)|^2\mbox{ periodic.}
\end{equation}
It turns out that for $z\in\{-\omega,\omega\}$, including the periodic case, 
$\Psi_z(x)$ is exponentially localised to one constituent monopole, 
whereas for $z\in\{\omega,1-\omega\}$, including the antiperiodic case, it is 
localised to the other one, see Fig.\ \ref{fig_cal_zero}.

At $z=\pm\omega$ the zero mode sees both monopoles, but decays only algebraically.
A completely analogous scenario is valid for higher charge \cite{bruckmann:03a}.

One can understand these facts by making the zero mode periodic,
\begin{equation}
\psi_z(x)=e^{-2\pi i z x_0/\beta}\,\Psi_z(x)\,,
\end{equation}
but $z$ now enters the Weyl-Dirac equation,
\begin{equation}
\sigma_\mu(D_\mu-2\pi i z\delta_{\mu 0})\psi_z(x)=0\,,
\end{equation}
as a mass term (in $4D$ as an imaginary chemical potential). 
It is known from the Callias index theorem \cite{callias:77} that each monopole 
supports a zero mode, when the mass is in `its Higgs range' 
and exactly this allocation takes place inside the caloron.

As can be seen by comparing with Eq.\ (\ref{eqn_nahm_pre}), the zero mode $\psi_z(x)$ 
is the one used in the Nahm transform. 
The spatial components of the dual gauge field in the case of noncompact directions
are obtained by replacing $i\partial_{\vec{z}}$ by $\vec{x}$ in Eq.~(\ref{eqn_nahm_gaugefield}),
\begin{equation}
\hat{\vec{A}}(z)=\int\!\!d^4x\,\psi^\dagger_z(x)\,\vec{x}\,\psi_z(x)
=\langle \vec{x}\rangle_{\psi_z}\,,
\end{equation}
and the localisation of $\psi_z$ is perfectly compatible with the Nahm picture of Fig.\ \ref{fig_cal_nahm},
where $\hat{\vec{A}}(z)$ is piecewise constant at the monopole locations $\vec{y}^{(1),(2)}$.

Does the zero mode notice the `other' zero mode at all? 
Yes, it detects it by a zero in its profile near the other core \cite{bruckmann:05a},
see Fig.~\ref{fig_cal_zero} right. 
This novel property again is of topological origin and might help in the detection of the constituents (see below). 

\section{(Some) Models for QCD}

After having presented the topological objects and in particular the new features of the caloron, 
I want to spend the remaining lecture on discussing their role in the physics of continuum models 
and lattice configurations of QCD.

As nonperturbative objects, magnetic monopoles and instantons 
(and vortices, which I have no time to discuss, see the proceedings of the preceding school \cite{greensite:07a}) are 
{\em natural candidates to explain the infrared phenomena of QCD}
(where the coupling is not small).
The latter have been of interest ever since the advent of QCD and still lack a derivation from first principles.

{\em Confinement} is the fact that quarks and gluons are not observed freely, rather in colourless bound states.
In pure Yang Mills-theory the interquark potential grows linearly with their distance, 
$V_{q\bar{q}}(R)\to \sigma R$, 
with a string tension $\sigma\simeq 1\,GeV/fm$.
The challenge for the theorist is to show an area law for large Wilson loops, 
\begin{equation}
\langle W(R\times T)\rangle\to \exp(-\sigma R T)\,.
\end{equation}
However, only below a critical temperature, as QCD is known to become a quark-gluon plasma 
at temperatures just available at current experiments like RHIC.
For the effect of confinement I will discuss the scenario of the dual supercondutor based on magnetic monopoles. 

Hadrons are massive and to reproduce the hadron spectroscopy data is an obvious task for a model of QCD. 
The very existence of a mass gap 
is one of the Millenium Prize problems.

Although in $\mathcal{L}_{QCD}$ at $m=0$ (the chiral limit, a good approaximation to reality) 
left-handed and right-handed quarks decouple, hadrons do not have parity doublers. 
This phenomenon is called {\em chiral symmetry breaking} and is due to the chiral condensate 
$\langle \bar{\psi}\psi\rangle\simeq - (240\, MeV)^3$. 
The famous Banks-Casher formula \cite{banks:80},
\begin{equation}
\langle \bar{\psi}\psi\rangle=-\fra{\pi}{V}\rho(\lambda=0)\,,
\end{equation}
relates it to the density of eigenvalues at zero virtuality of the Dirac operator. 
I will present how the instanton liquid generates this quantity.
 
Note that (massless) QCD is dimensionless. 
Therefore, all dimensionful observables emerge by quantum effects, the so-called `dimensional transmutation'.
The phenomena are widely believed to be caused by the dynamics of the gauge fields. 
But which nonperturbative degrees of freedom are the relevant ones? And what is their effective action?  
These important questions I will approach now, first in the continuum and later with the help of the lattice. 

\subsection{Semiclassics in QCD: the instanton liquid}

The semiclassical evaluation of the path integral -- 
as demonstrated for the kink in Sect.\ \ref{sect_kink_semi} --
will be repeated now for instantons, for more extensive reviews see 
\cite{schaefer:98,*bruckmann:00c,*diakonov:02}.
At the heart of the method lies again an expansion around classical fields,
$A_\mu(x)=A_\mu^{\rm cl}(x)+a_\mu(x)$,
plus a Gaussian integration.

However, there are several subtleties of this method in gauge theories.
First of all, the space of all gauge fields $A_\mu(x)$ is too big,
since it contains gauge equivalent configurations 
and one would implicitly integrate over the local gauge group.
A gauge must be fixed, and the Faddeev-Popov determinant, 
the so-called ghosts, needs to be included.
For practical reasons we use the background gauge $D_\mu(A^{\rm cl})a_\mu=0$
with Faddeev-Popov operator $-D_\mu^2(A^{\rm cl})$. 

The stationary points $A^{\rm cl}$ are superposed instantons and antiinstantons.
The topological charges of them cancel to typically a few units $Q=0,\,\pm 1, \ldots$
Let me stress that for these approximate solutions there is no
strict separation from perturbative fluctuations.

Concerning the diluteness of the building blocks, 
kinks are localised exponentially with the mass parameter $m$, 
see Eq.\ (\ref{eq_kink_soln}).
The instanton gauge fields of Eq.~(\ref{eq_inst_soln}), however, decay only
algebraically and a priori all values of $\rho$ occur 
(because classical Yang-Mills theory has no scale).
Hence finite density effects and interactions are expected to be more relevant.

The instanton moduli, locations $y_\mu$, sizes $\rho$ 
and colour orientations $U$, will be treated by explicit integration again.

The one instanton weight has been regularised and computed by 't Hooft to be
\begin{equation}
\int\!\! d^4\!y\, d\rho\, d\mu(U) \, J \, e^{-\fra{8\pi^2}{g^2}}
\fra{\det(-D_\mu^2(A^{\rm cl})}
{\sqrt{\det'(-D_\alpha^2(A^{\rm cl})+2i[F_{\mu\nu}^{\rm cl},..])}} 
=\int\!\! d^4\!y\, d\rho\cdot d(\rho)\,.
\end{equation}
The central object in these models is the instanton size distribution,
\begin{equation}
d(\rho)\sim\fra{1}{\rho^5}e^{-8\pi^2/g^2(\rho)}\,,
\end{equation}
which upon using the one-loop $\beta$-function becomes
\begin{equation}
d(\rho)\sim\rho^{b-5}\,,\qquad
b_{{\rm pure}\, SU(N)}=\fra{11N}{3}\,.
\end{equation}
Note that the classical scale invariance has been broken by quantum effects.

The size distribution suppresses small instantons, 
but it diverges for large $\rho$.
This could have been expected from the use of the perturbative $\beta$-function
in the infrared.
In most works about this model large instantons are cut-off empirically, 
but the problem is not fully clarified.

Instanton interactions (the deviation of the action 
of $n$ instantons and antiinstantons from the naive sum $n\cdot 8\pi^2/g^2$)
have been calculated by using a hard core \cite{ilgenfritz:81,*muenster:81} 
and a variational principle \cite{diakonov:84}.
The interactions depend on the relative colour orientations, 
but fortunately are repulsive on average.
The resulting size distribution,
\begin{equation}
d(\rho)\sim\rho^{b-5}\exp\left(-\,\#\,\sqrt{\fra{n}{V}}\rho^2\right)\,,
\end{equation}
is peaked around some $\bar{\rho}$. 
This has lead to the instanton liquid model proposed by Shuryak \cite{shuryak:81},
where the following values of the average instanton size and separation are used
\begin{equation}
\bar{\rho}\simeq\fra{1}{3}fm\,,\qquad
\bar{R}=\left(\fra{n}{V}\right)^{-\fra{1}{4}}\simeq 1 fm\,.
\label{eqn_phen_Rrho}
\end{equation}
The packing fraction $\pi^2\bar{\rho}^4/\bar{R}^4\simeq 1/8$ indicates 
that the system is fairly dilute.

This model is successful in predicting the chiral condensate 
(see the next subsection) 
and hadronic properties. 
However, to make contact with confinement, unphysically large instantons or 
particular arrangements of their colour orientations 
or strong overlap effects in regular gauge are needed 
\cite{diakonov:95b,*gonzalez-arroyo:96c,*negele:04}.
As I will discuss below, the scenario of a finite temperature 
with its calorons may improve the situation.

The topological susceptibility $\chi_{\rm top}=\lim_{V\to\infty}\fra{\langle Q^2\rangle}{V}$
can be estimated by a simple argument. 
The average topological charge vanishes due to CP invariance.
Therefore, the number of instantons in the instanton liquid 
equals the number of antiinstantons on average. 
Then the topological susceptibility equals the number variation 
in this grandcanonical ensemble, 
roughly
\begin{equation}
\chi_{\rm top}\simeq\fra{\langle (n_I-n_{\bar{I}})^2\rangle}{V}
\simeq\fra{\langle n\rangle}{V}
\simeq(1\, fm)^{-4}\,,
\end{equation}
which is in good agreement with the value of $(180\, MeV)^{-4}$
from the Witten-Veneziano relation \cite{witten:79a,*veneziano:79}.

\subsubsection{Instantons and $\rho(0)$}

The instanton liquid model generates a finite density of fermionic modes at eigenvalue 0
in the following way. Consider, say, 3 instantons and 2 antiinstantons of any size and colour orientation, 
but all well separated. 
Each individual object brings its own zero mode (cf.\ Sect.\ \ref{sect_inst_zero}). 
These 5 modes arrange themselves into one exact (and chiral) zero mode 
according to the index theorem and the total charge 1.
In addition, there will be 4 near zero modes.

The analogy from condensed matter physics are atoms that have a bound state for electrons 
(which are then localised). 
A finite density of these sources generates a band in the spectrum (with delocalised wave functions, conductivity and so on).

To get a quantitative handle on this band, 
we need the zero eigenvalue splitting in the background of an instanton/antiinstanton pair.
In degenerate perturbation theory, the new $\lambda$'s are 
(zero plus) the eigenvalues of the perturbation sandwiched between the unperturbed states. 
Here we obtain the quasi zero eigenvalues in terms of the overlap integral $T$,
\begin{equation}
\lambda=\pm T_{I\bar{I}}\,,\qquad
T_{I\bar{I}}=\int\!\!d^4x\,\psi^\dagger_{I}(x-y_I)i\gamma^\mu\partial_\mu\psi_{\bar{I}}(x-y_{\bar{I}})\,.
\end{equation}
The spread of the band follows as the average splitting,
\begin{equation}
\langle |T_{I\bar{I}}|^2\rangle_{\rm locations,\,orientations}
\sim \fra{n}{V}\rho^2\,.
\end{equation}
Involving knowledge about the band's shape from random matrix theory,
\begin{equation}
\fra{1}{V}\rho(\lambda=0)
\sim\sqrt{\fra{n}{V}}\fra{1}{\rho}
\sim\fra{1}{\bar{R}^2}\fra{1}{\bar{\rho}}\,,
\end{equation}
and using the phenomenological values from Eq.~(\ref{eqn_phen_Rrho}),
one arrives at a chiral condensate quite close to the phenomenological value.

Let me remark that chiral symmetry breaking seems to be rather robust: 
Ensembles of basically any object with a zero mode attached 
(and even random matrices)
have the potential to generate $\rho(\lambda=0)$.
Therefore, monopoles and vortices are relevant for this effect as well.
One might even speculate that the QCD vacuum is `democratic' in the sense 
that it contains all possible topological objects,
intertwined and equally important.

\subsubsection{New aspects by calorons}
\label{sect_cal_new}

At finite temperature, instanton ensembles will undergo interesting physical modifications.
As we have seen in Sect.~\ref{sect_cal}, 
large calorons are pairs of magnetic monopoles.
Beside possible relations to the Dual Superconductor picture 
discussed in the next section,
this could lead to a different suppression mechanism in the moduli integrals of the semiclassical treatment.
Furthermore, the properties of the constituents depend on the asymptotic Polyakov loop,
which is sensitive to the order parameter $\langle\tr\mathcal{P}\rangle$. 
In particular, the equal mass monopoles of traceless holonomy 
should be more relevant for the confined phase, whereas in the deconfined phase one of the monopoles becomes light.

The specific properties of calorons have been used to compute 
the gluino condensate in supersymmetric gauge theories 
\cite{davies:99,*diakonov:03}.
The quantum weight of the caloron has been calculated by Diakonov et al.\ \cite{diakonov:04a}.
It has an interesting consequence for the one-loop effective potential as a function of $\tr \mathcal{P}$. 
The trivial values $\mathcal{P}=\pm\Eins$, which are favoured perturbatively at high temperatures,
become unstable when the nonperturbative contribution of a caloron ensemble is added.
Hence calorons indicate at least the onset of confinement. 

A numerical simulation of a caloron ensemble \cite{gerhold:06} gave evidence 
for the influence of the holonomy on the physics at finite temperature, too:
calorons of (fixed) nontrivial holonomy give rise to a linearly rising potential, 
while trivial holonomy ones do not.

\subsection{The Dual Superconductor picture}

\begin{figure}[h]
\begin{minipage}{0.48\linewidth}
\psfrag{B}{\large $\!\vec{B}$}
\includegraphics[width=\linewidth]{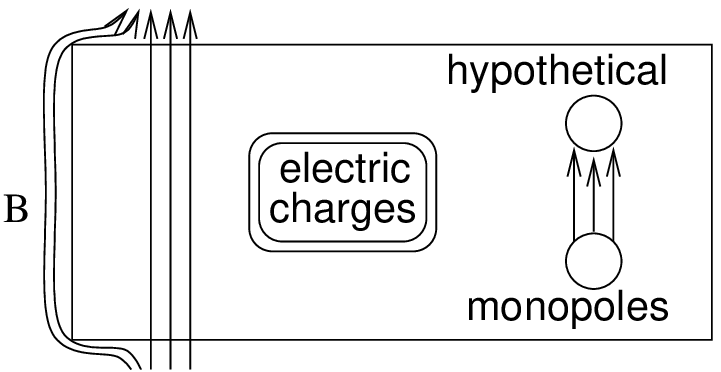}\hfill
\end{minipage}\hfill\begin{minipage}{0.42\linewidth}
\psfrag{q}{\large $q$}
\psfrag{p}{\large $\bar{q}$}
\includegraphics[width=\linewidth]{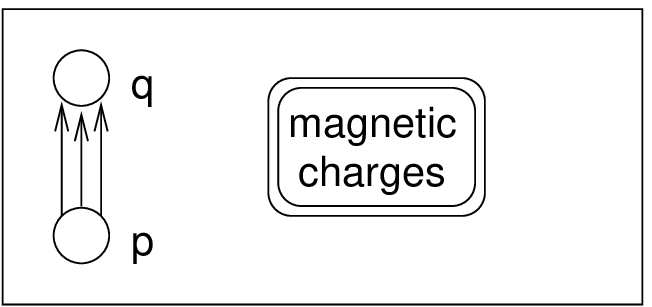}
\end{minipage}
\caption{The Meissner effect in the superconductor (left) and its analogy in the QCD vacuum (right).}
\label{fig_super}
\end{figure}

In a conventional superconductor, Cooper pairs (made of two electrons) condense
and squeeze the magnetic field --
if it penetrates the sample at all --
into flux tubes.
This so-called Meissner effect would also connect hypothetical magnetic monopoles, 
see Fig.\ \ref{fig_super} left.

The idea of viewing the QCD vacuum as a {\em Dual Superconductor} goes back to the 70's
\cite{nambu:74,*parisi:75,*mandelstam:76,*thooft:76a}.
One replaces the magnetic flux tube by a chromoelectric one.
This then connects quarks and antiquarks, see Fig.\ \ref{fig_super} right, 
and thereby generates a constant force respectively a linearly rising interquark potential, 
the signature of confinement.

The big question remains how to obtain magnetic monopoles, 
the dual Cooper pairs, in an unbroken gauge theory like QCD?
The proposal to use gauge fixing came again from 't~Hooft \cite{thooft:81a}.
Let us choose an auxiliary field $\varphi(x)$ transforming in the adjoint representation.
The Abelian gauge uses a gauge transformation $\Omega$ that makes $\varphi$ diagonal, 
for $SU(2)$ just $\Omega^\dagger\varphi\Omega\sim\sigma^3$.
The residual gauge freedom consists of a local $U(1)$ around $\sigma^3$, 
for $SU(N)$ it is the maximal Abelian or Cartan subgroup $U(1)^{N-1}$.

Splitting the gauge field $A_\mu=\{A_\mu^{1,2},A_\mu^3\}$ 
we have got fields transforming like adjoint matter
plus a residual `photon'.
Now the {\em Abelian projection} means to neglect $A_\mu^{1,2}$, 
which is said to become massive by quantum fluctuations. 
This would lead to a local $U(1)$ theory if there were no {\em defects}, 
remnants of the non-Abelian nature of the original gauge theory.

\enlargethispage{2\baselineskip}

Obviously, the gauge fixing procedure is ambiguous at $\phi^a(x)=0$. 
In Sect.~\ref{sect_mon_abel} we have come across an example of this, 
namely the BPS monopole as a static Yang-Mills configuration in unitary gauge, 
i.e.\ identifying $\varphi$ with $\phi=A_0$.
`Combing' $\varphi^a(x)$ to a diagonal form fails at the monopole location
$\varphi(\vec{x}=\vec{y})=0$ because of the hedgehog structure. 
Generalising this observation it follows that 
{\em the Abelian gauge fixing induces worldlines of magnetic monopoles}, 
closed due to charge conservation.
Having produced the magnetic monopoles, the analogue of the condensation is
that in the confined phase of QCD the monopole worldlines are expected to {\em percolate}.\\

By construction there are many Abelian gauges, 
the most popular being the Maximal Abelian gauge \cite{thooft:81a} 
and the Laplacian Abelian gauge \cite{vandersijs:97}.
They can induce different defects on the same configuration
(there is no reason why they should agree), 
which shows the ambiguity of this procedure for the first time.

In the continuum, topology predicts the existence (of a minimal number), 
but not the precise realisation of defects \cite{jahn:00,*bruckmann:02aa}. 
Instantons for example induce small monopole loops around their centers,
which tend to become larger in an instanton ensemble \cite{hart:96,*brower:97b}.

On the lattice there exists a procedure to identify monopoles \cite{degrand:80}
as well as lattice variants of Abelian gauges \cite{kronfeld:87a}. 
The empirical findings of {\em Abelian and monopole dominance} 
support the mechanism of the Dual Superconductor: 
the Abelian field $A_\mu^3$ alone generates 92\% of the original string tension \cite{suzuki:90}, 
likewise a further restriction to the monopole (singular) part of this gauge field 
keeps 95\% of the Abelian string tension \cite{stack:94}.

However, there are several drawbacks of this method. 
First of all, the physical results depend on the choice of gauge. 
Secondly, the Monte Carlo-sampling is done with the full field 
and the reduction described above is performed only in observables.
Therefore, this is not an effective theory and
it is not settled what would be the guiding principle and small parameter for the latter. 
Moreover, the representation-dependence of the string tension comes out wrongly \cite{greensite:07a}.

\section{Topological objects in lattice gauge theory}

Numerical simulations of gauge theories on a space-time lattice have delivered 
many important quantitative results to date.
Naturally, the next task is to {\em understand} these effects, 
e.g.\ in terms of continuum objects. 

The lattice data do have the potential to lead support to physical models. 
However, their interpretation is hampered by the fact, 
that a typical lattice configuration is dominated by UV (i.e.\ order lattice spacing) fluctuations.
In the following I will address means to get access to the underlying IR degrees of freedom.

\subsection{Cooling}
\label{sect_cooling}

Cooling \cite{berg:81,*hoek:86,*ilgenfritz:86} is an iterative procedure, 
where each link $U_\mu$ is replaced by the corresponding sum of staples $\tilde{U}_\mu^{(\nu)}$,
\vspace*{-1cm}

\begin{equation}
U_\mu(x)\rightarrow P\left(\sum_{\nu\neq\mu}\tilde{U}_\mu^{(\nu)}(x)\right)\,,\quad
\tilde{U}_\mu^{(\nu)}(x)=U_\nu(x)U_\mu(x+\hat{\nu})U_\nu^\dagger(x+\hat{\mu})\,,\qquad
\psfrag{m}{$\mu$}
\psfrag{n}{$\nu$}
\psfrag{U}{$U_\mu$}
\psfrag{u}{$\tilde{U}_\mu^{(\nu)}$}
\includegraphics[width=0.15\linewidth]
{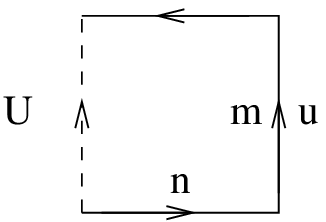}
\end{equation}
projected back onto the gauge group, denoted by $P$
(for $SU(2)$ just a multiplication with a scalar).
This process is local and reduces the action with as fixed points the solutions of equations of motion\footnote{
on the lattice, but reflecting continuum solutions quite well}.
Variants of cooling include smearing (which averages the staple with the old link 
and is equivalent to RG cycling \cite{degrand:98b}) 
and the use of improved actions \cite{iwasaki:83,*garciaperez:93}.

\begin{figure}[h]
\begin{center}
\psfrag{S}{$S$}
\psfrag{Q}{$Q$}
\psfrag{cooling}{cooling sweeps}
\includegraphics[width=0.4\linewidth]
{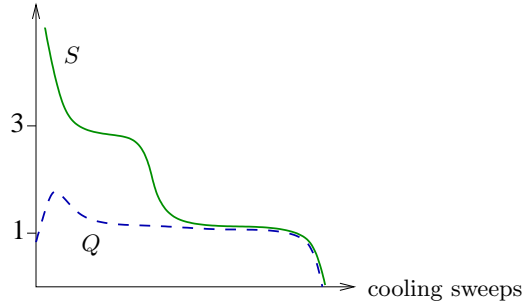}
\caption{A typical cooling history with a plateau at $3 \simeq S > Q \simeq 1$, 
probably made of 2 instantons and 1 antiinstanton, 
and a plateau at $ S \simeq Q \simeq 1$, a single instanton.}
\label{fig_cool_hist}
\end{center}
\end{figure}

A typical cooling history is shown in Fig.\ \ref{fig_cool_hist}.
In the early stage quantum fluctuations are removed and the action is typically reduced by several orders of magnitude.
As for the to\-po\-lo\-gi\-cal charge, there exist gluonic definitions 
discretising the density $\tr F_{\mu\nu}\tilde{F}_{\mu\nu}$ of Eq.~(\ref{eqn_def_Q}).
These usually yield integers for the total charge $Q$ at the plateaus that follow in the cooling history,
because the configurations there are smooth enough.
At these plateaus the configurations consist of selfdual and antiselfdual objects locally.

Inbetween the plateaus, instantons and antiinstantons 
(at finite temperature constituents of them \cite{bruckmann:04b}) annihilate.
In the late stage of cooling one obtains completely selfdual or anti\-self\-dual solutions, 
which finally might fall through the mesh 
(depending on the details of the cooling).

Cooling can be used as a tool to investigate classical solutions. 
For instance, 
the plots shown as appetiser were obtained by long overimproved cooling
on a finite temperature ($16^3\cdot4$) lattice and resemble the continuum charge 2 solutions 
of Fig.~\ref{fig_cal_higher}, bottom right panel, very well.

The main question is whether cooling (or smearing) also gives insight into the QCD vacuum.
Cooling is widely trusted w.r.t.\ global observables like topological charge $Q$ 
and its susceptibility, 
respectively.
When it comes to local objects, one has to keep in mind that the density and sizes 
of e.g.\ instantons are modified in the cooling process. 
Moreover, the method is biased to classical solutions. 
So the fact that the findings of cooling are consistent 
with the instanton liquid model is not a proof of the latter.
For the determination of when to stop cooling infrared features of the system like the string tension should be monitored. 
I will come back to this point in a moment.

\subsection{Fermionic techniques}

An alternative to the described gluonic methods is to use Dirac operators with good chiral properties
(i.e. fulfilling the Ginsparg-Wilson relation as discussed in Tom deGrand's lecture at this school)
as a tool to investigate lattice configurations.
This approach mainly relies on the phenomenon of {\em localisation}.
Low-lying Dirac eigenmodes are fairly smooth, 
because the small `energy' forbids large (covariant) momenta.
Moreover, zero modes of instantons and monopoles are localised to the cores of the latter.
Hence the low-lying fermionic modes shall find the relevant continuum objects\footnote{
However, the discrimination of topological modes from those caused by lattice dislocations can be delicate.}.

\begin{figure}[t]
\begin{center}
\psfrag{10}{}
\psfrag{15}{}
\epsfig{file=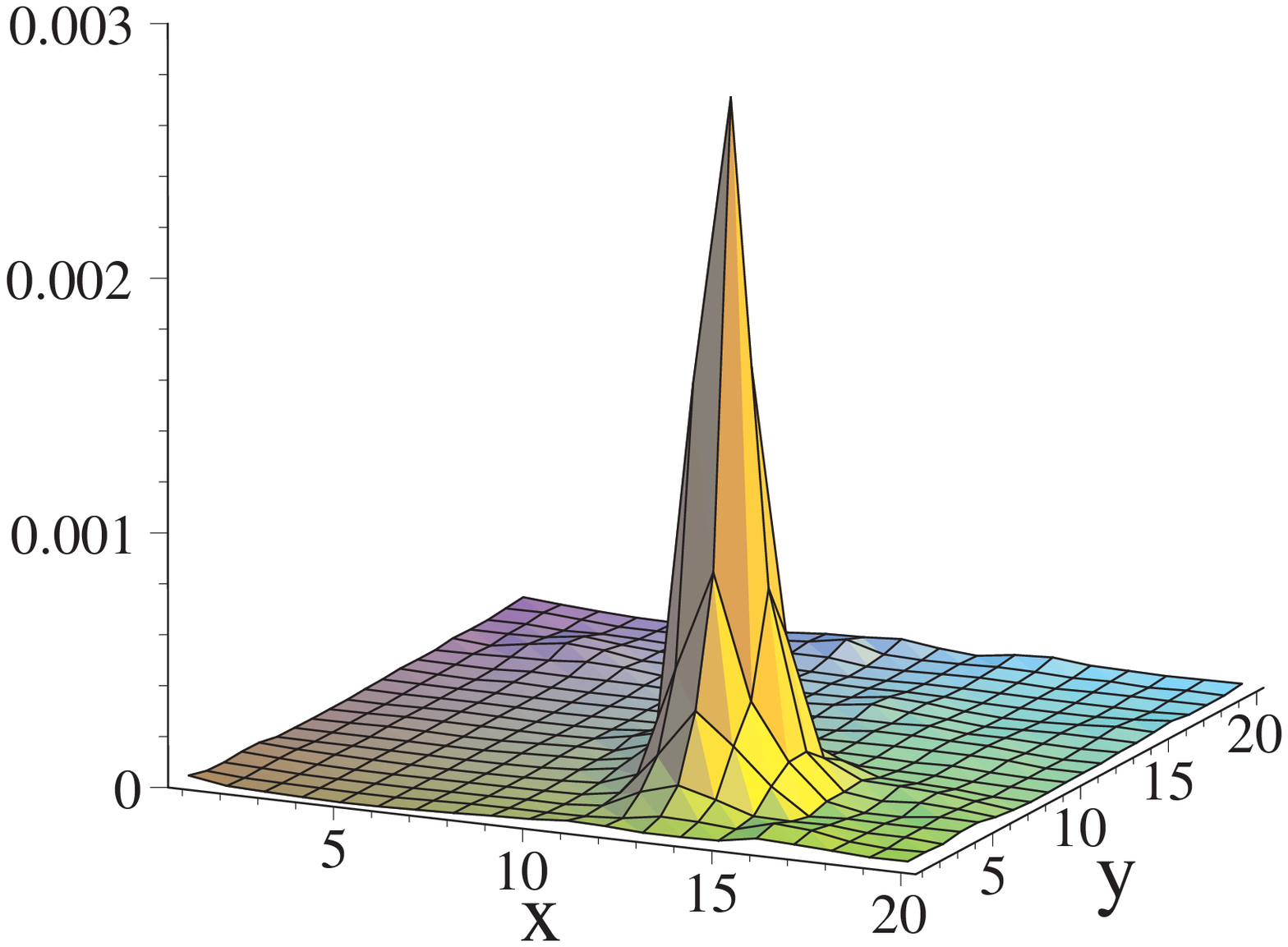,height=3.0cm,clip}
\epsfig{file=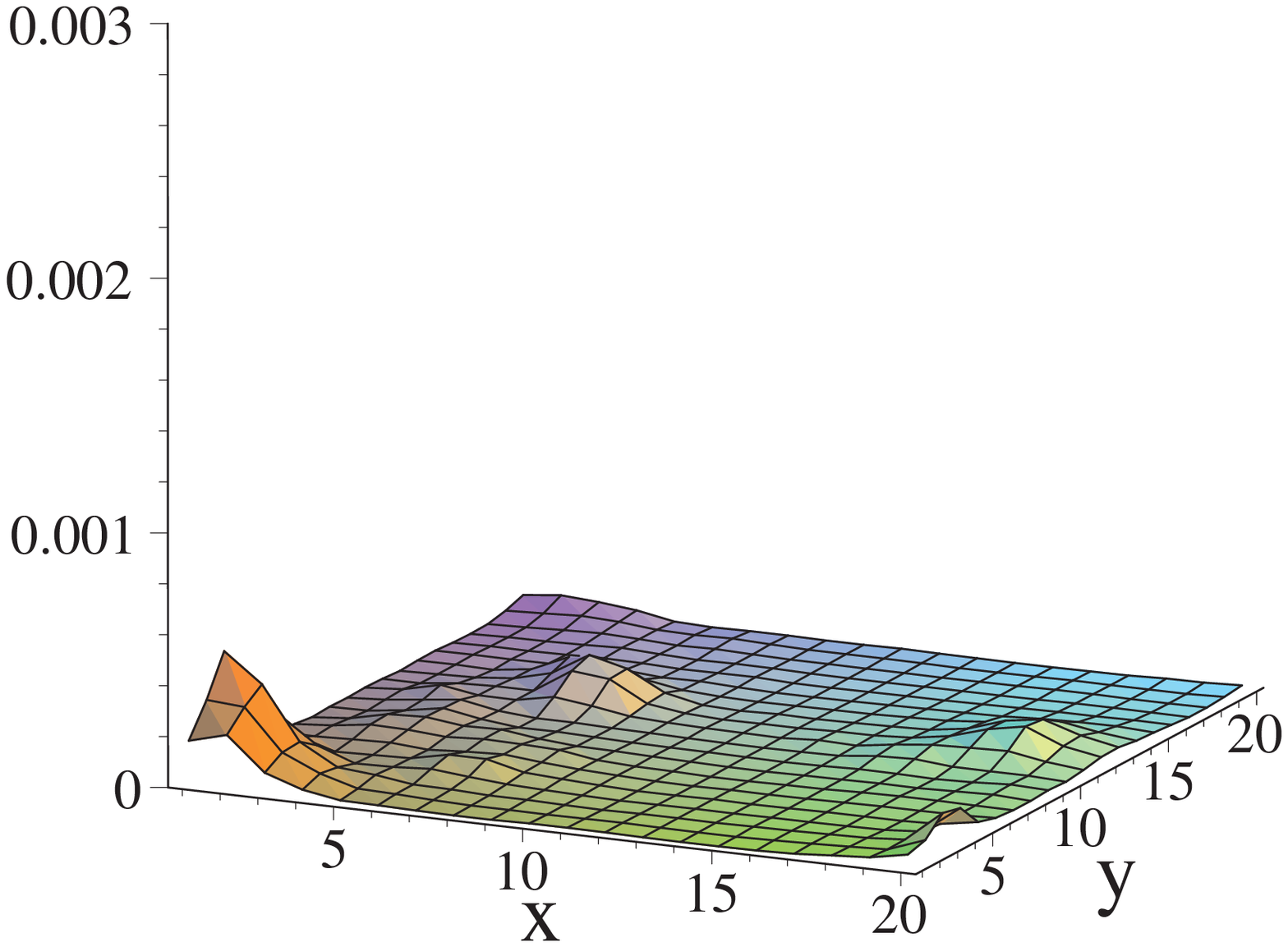,height=3.0cm,clip}
\epsfig{file=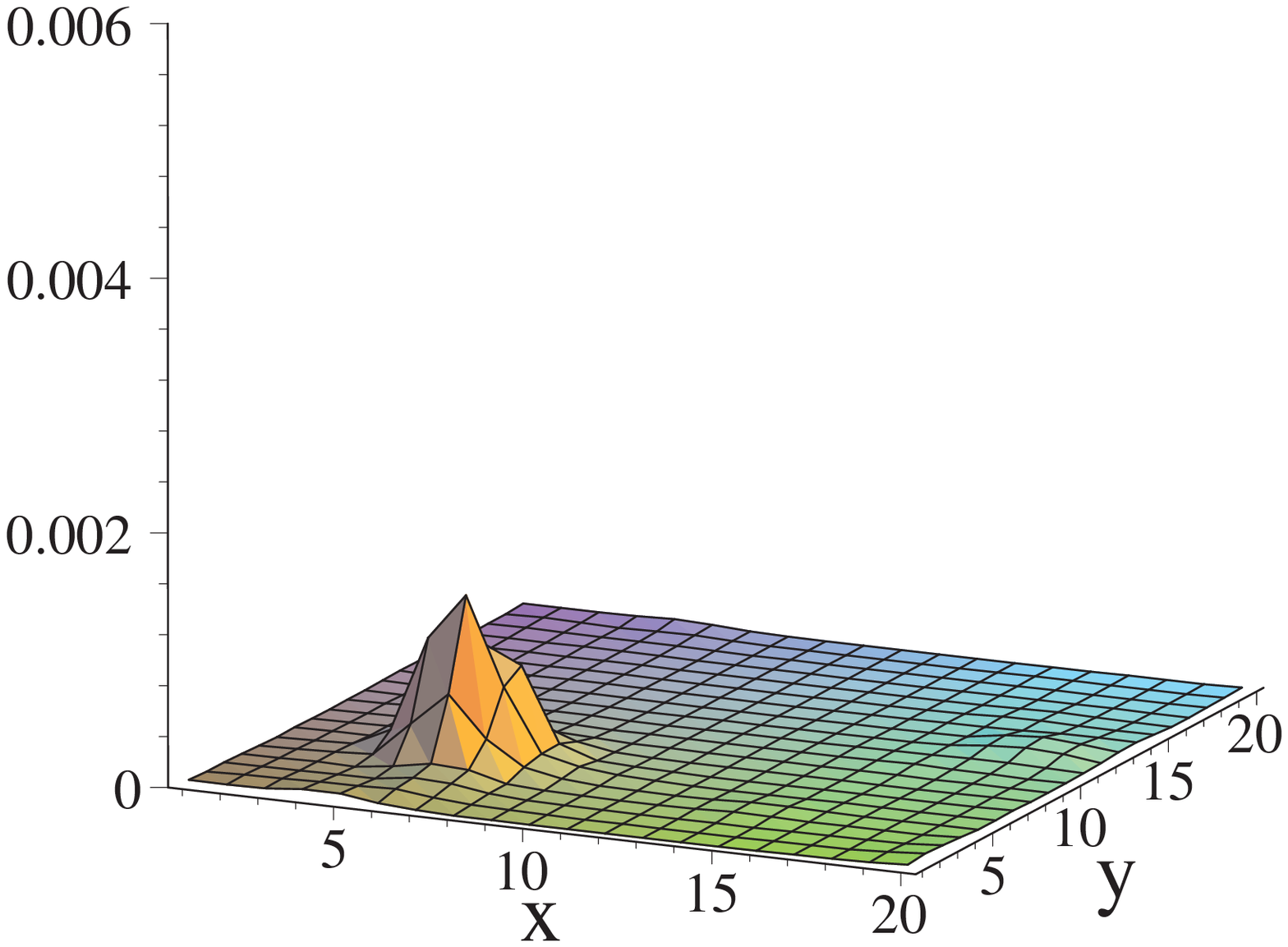,height=3.0cm,clip} 
\\
\epsfig{file=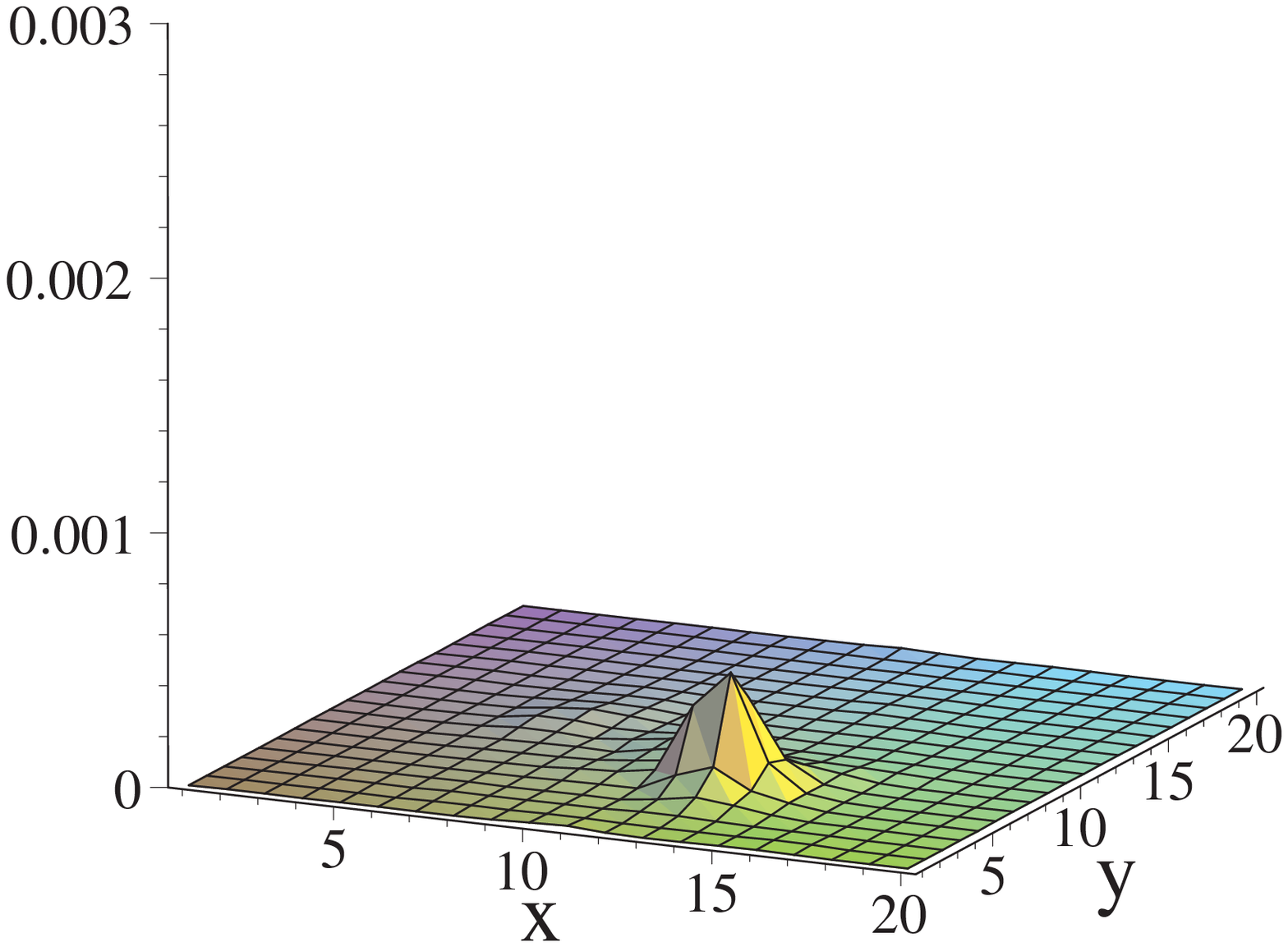,height=3.0cm,clip}
\epsfig{file=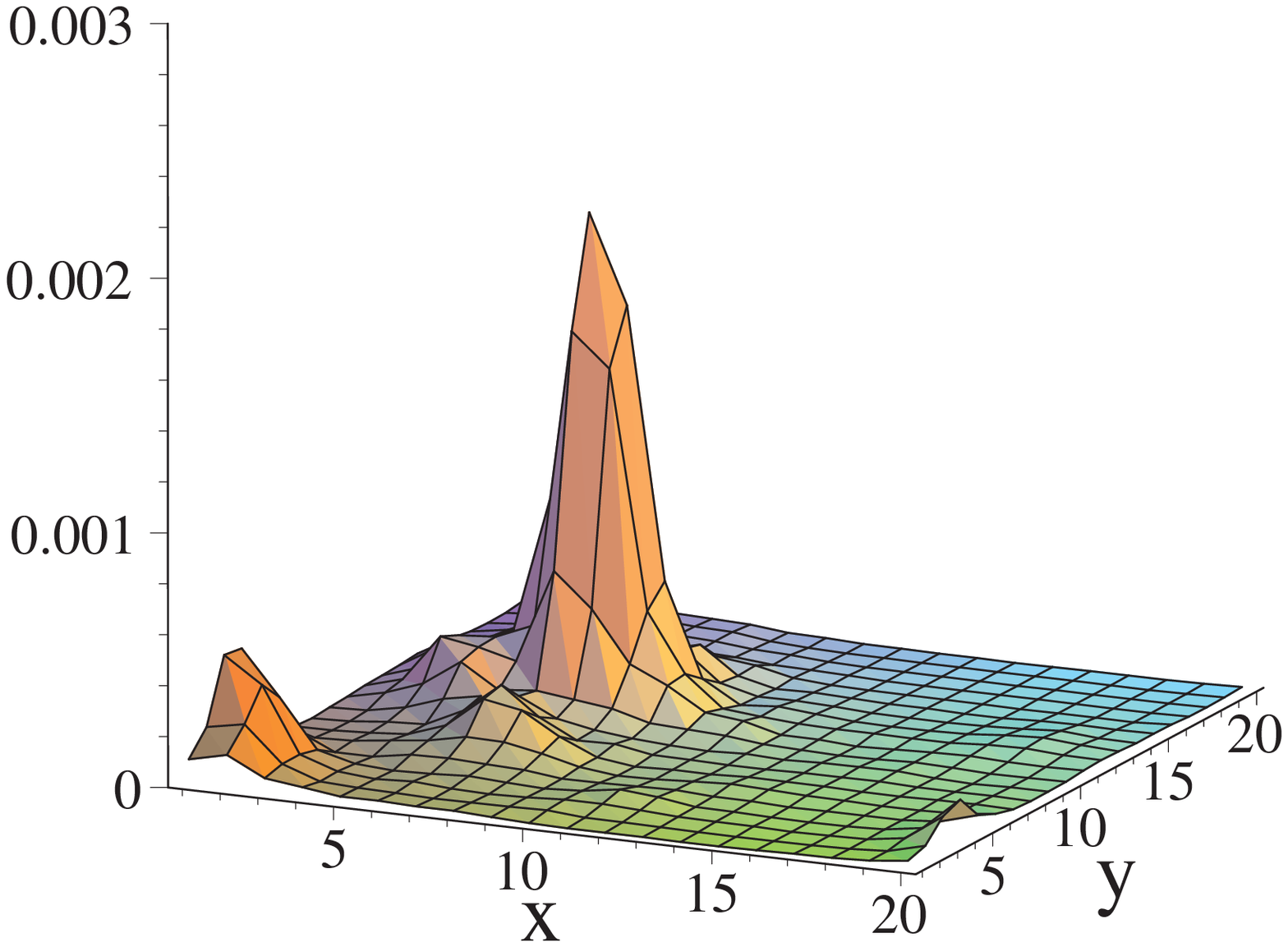,height=3.0cm,clip}
\epsfig{file=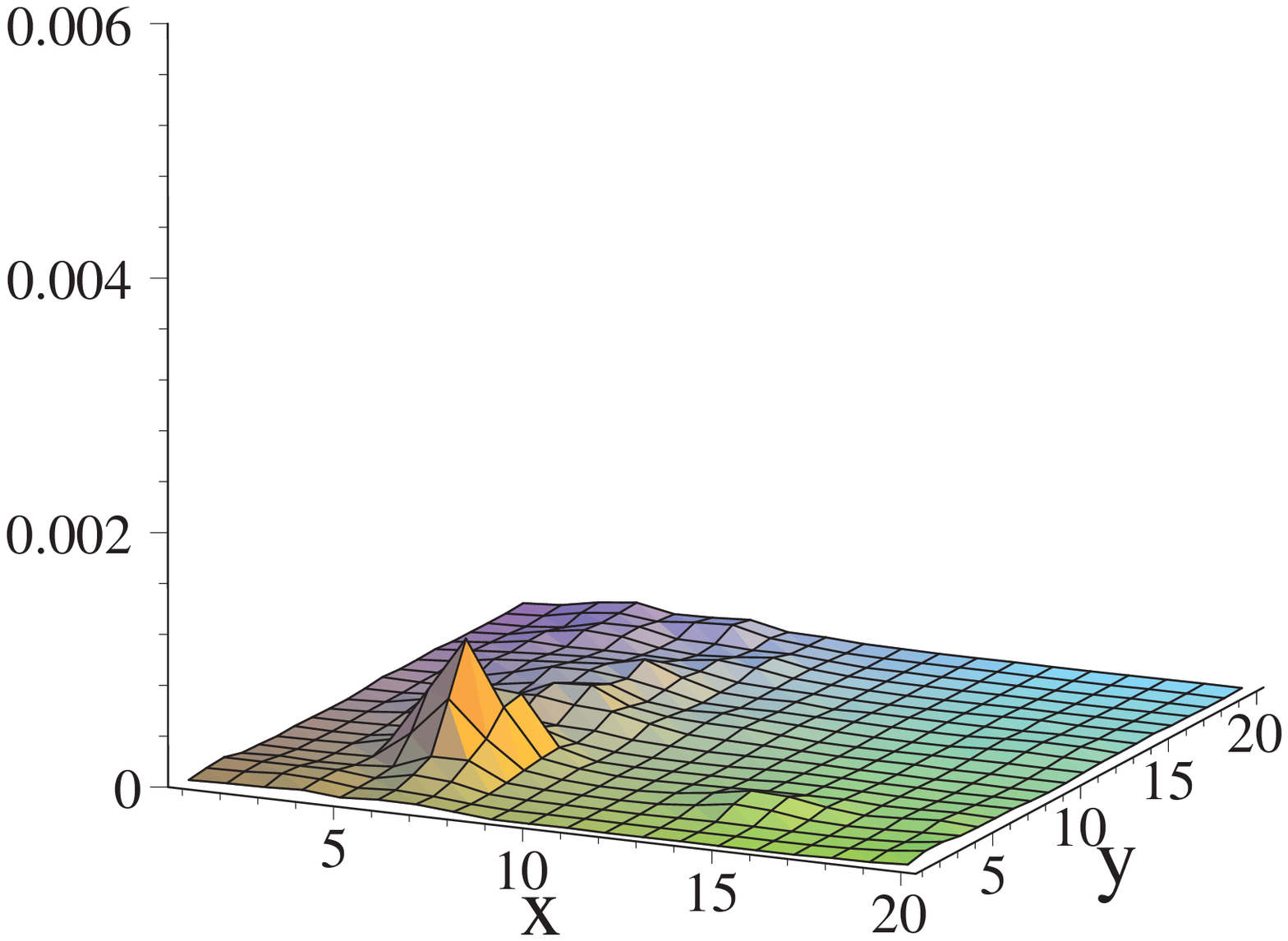,height=3.0cm,clip} 
\\
\epsfig{file=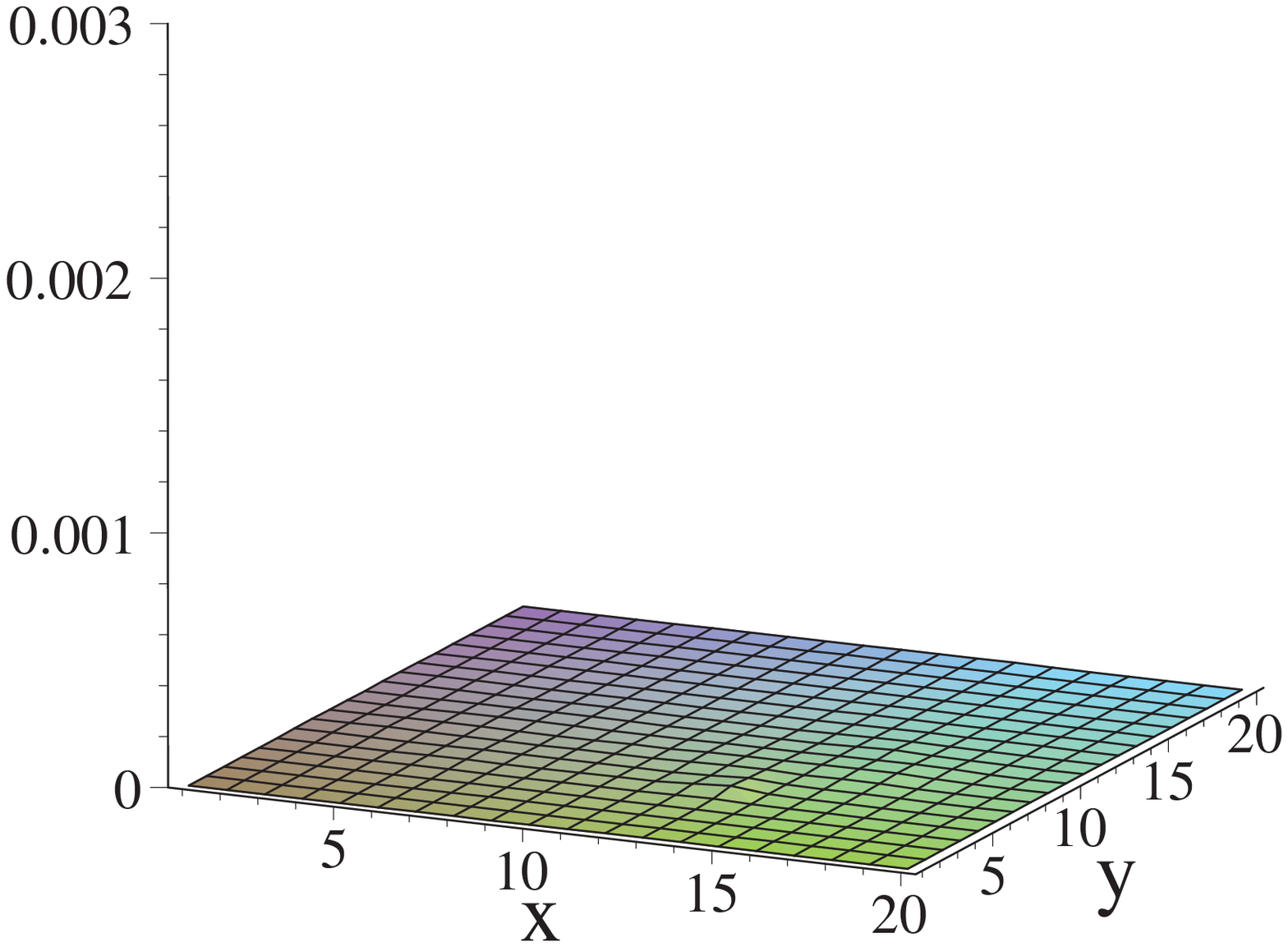,height=3.0cm,clip}
\epsfig{file=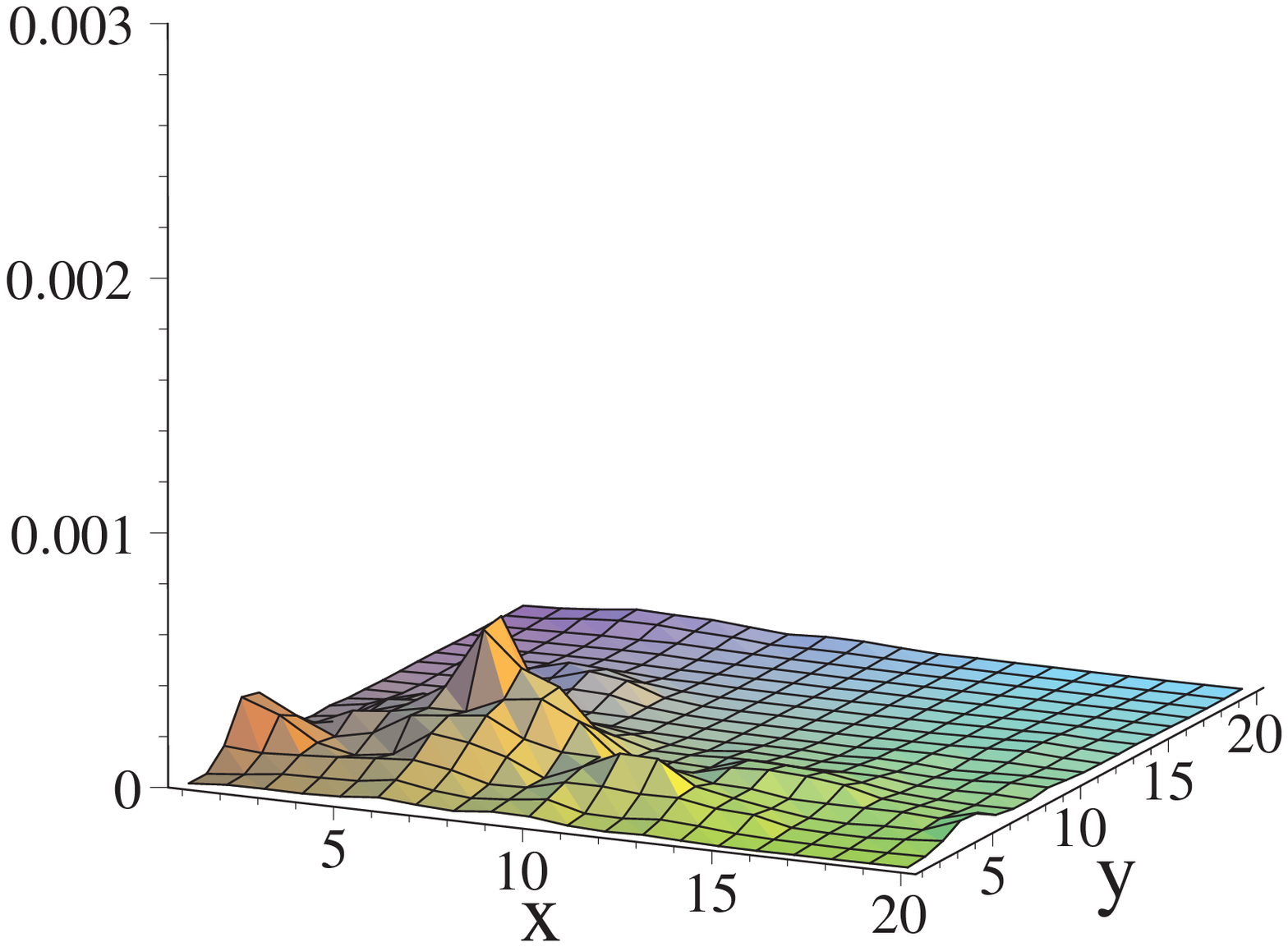,height=3.0cm,clip}
\epsfig{file=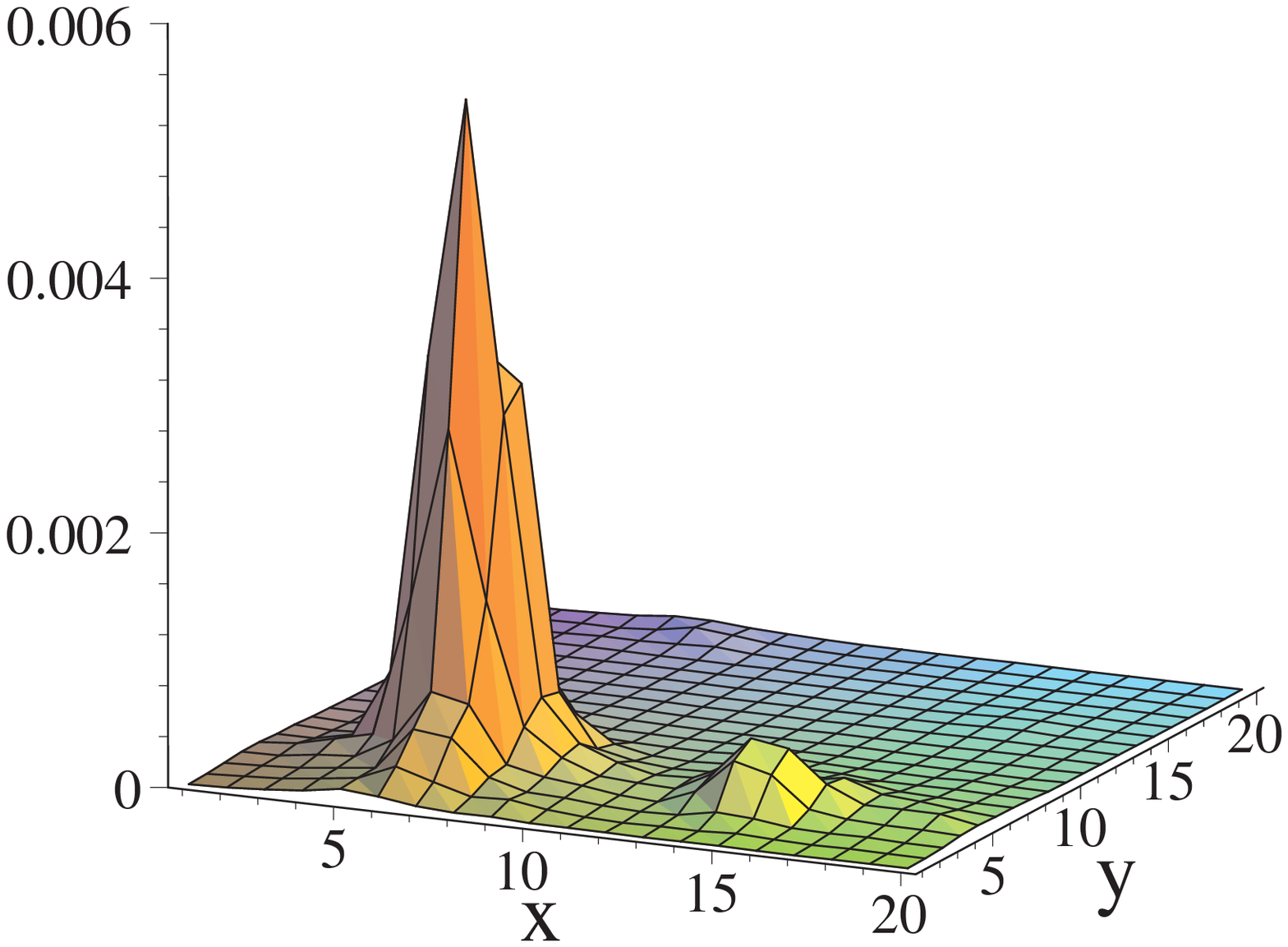,height=3.0cm,clip} 
\\
\caption{
Zero mode for a quenched configuration at finite temperature 
from \protect\cite{gattringer:02b}: 
the profile $|\psi_0(x)|^2$ for different boundary conditions, top to bottom, 
is localised in different lattice planes, left to right.
}
\label{fig_zero_thermalised}
\end{center}
\end{figure}

Indeed, it has been found by the director of this school that the zero modes at finite temperature 
hop with the phase boundary conditions in the compact direction just like the zero modes for calorons do
\cite{gattringer:02b}, see Fig.~\ref{fig_zero_thermalised}.

Another approach tries to decide on the dimensionality of the underlying gluonic structures
by investigating the scaling of the zero mode profile (its `inverse participation ratio') 
with the lattice spacing $a$, but till now is inconclusive \cite{aubin:04,*gubarev:05a,*deforcrand:06}.

As there is no strict topology on the lattice 
(unless the configurations are smooth, e.g.\ after cooling), 
one can define a topological charge through the index theorem as $Q_{\rm ferm}\equiv n_L-n_R$.
Such a prescription exists even locally \cite{niedermayer:98},
\begin{equation}
q_{\rm ferm}(x)\equiv\tr\, \gamma_5(\frac{1}{2}D_{x,x}-1)
=\sum_{n=1}^{4N_c\cdot{\rm Vol}} (\frac{\lambda_n}{2}-1)\psi^\dagger_n(x)\gamma_5\psi_n(x)\,,
\label{eqn_q_ferm}
\end{equation}
with $D_{x,x}$ the lattice Dirac operator and $\psi_n(x)$ its eigenmodes.

This brings me to the concept of a {\em spectral filter}. 
Truncating a sum like Eq.~(\ref{eqn_q_ferm}) at a small number $n$ of modes 
should remove short scale fluctuations, just like a Fourier transform.
With this method evidence for three-dimensional topological structures has been found \cite{horvath:03a}.
A later study revealed other dimensionalities of topological objects
at high cut-offs in the topological charge density \cite{ilgenfritz:07a}.

A similar filtering method has been given for the links themselves, 
based on a truncation in the eigenmodes of the lattice Laplace operator \cite{bruckmann:05b}
(and very recently also for the Polyakov loop in terms of fermionic modes 
\cite{gattringer:06b,*bruckmann:06b,*synatschke:07a}).

\begin{figure*}[h]
\begin{center}
\begin{minipage}{0.24\linewidth}
\includegraphics[width=\linewidth]{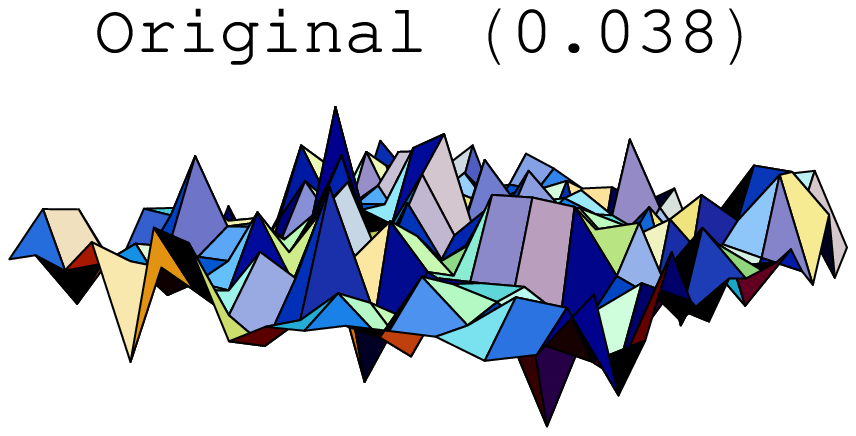}
\includegraphics[width=\linewidth]{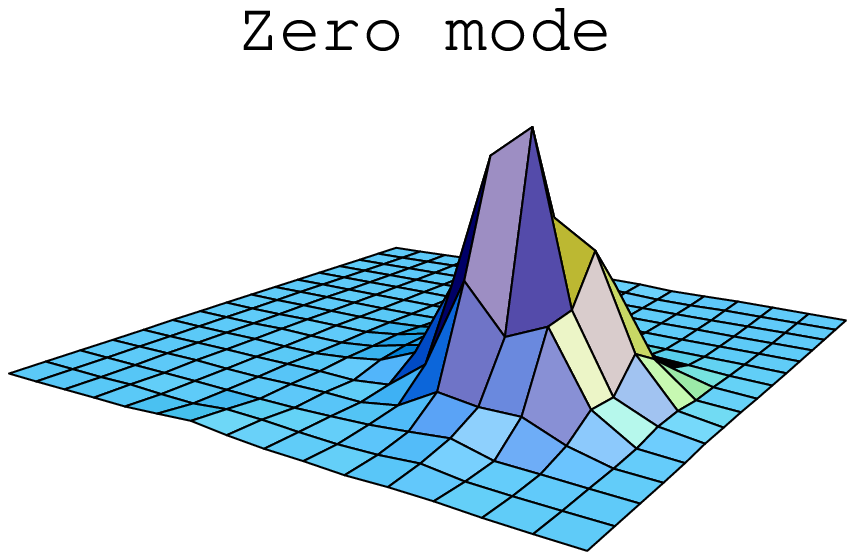}
\end{minipage}
\put(0.0,-66.0){\line(0,1){135}}
\begin{minipage}{0.72\linewidth}
\includegraphics[width=0.32\linewidth]{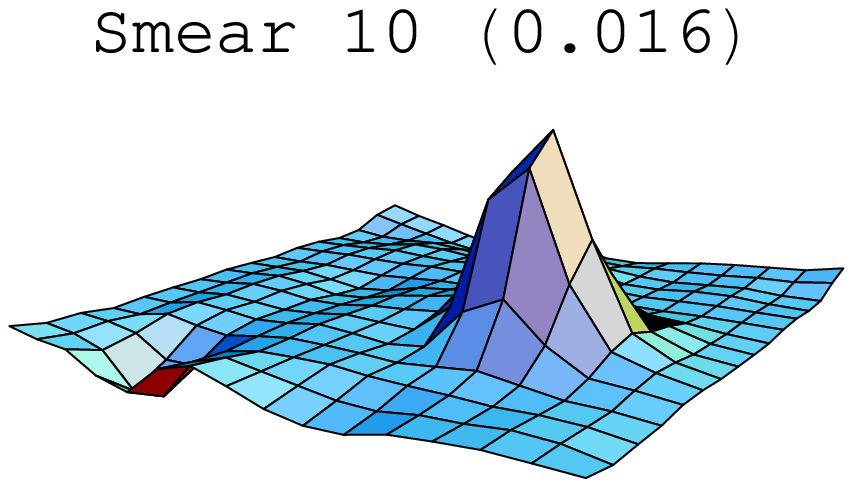}
\includegraphics[width=0.32\linewidth]{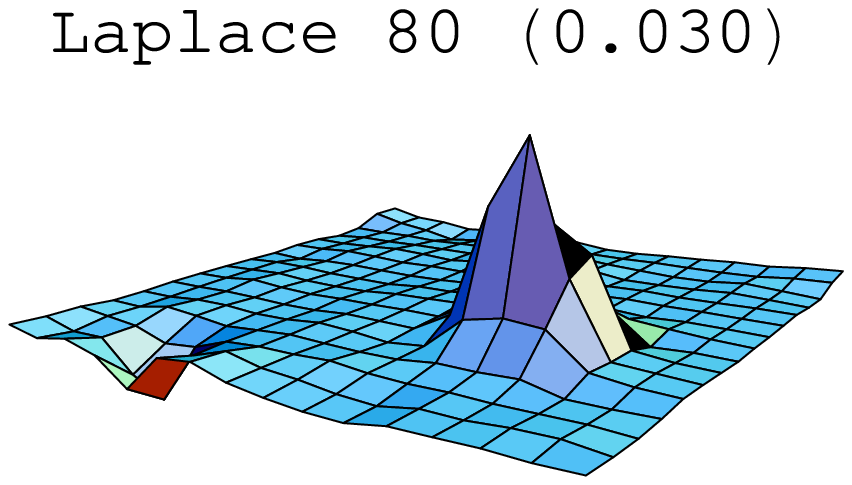}
\includegraphics[width=0.32\linewidth]{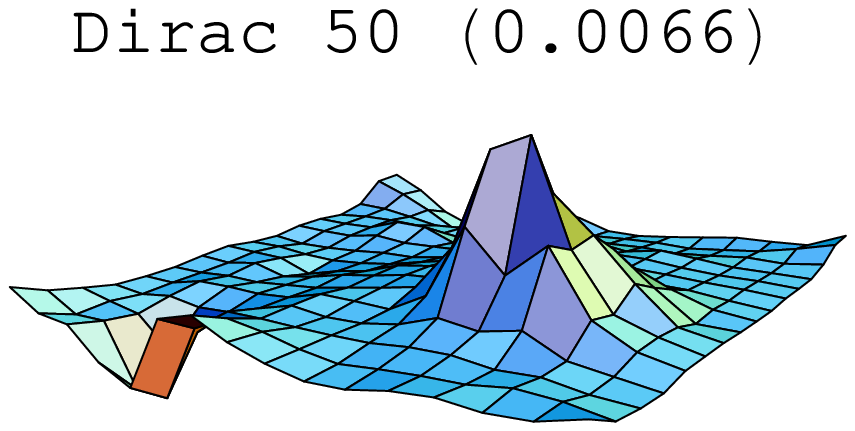}
\\

\includegraphics[width=0.32\linewidth]{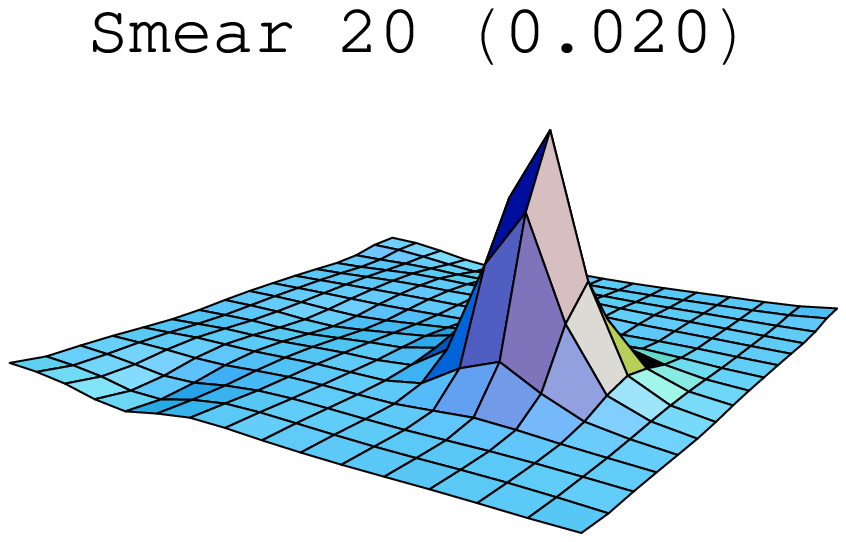}
\includegraphics[width=0.32\linewidth]{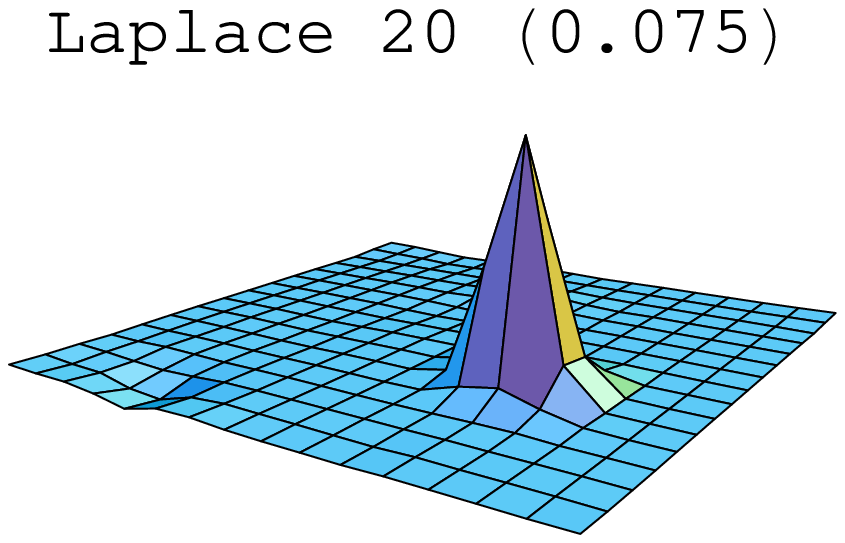}
\includegraphics[width=0.32\linewidth]{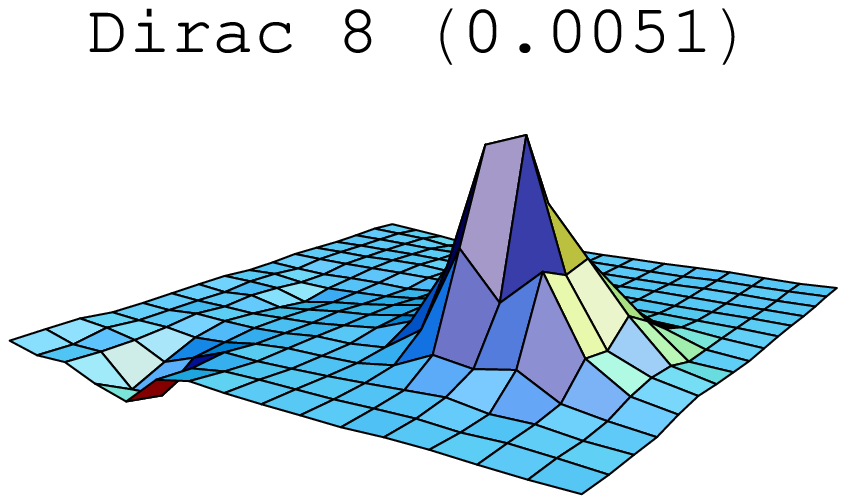}
\end{minipage}
\caption{
Effect of different filtering methods on the topological density 
for a 
$Q=1$ configuration in a fixed lattice plane, from \protect\cite{bruckmann:06a}. 
On the far left 
is shown 
the original topological density (top) 
and the profile of the chiral zero mode (bottom), to be compared with
the improved gluonic topological density after smearing 
and Laplace filtering 
and the topological density in terms of Dirac eigenmodes, Eq.~(\protect\ref{eqn_q_ferm}).
Results corresponding to mild and strong filtering are shown in the top and bottom row,
respectively.
} 
\label{fig_filters}
\end{center}
\end{figure*}

The local procedures like cooling/smearing and the spectral filters are of quite different nature 
and have their own ambiguities (like the choice of the specific parameters).
Therefore, it is an important finding 
that these methods do see the same structures of topological charge \cite{bruckmann:06a}.
As Fig.~\ref{fig_filters} shows this agreement holds at different levels of filtering.
It should help to finally identify the relevant degrees of freedom uniquely.

\section{Summary and acknowledgments}

In these lecture I explained the properties of topological objects 
and the physical mechanisms caused by them, 
from the rather pedagogical example of the kink
to some of the latest research results in QCD.
I tried to concentrate on the main mechanisms skipping many details and subtleties 
(as can be seen from the extensive use of parentheses\footnote{and footnotes}).

Although several infrared aspects of QCD are fairly well understood thanks to instantons, monopoles or vortices, 
I will not give extensive conclusions here.
Rather the QCD vacuum remains an interesting subject of investigations
(also in view of the QCD phase diagram discussed in Owe Philipsen's lecture) 
and we still do not know precisely, who is pulling the strings in this strongly interacting system. Stay tuned!\\

I would like to thank the organisers for inviting me to this very nice Winter school
and I wish them success for future schools.
I am also grateful to my collaborators over the years, in particular to Pierre van Baal, 
for many discussions on the subject. 
My work is supported by DFG (BR 2872/4-1).


\end{document}